\documentclass[10pt,letterpaper]{article}

\usepackage[top=0.85in,left=2.75in,footskip=0.75in]{geometry}
\usepackage{amsmath,amssymb}

\usepackage{changepage}
\usepackage[utf8x]{inputenc}

\usepackage{textcomp,marvosym}
\usepackage{url}

\usepackage{nameref}
\usepackage[table]{xcolor}
\usepackage[pdftex,colorlinks=true,linkcolor=blue,allcolors=blue, citecolor=blue,urlcolor=blue, allbordercolors={0 0 0}, pdfborderstyle={/S/U/W 1}]{hyperref}          

\makeatother
\usepackage[right]{lineno}

\usepackage{microtype}
\DisableLigatures[f]{encoding = *, family = * }

\usepackage{array}

\newcolumntype{+}{!{\vrule width 2pt}}

\newlength\savedwidth

\raggedright
\usepackage[aboveskip=1pt,labelfont=bf,labelsep=period,justification=raggedright,singlelinecheck=off]{caption}

\makeatletter
\renewcommand{\@biblabel}[1]{\quad#1.}
\makeatother

\usepackage{lastpage,fancyhdr,graphicx}
\usepackage{epstopdf}
\fancyheadoffset[L]{2.25in}
\fancyfootoffset[L]{2.25in}
\usepackage{mathtools}
\usepackage{amsmath,amsthm,amssymb} 
\usepackage{adjustbox}
\usepackage{array}
\usepackage{subcaption}

\usepackage{enumerate}
\usepackage{graphicx}
\usepackage{booktabs}
\usepackage{tikz}

\usepackage{authblk}
\usepackage{multirow}

\newcommand*{\figref}[2][]{%
  \hyperref[{fig:#2}]{%
    Fig~\ref*{fig:#2}%
    \ifx\\#1\\%
    \else
      \,#1%
    \fi
  }%
}

\newcommand*{\tableref}[2][]{%
  \hyperref[{tab:#2}]{%
    Table~\ref*{tab:#2}%
    \ifx\\#1\\%
    \else
      \,#1%
    \fi
  }%
}

\newcolumntype{M}[1]{>{\centering\arraybackslash}m{#1}}

\newcommand{\E}{\mathbb{E}}

\newcommand{\Reff}{R_{\rm eff}}

\newtheorem{theorem}{Theorem}

\date{\today}

\newcommand{\weq}{\ = \ }

\begin{document}

\vspace*{0.2in}

\begin{flushleft}
{\Large
\textbf\newline{Adaptive and optimized COVID-19 vaccination strategies across geographical regions and age groups} 
}
\newline
\\
Jeta Molla\textsuperscript{1,*},
Alejandro Ponce de Le\'on Ch\'avez\textsuperscript{2},
Takayuki Hiraoka\textsuperscript{3},
Tapio Ala-Nissila\textsuperscript{4,5},
Mikko Kivel\"a\textsuperscript{3},
Lasse Leskel\"a\textsuperscript{2}
\\
\bigskip
\textbf{1} Department of Applied Physics, Aalto University, Otakaari 1, 02150 Espoo, Finland
\\
\textbf{2} Department of Mathematics and Systems Analysis, Aalto University, Otakaari 1, 02150 Espoo, Finland
\\
\textbf{3} Department of Computer Science, Aalto University, Konemiehentie 2, 02150 Espoo, Finland
\\
\textbf{4} Quantum Technology Finland Center of Excellence and Department of Applied Physics, Aalto University, Otakaari 1, 02150 Espoo, Finland
\\
\textbf{5} Interdisciplinary Centre for Mathematical Modelling and Department of Mathematical Sciences, Loughborough University, Loughborough, Leicestershire LE11 3TU, United Kingdom
\bigskip

* jeta.0.molla@aalto.fi

\end{flushleft}

\section*{Abstract}
We evaluate the efficiency of various heuristic
strategies for allocating vaccines against COVID-19 and compare them to strategies found using optimal control theory. Our approach is based on a mathematical model which tracks the spread of disease among different age groups and across different geographical regions,
and we introduce a method to combine age-specific contact data to geographical movement data. 
As a case study, we model the epidemic in the population of mainland Finland
utilizing mobility data from a major telecom operator. 
Our approach allows
to determine which geographical regions and age groups should be targeted first in order to minimize the number of deaths. 
In the scenarios that we test, we find that distributing vaccines demographically and in an age-descending order is not optimal for minimizing deaths and the burden of disease.  Instead, more lives could
be saved by using strategies which emphasize high-incidence regions and distribute vaccines in parallel to multiple age groups. The level of emphasis that high-incidence regions should be given depends on the overall transmission rate in the population.
This observation highlights the importance of updating the vaccination strategy when the effective reproduction number changes due to the general contact patterns changing and new virus variants entering.


\section*{Author summary}
The COVID-19 vaccines are now available worldwide and many countries follow the practice of distributing them heuristically e.g. in age-descending order and demographically.  
Here we evaluate the effectiveness of such strategies by comparing them with optimized ones
from an age and spatially-structured mathematical model of COVID-19 transmission.  We find that vaccinating multiple age groups simultaneously and targeting regions with the the highest incidence can save more lives than heuristic strategies. Our work also reveals the importance of assessing the vaccination strategy at different stages of the epidemic.  


\section*{Introduction}

With reports of around three million deaths and 160 million cases worldwide \cite{covidstats}, the COVID-19 pandemic has caused a global public health crisis with far-reaching consequences to the economy and lives of people. Vaccines promise a way out of this situation, but due to limited supply and finite rate of vaccination they are not immediately effective in eradicating the epidemic. Health officials and governments around the world are thus faced with decisions on which order to vaccinate the population. This can be a matter of life and death to a large number of people and determine the speed at which we steer out of the crisis. The problem at hand is complicated by different mortality rates and activity levels in different age groups, localised incidence rates, and mobility patterns between regions, making it difficult to find an optimal solution on how to vaccinate using heuristic arguments. Given the scope of the crisis, even a small change in the relative efficiency of a strategy can have a large impact at the absolute scale in terms of saving lives. Therefore, critical evaluation on different vaccination strategies is imperative. 

Several studies have previously explored the effectiveness of different age-structured vaccination strategies against the COVID-19 \cite{Bubar_etal_2021, Goldenbogen2020, Goldstein2021, Matrajt2021, Matrajt_etal_2021,Moore_etal_2021, Wang2020}. Most of them agree that for minimizing cumulative incidence, i.e., the number of individuals who experience infection by the end of the epidemic, it is optimal to give priority to younger generations, as their higher activity accounts for a large part of the transmission. However, if the minimization of deaths and hospitalizations is targeted, it is often preferable to allocate vaccines first to the elderly who have a higher risk of severe illness and death. 
%
The set of strategies considered in the aforementioned studies is 
limited to sequential vaccinations of different age groups. 
They do not take into account parallel vaccination across age groups nor other factors such as the 
mobility and contact patterns of individuals.
Further,  suitable geographical distribution of vaccines is important especially when prevalence is inhomogeneously distributed across different geographical regions. Bertsimas et al. \cite{bertsimas2020optimizing} and Grauer et al. \cite{grauer2020strategic} have shown that allocating  vaccines to regions with high incidence 
can reduce the number of deaths compared to the strategy of distributing vaccines demographically.Further, Lemaitre et al. \cite{lemaitre2021optimizing} have studied optimal spatial allocation of COVID-19 vaccines via an optimal control framework taking into account the mobility network and the
spatial heterogeneities.
Ideally, all aforementioned factors should be optimized simultaneously, but
once we start to take into account such parallel and region-based prioritization strategies, the space of possible strategies becomes so large that a  brute-force search 
for an optimal strategy 
is no longer feasible; hence we need an efficient algorithm for finding a strategy that optimizes the given objective function.


To this end, we here construct an epidemic model that takes into account the various factors mentioned above. We use the model to study the effectiveness of different vaccination strategies by nonlinear optimization methods. The epidemic progression is described by a deterministic compartmental model adapted to COVID-19. As a case study, we adjust the model parameters to the recent epidemic situation on mainland Finland. Based on census data, age-structured contact patterns, and mobility patterns from a mobile phone operator, we infer contact patterns between individuals in different regions and age groups. Based on the available data of reported cases and vaccination counts, the performance of several vaccination strategies that are implemented or considered by health authorities is evaluated by means of a nonlinear programming framework. This framework allows us to optimize age-based and region-based vaccination schedules. As our main result, we find that the heuristic strategy of vaccinating the high-risk groups serially and distributing vaccines uniformly based on the local population density may not be optimal in minimizing deaths and mitigating the disease burden. Instead, better results can be obtained by parallel vaccination of different age groups and geographically targeted distribution of vaccines in a way that adapts to the ongoing incidence over time and takes into account demographic and behavioral differences across different regions.
This calls for re-evaluation of the details of any chosen vaccination strategy during the course of vaccinating the population.

\section*{Methods}
\label{sec:Methods}

The level of detail in modelling epidemic spreading dynamics depends both on the questions that need to be answered and the availability of relevant data. One of the characteristic features of the COVID-19 epidemic is the large heterogeneity in mortality across different age groups. For evaluating vaccination strategies, we also need to include the initial state of the epidemic at a given time, the arrival rate of new vaccine doses and their efficacy, and contact patterns between individuals of different ages for transmission rates. The final complication comes from geographic heterogeneity which requires local population densities and accurate mobility data between different regions.


\subsection*{Region- and age-based epidemiological model}

We introduce a deterministic compartmental model of COVID-19 transmission and vaccination which takes into account both heterogeneities across age groups and mobility across geographical regions. 
 We assume that new vaccine doses arrive at a constant rate and all types of vaccines have equal efficacy.   We consider an extension of the all-or-nothing model \cite{Bubar_etal_2021,Sjodin_Rocklov_Britton_2021-04} in order to take into account
 individual variations in immunity.
To this end some vaccinated individuals develop full immunity while others have only partial protection against transmission and severe illness  after receiving the first dose.
The proportion of individuals accepting to be vaccinated is assumed to be constant across the population. 
 
 The population of a country is modelled as a closed system of $N$ individuals, divided into regions  $k=1,\dots, K$ and age groups $g=1,\dots, G$. An individual resident in  region $k$ in age group $g$ is called a $kg$-individual, and the number of such individuals is denoted by $N_{kg}$. In what follows, $g,h$ always refer to age, and $k,\ell,m$ to regions. 
The population size of region $k$ is denoted by $N_k = \sum_g N_{kg}$, and 
the size of age group $g$ by $N_g = \sum_k N_{kg}$. 

 The population in each stratum is divided into 16 time-dependent epidemiological compartments described in \tableref{Compartments}. 

\begin{table}[ht]

\caption{\label{tab:Compartments} { \bf Epidemiological compartments.} There are $KG$ copies of each compartment, denoted $S^u_{kg}, S^v_{kg}, \dots, V_{kg}$ for regions $k=1,\dots,K$ and age groups $g=1,\dots,G$.}
\centering
\small
\begin{tabular}{ll}
\toprule
\bf Symbol & \bf Description \\
\midrule
$S^u$ & Susceptible, unvaccinated \\
$S^v$ & Susceptible, invited for vaccination \\
$S^x$ & Susceptible, unable or unwilling to be vaccinated \\
$S^p$ & Susceptible, developed weak immunity after vaccination \\
$E$ & Infected but not yet infectious \\
$E^v$ & Vaccinated infected but not yet infectious \\ 
$I$ & Infected and infectious \\
$I^v$ & Vaccinated infected and infectious\\
$Q^0$ & Quarantined at home, mild disease \\
$Q^1$ & Quarantined at home, severe disease \\
$H^w$ & Hospitalized, in general ward \\
$H^c$ & Hospitalized, in critical care \\
$H^r$ & Hospitalized, in recovery ward \\
$D$   & Deceased \\
$R$   & Recovered with full immunity \\
$V$   & Vaccinated with full immunity \\
\bottomrule
\end{tabular}
\end{table}
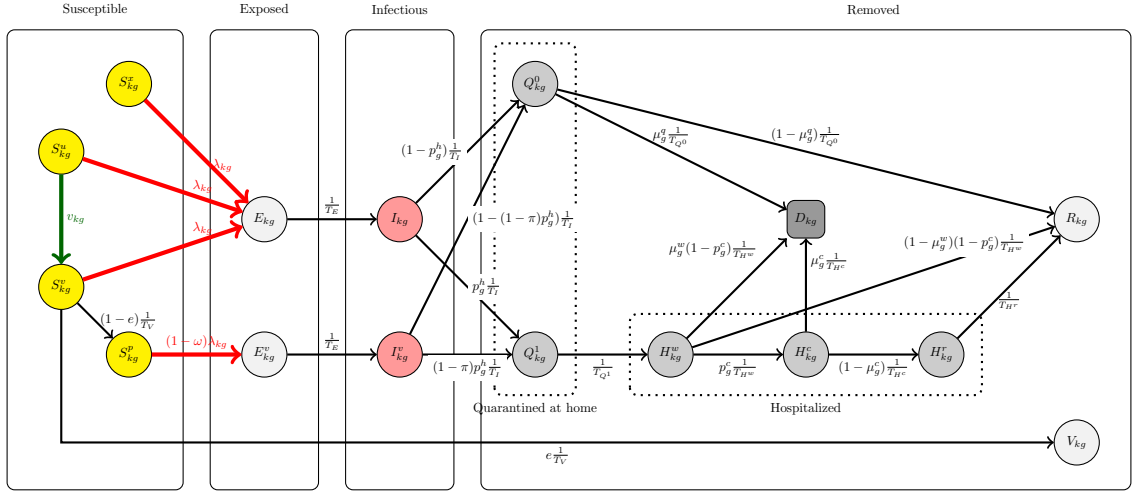
\begin{figure}[ht]
\begin{adjustwidth}{-2.25in}{0in} 
\centering
\begin{tikzpicture}
[
scale=0.9,
every node/.style={scale=0.5},
snode/.style={draw, circle, fill=yellow, minimum size=12mm}, 
enode/.style={draw, circle, fill=black!05, minimum size=12mm}, 
inode/.style={draw, circle, fill=red!40, minimum size=12mm}, 
hnode/.style={draw, circle, fill=black!05, minimum size=12mm}, 
rnode/.style={draw, circle, fill=black!20, minimum size=12mm}, 
dnode/.style={draw, rectangle, fill=black!40, minimum size=10mm, rounded corners=1mm}, 
bbox/.style={black, rounded corners=1mm},
sbox/.style={black, thick, dotted, rounded corners=1mm},
psi/.style = {->, draw, thick, color=black},
vac/.style = {->, draw, ultra thick, color=black!60!green}, 
inf/.style = {->, draw, ultra thick, color=red}, 
]
\newcommand{\pd}{0.8}
\newcommand{\pds}{0.6}

\newcommand{\toff}{0.8}
\newcommand{\boff}{1.0}
\newcommand{\loff}{0.8}
\newcommand{\roff}{0.8}
\newcommand{\loffs}{0.6}
\newcommand{\roffs}{0.6}

\draw[bbox] (-1-\loff, -1-\boff) rectangle (0+\roff, 4+\toff);
\draw[bbox] (2-\loff, -1-\boff)  rectangle (2+\roff, 4+\toff);
\draw[bbox] (4-\loff, -1-\boff)  rectangle (4+\roff ,4+\toff);
\draw[bbox] (6-\loff, -1-\boff)  rectangle (14+\roff, 4+\toff);
\draw[sbox] (6-\loffs, 0-\pds) rectangle (6+\roffs, 4+\pds);
\draw[sbox] (8-\loffs, 0-\pds) rectangle (12+\roffs, 0+\pds);

\node at (-0.5, 5.1) {Susceptible};
\node at (2, 5.1) {Exposed};
\node at (4, 5.1) {Infectious};
\node at (11, 5.1) {Removed};
\node at (6, -0.8) {Quarantined at home};
\node at (10, -0.8) {Hospitalized};

\node[snode] at (0, 4)  (SX) {$S^x_{kg}$};
\node[snode] at (-1, 3) (SU) {$S^u_{kg}$};
\node[snode] at (-1, 1) (SV) {$S^v_{kg}$};
\node[snode] at (0, 0)  (SP) {$S^p_{kg}$};

\node[enode] at (2, 2) (E) {$E_{kg}$};
\node[enode] at (2,0)  (E2) {$E^v_{kg}$};
\node[inode] at (4, 2) (I) {$I_{kg}$};
\node[inode] at (4, 0) (I2) {$I_{kg}^v$};

\node[rnode] at (6, 4)     (Q0) {$Q^0_{kg}$};
\node[rnode] at (6, 0)     (Q1) {$Q^1_{kg}$};
\node[rnode] at (8, 0)     (HW) {$H^w_{kg}$};
\node[rnode] at (10, 0)    (HC) {$H^c_{kg}$};
\node[rnode] at (12, 0)    (HR) {$H^r_{kg}$};
\node[hnode] at (14, 2)    (R)  {$R_{kg}$};
\node[hnode] at (14, -1.3) (V)  {$V_{kg}$};
\node[dnode] at (10, 2)    (D)  {$D_{kg}$};

\draw[vac] (SU) -> node [midway, right] {$v_{kg}$} (SV);
\draw[psi] (SV) -> node [midway, right] {$(1- e) \frac{1}{T_{V}}$} (SP);

\draw[inf] (SU) -> node [near end, above] {$\lambda_{kg}$} (E);
\draw[inf] (SV) -> node [near end, above] {$\lambda_{kg}$} (E);
\draw[inf] (SX) -> node [near end, above] {$\lambda_{kg}$} (E);
\draw[inf] (SP) -> node [above] {$ (1-\omega) \lambda_{kg}$} (E2);

\draw[psi] (E) -> node [midway, above] {$\frac{1}{T_E}$} (I);
\draw[psi] (E2) -> node [midway, above] {$\frac{1}{T_E}$} (I2);

\draw[psi] (I) -> node [midway, left, fill=white] {$(1-p^h_g) \frac{1}{T_I}$}  (Q0);
\draw[psi] (I) -> node [midway, right, fill=white] {$p^h_g \frac{1}{T_I} $}     (Q1);
\draw[psi] (I2) -> node [midway, right, fill=white] {$(1-(1-\pi)p^h_g) \frac{1}{T_I}$}  (Q0);
\draw[psi] (I2) -> node [midway, below, fill=white] {$(1-\pi) p^h_g \frac{1}{T_I} $}     (Q1);
\draw[psi] (Q0) -> node [midway, above] {$(1-\mu^q_g) \frac{1}{T_{Q^0}}$} (R);
\draw[psi] (Q1) -> node [midway, below] {$\frac{1}{T_{Q^1}}$} (HW);
\draw[psi] (HW) -> node [midway, below] {$p^c_g \frac{1}{T_{H^w}}$} (HC);
\draw[psi] (HW) -> node [near end, above, fill=white] {$(1-\mu^w_g) (1-p^c_g) \frac{1}{T_{H^w}}$} (R);
\draw[psi] (HC) -> node [midway, below] {$(1-\mu^c_g) \frac{1}{T_{H^c}}$} (HR);
\draw[psi] (HR) -> node [midway, below] {$\frac{1}{T_{H^r}}$} (R);
\draw[psi] (Q0) -> node [midway, above] {$\mu^q_g \frac{1}{T_{Q^0}}$} (D);
\draw[psi] (HW) -> node [near end, above left] {$\mu^w_g (1-p^c_g) \frac{1}{T_{H^w}}$} (D);
\draw[psi] (HC) -> node [near end, right] {$\mu^c_g \frac{1}{T_{H^c}}$} (D);

\draw[psi] (SV) |- node[pos=0.75, name=5, below] {$ e \frac{1}{T_V}$} (V);

\end{tikzpicture}
\caption{\label{fig:SEIRLabelsSimple} {\bf Disease transmission dynamics.} Each node in the diagram corresponds to one differential equation with the time derivative of the associated variable on the left side, the values of the source nodes of incident arrows on the right side, each incoming arrow equipped with a plus sign, and each outgoing arrow equipped with a minus sign. }
\end{adjustwidth}
\end{figure}

The dynamics of the disease is modelled using a deterministic nonlinear system of $16KG$ ordinary differential equations with structure shown in \figref{SEIRLabelsSimple}.  We treat the variables as expectation values, so they may take non-integer values.
This leads to a system where susceptible compartments evolve according to
\begin{equation}
\small
\label{eq:agespatial}
\begin{aligned}
 \frac{d}{dt} S^x_{kg}
 &\weq - \lambda_{kg} S^x_{kg},\\
 \frac{d}{dt} S^u_{kg}
 &\weq - \lambda_{kg} S^u_{kg} - v_{kg} S^u_{kg}, \\
 \frac{d}{dt} S^v_{kg}
 &\weq v_{kg} S^u_{kg} - \lambda_{kg} S^v_{kg} - \frac{1}{T_V} S^v_{kg} , \\
 \frac{d}{dt} S^p_{kg}
 &\weq (1- e)\frac{1}{T_{V}}S_{kg}^v - (1-\omega) \lambda_{kg}S_{kg}^x,
\end{aligned}
\end{equation}
infected but noninfectious compartments according to
\begin{equation}
\small
\label{eq:agespatialE}
\begin{aligned}
 \frac{d}{dt} E_{kg}
 &\weq \lambda_{kg}(S_{kg}^x + S_{kg}^u + S_{kg}^v) -\frac{1}{T_E}E_{kg},\\
 \frac{d}{dt} E_{kg}^v
 &\weq (1-\omega)\lambda_{kg}  S_{kg}^p -\frac{1}{T_E}E_{kg}^v,
\end{aligned}
\end{equation}
infectious compartments according to
\begin{equation}
\small
\label{eq:agespatialI}
\begin{aligned}
 \frac{d}{dt} I_{kg}
 &\weq \frac{1}{T_E} E_{kg} - \frac{1}{T_I} I_{kg},\\
 \frac{d}{dt} I_{kg}^v
 &\weq \frac{1}{T_E} E_{kg}^v - \frac{1}{T_I} I_{kg}^v
\end{aligned}
\end{equation}
and removed compartments according to
\begin{equation}
\small
\label{eq:agespatialR}
\begin{aligned}
 \frac{d}{dt} Q_{kg}^0
 &\weq (1-p_g^h)\frac{1}{T_I} I_{kg} + (1-(1-\pi) p_g^h)\frac{1}{T_I} I_{kg}^v - \frac{1}{T_{Q^0}} Q^0_{kg},\\
 \frac{d}{dt} Q_{kg}^1
 &\weq p_g^h\frac{1}{T_I} (I_{kg} + (1-\pi) I^v_{kg}) - \frac{1}{T_{Q^1}} Q^1_{kg},\\ 
 \frac{d}{dt} H_{kg}^w
 &\weq \frac{1}{T_{Q^1}} Q^1_{kg} - \frac{1}{T_{H^w}}H_{kg}^w,\\
 \frac{d}{dt} H_{kg}^c & \weq p_g^c \frac{1}{T_{H^w}} H_{kg}^w
   - \frac{1}{T_{H^c}} H_{kg}^c, \\ 
 \frac{d}{dt} H_{kg}^r
 &\weq (1-\mu_g^c) \frac{1}{T_{H^c}} H_{kg}^c - \frac{1}{T_{H^r}} H_{kg}^r, \\ 
 \frac{d}{dt} R_{kg}
 &\weq (1-\mu^q_g) \frac{1}{T_{Q^0}} Q^0_{kg}
    + (1-\mu^w_g) (1-p^c_g) \frac{1}{T_{H^w}} H^w_{kg}
    + \frac{1}{T_{H^r}} H^r_{kg}, \\
 \frac{d}{dt} D_{kg}
 &\weq \mu_g^q \frac{1}{T_{Q^0}} Q_{kg}^0 + \mu_g^w (1-p_g^c) \frac{1}{T_{H^w}} H_{kg}^w + \mu_g^c\frac{1}{T_{H^c}} H_{kg}^c, \\
 \frac{d}{dt} V_{kg}
 &\weq  e \frac{1}{T_V} S^v_{kg}.
\end{aligned}
\end{equation}
In formulae \eqref{eq:agespatial}--\eqref{eq:agespatialR}, the force of infection inflicted on $kg$ susceptibles $\lambda_{kg} = \lambda_{kg}(t)$ varies over time as a function of infectious states in all strata and additional parameters. The force of infection (per capita rate of infections) inflicted on susceptible $kg$ individuals equals 
\begin{equation}
 \label{eq:ForceOfInfectionDerived}   
 \lambda_{kg}(I)
 \weq \beta \sum_{m,\ell,h} \frac{\beta_{gh}}{\hat N_m} \theta_{km} (I_{\ell h} + I^v_{\ell h}) \theta_{\ell m},
\end{equation}
where $\beta$ is a constant used for adjusting the overall rate of infectious contacts, $(\beta_{gh})$ is a $9$-by-$9$ mobility-adjusted age contact matrix, $(\theta_{k\ell})$ is a $5$-by-$5$ baseline mobility, and $\hat N_m$ is the effective population size of region $m$. This corresponds to a model where $\beta\times  {\beta_{gh}}/{\hat N_m}$ is the contact rate between any unordered pair of individuals present in region $m$, with one individual belonging to age group $g$ and the other to age group $h$.

The per-capita rate of vaccines offered to residents of region $k$ in age group $g$ is a time-dependent function
$v_{kg} = v_{kg}(t)$
obtained as a solution of a minimization problem  or defined manually corresponding to vaccination strategies listed in \tableref{VaccinationStrategies}. The other model parameters are constant and are listed in \tableref{Parameters}.

\begin{table}[ht]
\begin{adjustwidth}{-1.25in}{0in} 
\centering
\caption{\label{tab:Parameters} {\bf Parameters for the epidemic model.} The parameters here have been taken from Ref. \cite{Sjodin_Rocklov_Britton_2021-04} except for the vaccine efficacy $e$ which depends on several factors including the vaccination type, disease variant, number of doses and time from the vaccination \cite{madhi2021efficacy,thompson2021interim,vasileiou2021effectiveness}. Here we set $e$ following Ref.~\cite{NEJM-vacc-efficacy}.}
\tiny
\begin{tabular}{cm{10em}rrrrrrrrr}
\toprule
\bf & \bf Description  & 0--9 & 10--19 & 20--29 & 30--39 & 40--49 & 50--59 & 60--69 & 70--79 & 80+ \\
\midrule
$T_E$  & Latent period(days) & 3 & 3  & 3 & 3 & 3 & 3 & 3 & 3 & 3 \\
$T_I$     & Transmission period (days) & 4 & 4 & 4 & 4 & 4 & 4 & 4 & 4 & 4 \\
$T_{Q^0}$ & Quarantine  period with mild symptoms (days) & 5 & 5 & 5& 5 & 5 & 5 & 5 & 5 & 5 \\
$T_{Q^1}$ & Quarantine period with severe symptoms (days) & 3 & 3 & 3 & 3 & 3& 3 & 3 & 3 & 3\\
$T_{H^w}$ & Hospital ward period (days) & 5 & 5 & 5 & 5 & 5 & 5 & 5 & 5 & 5 \\
$T_{H^c}$ & Critical care period (days) & 9 & 9 & 9 & 9 & 9& 9 & 9 & 9 & 9 \\
$T_{H^r}$ & Post-critical care period (days) & 1 & 1 & 1 & 1 & 1 & 1 & 1 & 1 & 1 \\
$T_V$     & Vaccination immunity delay (days) & 10 & 10 & 10 & 10 & 10 & 10 & 10& 10 & 10 \\
$p^h_g$   & Fraction of severe cases  & 0 & 0 & 0.02 & 0.03 & 0.04 & 0.08 & 0.16 & 0.43 & 0.52 \\
$p^c_g$   & Fraction of critical cases among severe & 0 & 0 & 0 & 0 & 0 & 0.01 & 0.03 & 0.05 & 0.01 \\[1ex]
$\mu^q_g$ & Fraction of non-hospitalized that die& 0 & 0 & 0 & 0 & 0 & 0 & 0 & 0.08 & 0.2 \\
$\mu^h_g$ & Fraction of hospitalized that die & 0 & 0 & 0 & 0 & 0 & 0 & 0 & 0.2 & 0.4 \\
$\mu^c_g$ & Fraction of inds. In
critical care that die  & 0.35 & 0.1 & 0.1 & 0.15 & 0.15 & 0.22 & 0.46 & 0.49 & 0.52 \\[1ex]
$e$ & Vaccine efficacy & 0.7 & 0.7 & 0.7 & 0.7 & 0.7 & 0.7 & 0.7 & 0.7 & 0.7 \\
$\omega$ & Reduction in susceptibility   & 0 & 0 & 0 & 0 & 0 & 0 & 0 & 0 & 0\\ 
$\pi$ & Protection  against severe illness & 0 & 0 & 0 & 0 & 0 & 0 & 0 & 0 & 0\\
$\tau$ & Fraction of daily activity spent in a region & 0.5 & 0.5 & 0.5 & 0.5 & 0.5 & 0.5 & 0.5 & 0.5 & 0.5\\
\bottomrule
\end{tabular}
\end{adjustwidth}
\end{table}

In our numerical investigations, the population is stratified into 9 age groups and 5 geographical regions (\tableref{PopulationErva}), giving us total of 45 age-region strata. Per each stratum, there are 16 epidemiological compartments, including three susceptible compartments (unvaccinated, vaccinated with developing immunity, and vaccinated without developing immunity) and two tracks (mild and severe) of infected individuals. This leads to a full model with 720 age-region-compartment combinations.  

\subsection*{Mobility}
\label{sec:Mobility}
Mobility of individuals is modelled using a Lagrangian approach \cite{brauer2019} using a $K$-by-$K$ probability matrix where entry $\theta_{k\ell}$ equals the fraction of time that a typical resident of region $k$ spends in region $\ell$. Then
\begin{equation}
 \label{eq:EffPopulation}
 \hat N_{\ell g} \weq \sum_{k} N_{kg} \theta_{k\ell}
\end{equation}
equals the mean number of individuals of age group $g$ present in region $\ell$, and
\[
 \hat N_\ell \weq \sum_g \hat N_{\ell g}
\]
represents the mean number of individuals present in region $\ell$.

The baseline mobility matrix representing typical mobility in Finland during normal times without pandemic is a 5-by-5 matrix with entries estimated from available data on cross-region travels as
\begin{equation}
 \label{eq:MobilityMatrix}
 \theta_{km}
 \weq \left( (1-\tau) + \tau\Big( 1 - \frac{\psi_{k+}}{N_k}\Big) \right) \delta_{km}
 + \tau \frac{\psi_{km}}{N_k}  (1-\delta_{km}),
\end{equation}
where $\psi_{k+} = \sum_{m \ne k} \psi_{km}$, $(\psi_{km})$ is an estimated trip matrix with $\psi_{km}$ telling the daily number of trips that residents of region $k$ make to region $m$ in Table 1 of the Supplementary material, $N_k$ is the number of residents in region $k$ obtained from in Table 2 of the Supplementary material, and $\delta_{km}$ is the Kronecker delta. The parameter $\tau$ represents the fraction of daily activity time that a typical commuter spends in a remote region. In our numerical simulations we set $\tau= \{0, 0.5, 1 \}$ due to lack of reliable data for estimating this factor. Equation~\eqref{eq:MobilityMatrix} can be interpreted as the expected fraction of active day time that a resident of region $k$ spends in region $m$, with ${\psi_{k+}}/{N_k}$ being the probability that a randomly selected resident of region $k$ commutes outside the home region on a given day.

\subsection*{Calibration of the overall infectious contact rate}
\label{sec:sup-model-calibrating}

The overall infectious contact rate parameter $\beta$ is parameterised in terms of an effective reproduction number $\Reff$ as follows.  Denote by $K^{(\beta)}$ a $KG$-by-$KG$ matrix with entries
\[
 K^{(\beta)}_{kg, \ell h}
 \weq \beta T_I S_{kg}(0) M_{kg, \ell h},
\]
where
\[
 M_{kg, \ell h}
 \weq \beta_{gh} \sum_m \frac{\theta_{km} \theta_{\ell m}}{\hat N_m},
\]
and $S_{kg}(0) = S_{kg}^u(0) + S_{kg}^v(0) + S_{kg}^x(0)$ is the number of $kg$ susceptibles at time zero. The variable $K^{(\beta)}_{kg,\ell h}$ indicates the expected number of new infections among $kg$ individuals caused by an infectious $\ell h$ individual 
who got infected at time zero. Then we set
\[
 \beta \weq \frac{\Reff}{\rho(K^{(1)})},
\]
where $\rho(K^{(1)})$ is the spectral radius of the matrix $K^{(1)}=T_I S_{kg}(0) M_{kg, \ell h}$, and $\Reff$ is set to values 0.75, 1.0, 1.25, 1.50 in different scenarios. With this choice, the spectral radius of $K^{(\beta)}$ equals $\rho(K^{(\beta)}) = \Reff$, and $\Reff < 1$ (resp.\ $\Reff > 1$) indicates the convergence to zero (resp.\ divergence) of a subsystem of differential equations
\begin{align*}
 \frac{d}{dt} E_{kg}
 &\weq \beta S_{kg}(0) \sum_{\ell h} M_{kg, \ell h} I_{\ell h} - \frac{1}{T_E} E_{kg}, \\
 \frac{d}{dt} I_{kg}
 &\weq \frac{1}{T_E} E_{kg} - \frac{1}{T_I} I_{kg},
\end{align*}
only containing the infectious compartments, linearised in a neighbourhood of a stable initial state where $S_{kg}^u(0), S_{kg}^v(0), S_{kg}^x(0)$ are fixed to their current states, and $E_{kg} = I_{kg} = 0$ for all $kg$, see \cite{diekmann2012mathematical, Diekmann_Heesterbeek_Roberts_2010}. Hence $\Reff < 1$ indicates that all infectious compartments would decrease locally in time even without future vaccinations. In the special case where $S_{kg}(0) = N_{kg}$ for all $kg$, $\Reff$ reduces to the basic reproduction number. In general this is not the case because $\Reff$ also takes into account the accumulated immunity at time zero due to prior vaccinations and recovery.

\subsection*{Pair contact rates}

\label{sec:AgeContactMatrix}
Contacts between individuals are modelled so that $\beta_{gh}/{\hat N_m}$ denotes the mean contact rate (unnormalized, corresponding to no pandemic) in region $m$ between any unordered pair of individuals present in region $m$, such that one individuals is in age group $g$ and the other in age group $h$. 
For $g \ne h$ we find that $\hat E^{(m)}_{gh} = \hat N_{mg} \hat N_{mh}$ with the terms on the right given by \eqref{eq:EffPopulation}. For $g=h$, we note that $\hat E^{(m)}_{gg} = \E Y^{(m)}_{gg}$ is the expectation of a random integer
\[
 Y^{(m)}_{gg}
 \weq \sum_{k=1}^K \sum_{1 \le i < j \le N_{kg}} \!\!\!
 B_{ki} B_{kj} \ +
 \sum_{1 \le k < \ell \le K}
 \sum_{i=1}^{N_{kg}} \sum_{j=1}^{N_{\ell g}} B_{ki} B_{\ell j},
\]
where the random variables $B_{ki} \in \{0,1\}$ on the right are mutually independent and such that $\E B_{ki} = \theta_{km}$ for all $k,i$. Then a direct computation shows that
\begin{align*}
 \E Y^{(m)}_{gg}
 &\weq \sum_k \binom{N_{kg}}{2} \theta_{km}^2 + \frac12 \sum_k \sum_{\ell \ne k} N_{kg} \theta_{km} N_{\ell g} \theta_{\ell m} \\
 &\weq \sum_k \binom{N_{kg}}{2} \theta_{km}^2 + \frac12 \left( \sum_k \sum_{\ell} N_{kg} \theta_{km} \right)^2
 - \sum_k N_{kg}^2 \theta_{km}^2 \\
 &\weq \frac12 \hat N_{mg}^2 
 - \sum_k N_{kg} \theta_{km}^2.
\end{align*}
Then the expected number of such pairs equals
\begin{equation}
 \label{eq:MeanPairCounts}
 \hat E^{(m)}_{gh}
 \weq
 \begin{cases}
  \hat N_{mg} \hat N_{mh}, &\quad g \ne h; \\
  \frac12 \hat N_{mg}^2 - \frac12 \sum_k N_{kg} \theta_{km}^2, &\quad g=h,
 \end{cases}
\end{equation}
when we assume that each resident of each region $k$ is present in region $m$ with probability $\theta_{km}$, independently of the other individuals. Then the aggregate rate of contacts between age groups $g$ and $h$ is given by
$\beta_{gh} E_{gh}$, where
\[
 E_{gh}
 \weq \sum_m \frac{\hat E^{(m)}_{gh}}{\hat N_m} 
\]
is a mobility correction factor. The aggregate contact rate between age groups $g$ and $h$ can alternatively be computed as $(1-\frac12 \delta_{gh}) N_g C_{gh}$, where $N_g$ is the size of age group $g$ and $C_{gh}$ is the age contact matrix. By solving the balance equation $(1-\frac12 \delta_{gh}) N_g C_{gh} = \beta_{gh} E_{gh}$, we find that
\begin{equation}
 \label{eq:PairContactRates}
 \beta_{gh}
 \weq (1-\frac12 \delta_{gh}) \frac{N_g C_{gh}}{E_{gh}}.
\end{equation}
For baseline age contact matrix $C_{gh}$, we use the one in Table 3 of the Supplementary material, obtained from Finland 2006 POLYMOD matrix, then pairwise degree corrected, then extrapolated and density corrected, then time-corrected to represent an nonnormalised age-based contact structure in Finland in 2021 (assuming no pandemic), see \cite{Arregui_Aleta_Sanz_Moreno_2018}.

\subsection*{Data and initialization}

The model is initialized to the epidemic situation in mainland Finland on the day of origin set to 18 April 2021. The age-structured population sizes were retrieved from national statistics \cite{stats_finland_api}. The population sizes per region can be found at  \tableref{PopulationErva}, and further details are in Section 2 of the Supplementary material. We build an age-dependent contact structure by adjusting a questionnaire-based contact matrix \cite{Mossong_etal_2008} to a setting where the age structure can vary between the geographical regions. Mobility between regions is estimated using aggregate tracking data from a major mobile phone operator. 

The disease progression, vaccination, and hospitalization status in the age-region compartments is based mostly on data from Finnish health authorities \cite{thl_api}. With this data we initialize 8 out of 16 compartments for each age-region combination. The compartment related to deaths is set empty, so that the final results only consider new deaths after the initial date. Taking into account all age-region combinations, the model is initialized with 360 values.

\begin{table}[ht]
\small
\centering
\caption{\label{tab:PopulationErva} Population, incidence (7-day case notification rate per 100 000 individuals), and vaccine uptake (proportion of vaccinated with first dose per 100 individuals) in five regions (university hospital specific catchment areas) of mainland Finland on 18 April 2021.}
\begin{tabular}{lrrrrrr}
\toprule
Region & Population & Incidence & Vaccine uptake \\
\midrule
HYKS    & 2 198 182    & 53.6 & 23.4 \\
TYKS    &   869 004    & 39.9 & 26.9 \\
TAYS    &   902 681    & 24.9 & 25.2 \\
KYS     &   797 234    & 10.0 & 25.4 \\
OYS     &   736 563	   & 10.3 & 22.7 \\
\midrule
Total   & 5 503 664    & 34.7 & 24.4 \\
\bottomrule
\end{tabular}
\end{table}

\subsection*{Heuristic vaccination strategies}

We construct heuristic vaccination strategies which can depend on three variables for each region $k$ and given time $t$: the proportion of population $\hat{N}_k$, the proportion of new infections $\hat{I}_k^D(t)$ during the last $D$ days, and the proportion of hospitalized individuals $\hat{H}_k^D(t)$ during the last $D$ days in region $k$. Given that there are in total $v(t)$ vaccine doses to distribute on day $t$, the region $k$ will receive
\begin{equation}
 \label{eq:HeuristicStrategies}
 v_k(t)
 = v(t) \Big( w_1 \hat{N}_k + w_2 \hat{I}^D_k(t) + w_3 \hat{H}^D_k(t) \Big),
\end{equation}
vaccine doses.  Then we distribute the vaccines in each region in an age-descending order, i.e., first vaccinate the entire oldest group, and then the second oldest, etc. The choice of weights $w_1$, $w_2,$ and $w_3$ determines the relative allocation of vaccines across regions,
with $w_1+w_2+w_3=1$. Within regions, the $v_k(t)$ vaccine doses are distributed in an age-prioritized strategy from older to younger age groups, i.e., first vaccinate the entire oldest group, and then the second oldest, etc. We set $D=14$ and build 8 different vaccination strategies by setting the $w_i$ values as shown in  \tableref{VaccinationStrategies}. See Section 1 of the Supplementary material for further details. The feasibility of implementing strategy {\tt Pop}+{\tt Inc}+{\tt Hosp} corresponding to equal weights $w_1=w_2=w_3$ has been discussed by Finnish health authorities \cite{THL-2021-03-25}.

\begin{table}[ht]
\centering
\small
\caption{\label{tab:VaccinationStrategies} Adaptive vaccination strategies and their corresponding weights corresponding to \eqref{eq:HeuristicStrategies}. {\tt Pop}, {\tt Inc} and {\tt Hosp} refer to strategies where vaccines are distributed demographically, based on the regional incidence level only, and based on the number of hospitalized cases only, respectively.}
\begin{tabular}{lccc}
\toprule
Strategy & \bf $w_1$ & \bf $w_2$ & \bf $w_3$ \\
\midrule
{\tt Pop} & 1 & 0 & 0\\
{\tt Inc} & 0 & 1 & 0\\
{\tt Hosp} & 0 & 0 & 1\\
{\tt Pop}+{\tt Hosp} & 1/2 & 0 & 1/2\\
{\tt Pop}+{\tt Inc} & 1/2 & 1/2 & 0\\
{\tt Inc}+{\tt Hosp} & 0 & 1/2 & 1/2\\
{\tt Pop}+{\tt Inc}+{\tt Hosp} & 1/3 & 1/3 & 1/3\\
\bottomrule
\end{tabular}
\end{table}

\subsection*{Optimized vaccination strategies}

In order to obtain an optimized age-specific and time-dependent vaccination strategy, we formulate the problem in terms of optimal control theory with the aim of minimizing the total number of deaths while satisfying the constraints of a fixed  daily maximum amount of vaccines available over the course of a single pandemic wave. More specifically, our objective is to determine  optimal time-varying-per-capita rate of vaccines
$\nu: (k,g,t) \mapsto \nu_{kg}(t)$
that minimizes the cumulative number of deaths calculated by \eqref{eq:agespatial}. Thus, the objective functional to be minimized is given by
\begin{equation}
 J(\nu) =
 \int_{0}^{t_f} \sum\limits_{k=1}^K\sum\limits_{g=1}^G D_{kg}(t)dt, 
\end{equation}
where the instantaneous expected death rate $D_{kg}(t)$ is obtained as a solution of \eqref{eq:agespatial}--\eqref{eq:agespatialR}, and
$t_f$ is a sufficiently large time instant by which the full population is vaccinated.

The optimal control  formulation is: find 
$\nu^*: (k,g,t) \mapsto \nu^*_{kg}(t)$
such that 
\begin{equation}
\label{eq:nlpprob}
\begin{split}
   & J(\nu^*) = 
   \min\limits_{\nu}J(\nu) \,\,\,\, \text{subject to \eqref{eq:agespatial} and}\\
   &\sum\limits_{k=1}^{K}\sum\limits_{g=1}^{G} \nu_{kg}(t)S_{kg}(t) = \nu_{\max}, \\ 
    \end{split}
\end{equation}
where $\nu_{\max}$ is the maximum rate of available vaccines. 
To solve this control problem numerically,  we use Pontryagin’s Maximum Principle \cite{ kirk2004optimal, macki2012}.
This principle converts problem \eqref{eq:nlpprob} into the problem of minimizing the~Hamiltonian $\mathcal{H} = \sum_{k=1}^K \sum_{g=1}^G \mathcal{H}_{kg}$ given by
\begin{equation}
\begin{aligned}
 \mathcal{H}_{kg} \weq
 & D_{kg} \\ 
 & + 
  {\Lambda_{S_{kg}^u}}
  \left(- \lambda_{kg} S^u_{kg} - v_{kg} S^u_{kg}\right) -{\Lambda_{S_{kg}^x}} \lambda_{kg} S^x_{kg}  \\
 & + \Lambda_{S_{kg}^v}
 \left(v_{kg} S^u_{kg} - \lambda_{kg} S^v_{kg}
 - \frac{1}{T_V} S^v_{kg})\right)\\
 & + \Lambda_{S_{kg}^p}
 \left((1- e)\frac{1}{T_{V}}S_{kg}^v - \lambda_{kg}S_{kg}^p \right) \\ 
 & + \Lambda_{E_{kg}}\left(\lambda_{kg}(S_{kg}^u + S_{kg}^v + S_{kg}^x) -\frac{1}{T_E}E_{kg}\right)\\
 &+ \Lambda_{E^v_{kg}}\left((1-\omega)\lambda_{kg}S_{kg}^p -\frac{1}{T_E}E^v_{kg}\right)\\
 & + \Lambda_{I_{kg}}
 \left(\frac{1}{T_E}E_{kg} - \frac{1}{T_I} I_{kg} \right) + \Lambda_{I_{kg}^v}
 \left(\frac{1}{T_E}E_{kg}^v - \frac{1}{T_I} I_{kg}^v \right) \\ 
 & + \Lambda_{Q_{kg}^0} \left((1-p_g^h)\frac{1}{T_I} I_{kg} + (1-(1-\pi) p_g^h)\frac{1}{T_I} I_{kg}^v - \frac{1}{T_{Q^0}} Q^0_{kg} \right) \\ 
 & + \Lambda_{Q_{kg}^1}
 \left(p_g^h\frac{1}{T_I} (I_{kg} + (1-\pi) I^v_{kg}) - \frac{1}{T_{Q^1}} Q^1_{kg} \right) \\ 
 & + \Lambda_{H_{kg}^w} \left(\frac{1}{T_{Q^1}} Q^1_{kg}(t) - \frac{1}{T_{H^w}}H_{kg}^w(t)\right) \\
 & + \Lambda_{H_{kg}^c}
 \left( p_g^c \frac{1}{T_{H^w}} H_{kg}^w - \frac{1}{T_{H^c}} H_{kg}^c \right)  \\ 
 & + \Lambda_{H_{kg}^r}
 \left( (1-\mu_g^c) \frac{1}{T_{H^c}} H_{kg}^c - \frac{1}{T_{H^r}} H_{kg}^r \right)\\
 & + \Lambda_{R_{kg}}
 \left( (1-\mu^q_g) \frac{1}{T_{Q^0}} Q^0_{kg}
 + (1-\mu^w_g) (1-p^c_g) \frac{1}{T_{H^w}} H^w_{kg}
 + \frac{1}{T_{H^r}} H^r_{kg} \right)\\  
 & + \Lambda_{D_{kg}} \left( \mu_g \frac{1}{T_{Q^0}} Q_{kg}^0 + \mu_g^w (1-p_g^c) \frac{1}{T_{H^w}} H_{kg}^w
 + \mu_g^c\frac{1}{T_{H^c}} H_{kg}^c \right) \\ 
 & + \Lambda_{V_{kg}}
 \left(  e \frac{1}{T_V} S^v_{kg} \right),
\end{aligned}
\end{equation}
where $\Lambda_{S_{kg}^u}, \dots, \Lambda_{V_{kg}}$ appearing above are time-dependent
Lagrange multipliers \cite{nlpbook}. 
Then, we differentiate $\cal{H}$ with respect to $\nu_{kg}$ to obtain
\[ 
 \frac{\partial \cal{H}}{\partial \nu_{kg}}(t)
 \weq - \left(\Lambda_{S_{kg}^u}(t)
 - \Lambda_{S_{kg}^v}(t) \right)S_{kg}^u(t).
\]
Further, we differentiate $\cal{H}$ with respect to the state variables $S_{kg}^u$, $S_{kg}^v$, $S_{kg}^p$, $S_{kg}^x$, $E_{kg}$, $E_{kg}^v$  $I_{kg}$, $I_{kg}^v$, $Q_{kg}^0$, $Q_{kg}^1$, $H_{kg}^w$
$H_{kg}^c$, $H_{kg}^r$, $R_{kg}$, $D_{kg}$, $V_{kg}$ to derive a so-called adjoint system of equations. 
By collecting the state variables into a vector $Y = [S_{kg}^u, \dots, V_{kg}]$, and the Lagrange multipliers into a vector
$\Lambda_Y = [ \Lambda_{S_{kg}^u}, \dots, \Lambda_{V_{kg}}]$,
%
we have 
\[ 
\dot{\Lambda}_Y = - \frac{\partial \cal{H}}{\partial Y},
\]
with transversality conditions
$
\Lambda_Y(T_f) = 0.
$
We solve the adjoint system of equations backwards in time because we only have the final conditions. For more details see Section 2  of the Supplementary material.

\section*{Results}

We summarize our results by focusing on the medium-level mobility scenario, i.e., for $\tau = 0.5$. The relative performance of different vaccination strategies and their qualitative behavior is robust across different mobility levels (see Section 5 of the Supplementary material).

The code used for doing the analysis and producing the results in this paper is publicly available as a Github repository\footnote{\url{https://github.com/FINCoVID19/optimized_vaccination_finland}}.

\subsection*{Comparison of adaptive heuristic strategies}

\begin{figure}[ht]
\centering
\includegraphics[width=0.99\textwidth]{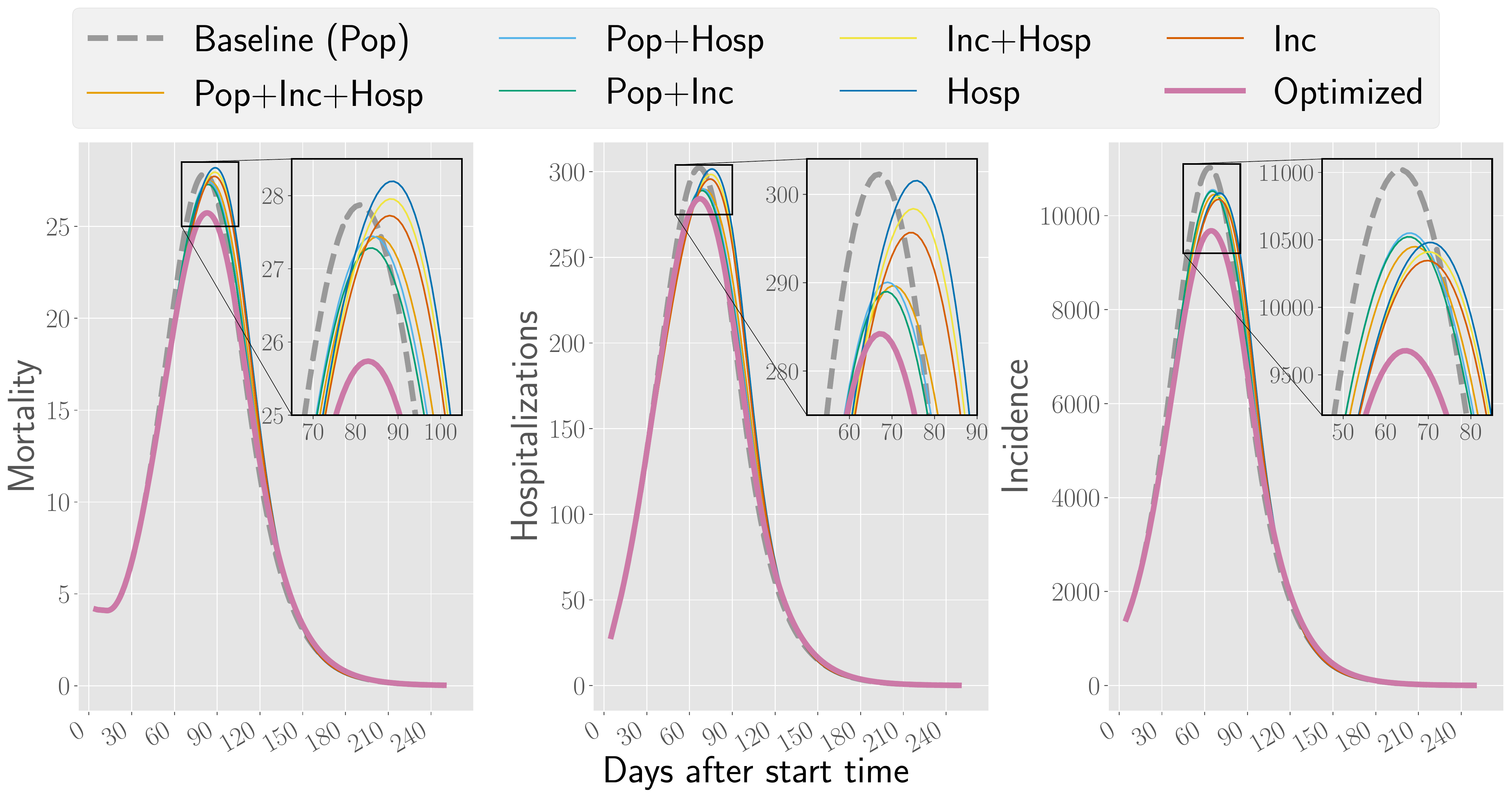}
\caption{Mortality, hospitalizations, and incidence for the vaccination strategies in Table 2.
In this scenario, the effective reproduction number is $\Reff = 1.5$ and the mobility value is $\tau = 0.5$. For other parameter combinations, see  Section 5 of the Supplementary material.}
\label{fig:Fig_1}
\end{figure}

We first compare different vaccination strategies at the level of the whole country (\figref{Fig_1}) to a baseline strategy,  in which vaccine doses are first allocated to regions weighted by population counts and then serially to age groups in descending order within each region. This static {baseline strategy} \texttt{Pop} differs from all other strategies which we call \emph{adaptive heuristic strategies} in a way that it does not try to adapt to the evolution of the epidemic in any way. The adaptive heuristic strategies allocate more vaccine doses to regions with more infections and/or hospitalizations, but are similarly age-prioritized within regions.

\begin{figure}[ht]
\centering
\includegraphics[width=0.99\textwidth]{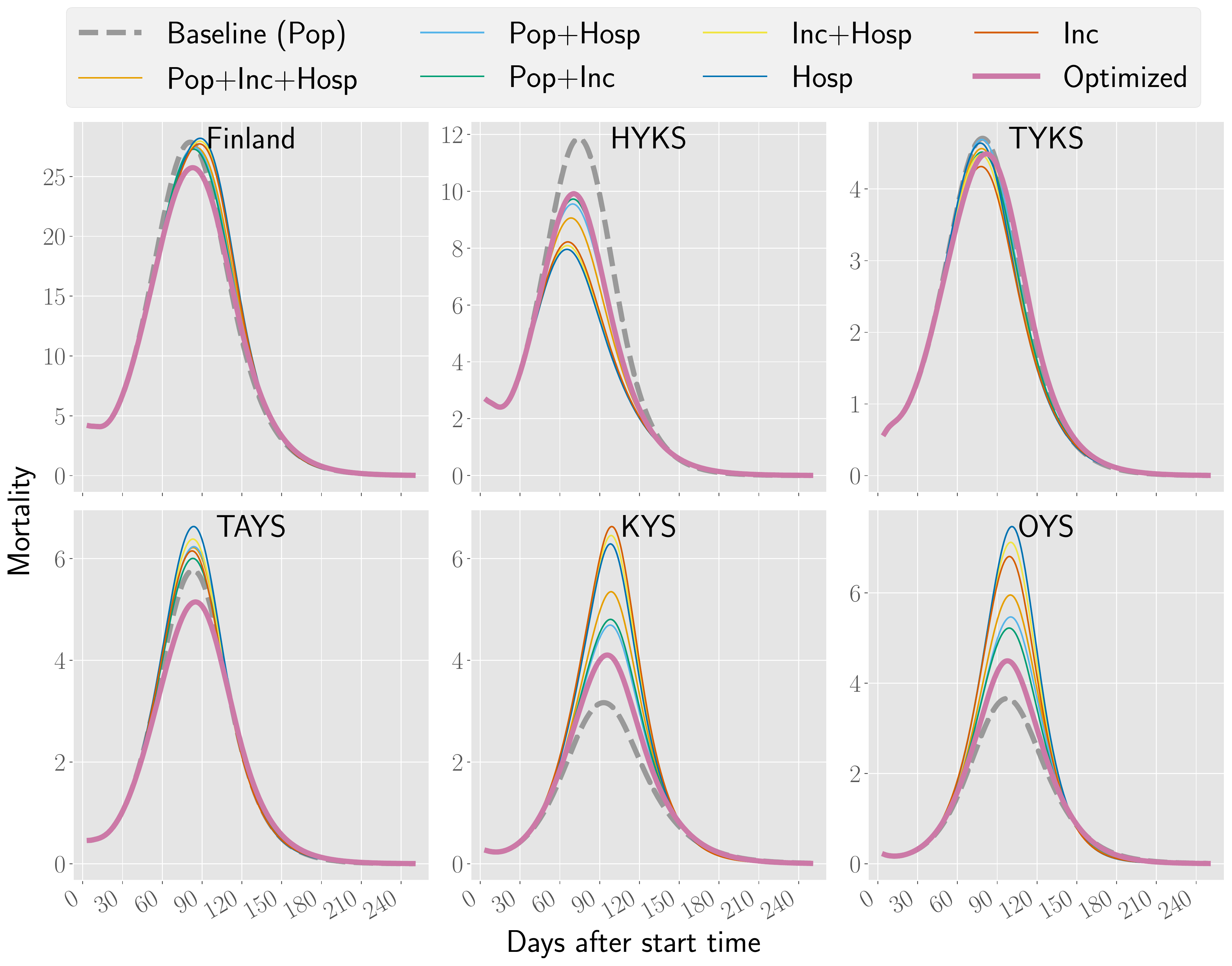}
\caption{Mortality in Finland and the five hospital catchment areas included here for the vaccination strategies in Table 2. For this scenario, the basic reproduction number $\Reff = 1.5$ and the mobility value $\tau = 0.5$. For other values of $\Reff$ and $\tau$, see  Section 5 of the Supplementary material.
}
\label{fig:Fig_2}
\end{figure}

\figref{Fig_1} describes the performance of different strategies over time.  All adaptive strategies succeed in lowering incidence compared to the baseline. For mortality and hospitalizations, the heuristic strategies outperform the baseline initially, but tend to lose most of their advantage in the long run.  This is because the adaptive heuristics delay the epidemic and its peak as compared to the baseline, and eventually the less-vaccinated regions in the adaptive heuristics will do worse than in the baseline strategy. This can be further seen in Fig 3 which shows the evolution of mortality in each region.  This can be further seen in \figref{Fig_2} which shows the evolution of mortality in each region. In contrast, the optimized strategy succeeds in keeping mortality and hospitalizations below baseline also after the peak.

%
Whether or not it pays off to delay the epidemic with adaptive strategies at the cost of allocating less vaccines to less affected regions depends on how fast the disease is progressing. Specifically, the total performance over the full time horizon depends on the transmission rates of the disease (see \tableref{ResultsRTauAbs_main}): In low-transmission scenarios the adaptive heuristics perform well and delaying the epidemic can be beneficial because there is time to develop additional immunity in the low-incidence regions to hinder future spreading. In high-transmission scenarios the adaptive heuristics put too much emphasis on the initially high-incidence regions and leave the low-incidence regions vulnerable to large future outbreaks.

As expected, none of the strategies can outperform the baseline in every region. The regions that have initially less incidence will suffer on the expense of the high-incidence regions when changing from the baseline strategy to adaptive strategies. However, as stated before, if all individuals in the country are treated equally regardless of their region of residence, the transmission rate will determine which strategy is best for minimizing the total disease-induced mortality in the country.

Among the adaptive vaccination strategies, the number of hospitalized individuals is not in general as good a measure as incidence when determining where to distribute the vaccines. This might be due to the delay in the hospitalization which means that vaccination continues in regions where the effective reproduction number is already low, at the expense of regions where incidence is on the rise but not yet reflected in hospitalizations. 

It should be noted that in our model the number of daily new infections is assumed to be accurately reported, which is not a realistic assumption. While it does not make any difference for the strategy if the total numbers are systematically lower due to underreporting, fluctuations in the numbers and systematic biases in the measurements across regions could have an impact.

\subsection*{Performance of optimized vaccination strategies}

\begin{figure}[ht]
\centering
{\includegraphics[width=\textwidth, height = 10cm]{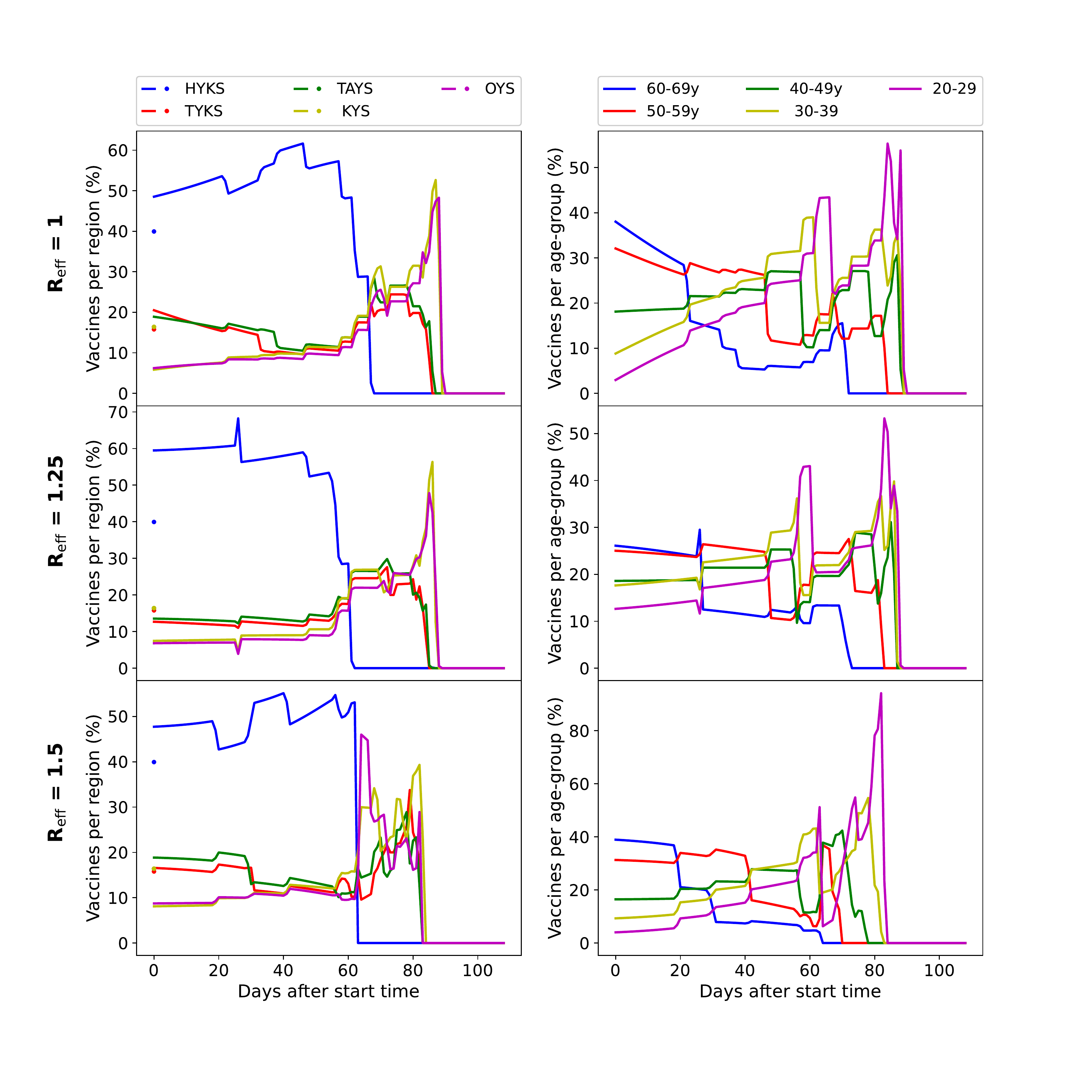}}
\caption{Percentage of vaccine doses allocated by the optimized strategy to
regions (left) and age groups (right) in three scenarios ($\Reff = 1, 1.25, 1.5$). On the left, dots  represent the percentage of vaccines which each region would receive with {\tt Pop} (baseline). 
}
\label{fig:Fig_3}
\end{figure}

We will next discuss the performance of an optimized vaccination strategy found by running the numerical algorithm  with the objective of minimizing the total disease-induced mortality over a 250-day time horizon. 
Our numerical investigations show that the optimization algorithm is robust for different levels of the vaccine efficacy, reduction in susceptibility  and protection against severe symptoms, see Section 3 of the  Supplementary material for further details.

Our numerical results indicate that the optimized strategy shares good features of both the static baseline strategy and the adaptive heuristic strategies: There is an initial drop in mortality similar to heuristic strategies, but in the long term the difference to baseline is not as large as for the heuristic strategies. In other words, at later times of the epidemic the optimized strategy demonstrates the highest reduction in mortality.
Overall, the optimized strategy shows reduction in mortality by up to 50 individuals for $\Reff = 1.5$ (see \tableref{ResultsRTauAbs_main}). The reason why the differences in mortality are not very large is because the majority of individuals in high-risk groups have already been vaccinated in the beginning of the calculations (18 April 2021). However, cumulative incidence can reach differences of up to tens of thousands, as \tableref{ResultsRTauAbs_main} shows.


The percentage of vaccine doses allocated by the optimized strategy to each geographical region and age group is shown in \figref{Fig_3} for three transmission scenarios.
Similarly to the heuristic strategies, the optimized strategy depends heavily on the disease parameters. The effective reproduction number does not just fine-tune the strategy, but there is a transition from one approach to another: 
For a low-transmission scenario ($\Reff = 0.75$) in which the epidemic is in clear decline, the optimized strategy does not preferentially target older age groups but tries to reduce the number of infections, and the optimized strategy is the one that follows the number of infected.
In scenarios with a high overall transmission rate, the optimized strategy favours older age groups having higher risk of severe illness and death.

Both the low-transmission and high-transmission scenarios lead to an optimized strategy that favours the initially high-incidence region, and this effect is stronger for low-transmission scenarios. Specifically, the optimized strategy initially targets the capital region (HYKS) with approximately 20 (resp.\ 8) percentage points higher share of available vaccine doses than the baseline strategy for $\Reff = 1.25$ (resp.\ 1.5). Interestingly, the optimization finds that the age prioritization is smaller and geography prioritization more aggressive in the scenario with 
$\Reff=1.25$ than in scenarios with $\Reff=1.0$ and $\Reff=1.5$.

\begin{table}[ht]
\begin{adjustwidth}{-1.25in}{0in} 
\centering
\small
\caption{\label{tab:ResultsRTauAbs_main} Absolute difference in mortality (expected number of deaths) and cumulative incidence (expected number of cases) during a 250-day time horizon resulting from different vaccination strategies with respect to baseline strategy ({\tt Pop}) for $\tau = 0.5 $. Highest reductions are indicated in boldface.  Results for different values of $\tau$ are shown in  Section 5 of the Supplementary material,  including hospitalizations.}
\begin{tabular}{cr|rrrrrrr}
\toprule
& $\Reff$& \tt Hosp & \tt Inc & \tt Inc+Hosp & \tt Pop+Hosp & \tt Pop+Inc & \tt Pop+Inc+Hosp & \tt Optimized \\ 
\midrule 
\parbox[t]{2mm}{\multirow{4}{*}{\rotatebox[origin=c]{90}{Mortality}}} 
& $0.75$ & -0.38 & \bf -0.42 & -0.40 & -0.24 & -0.26 & -0.31 & \bf -0.42  \\ 
& $1.00$ & -2.41 & -2.91 & -2.67 & -1.79 & -2.01 & -2.24 & \bf -3.82  \\ 
& $1.25$ & 9.20 & 2.00 & 5.47 & -0.79 & -3.94 & 0.02 & \bf -23.46  \\ 
& $1.50$ & 87.04 & 62.44 & 74.21 & 28.84 & 16.81 & 39.68 & \bf -50.07  \\ 
\midrule 
\parbox[t]{2mm}{\multirow{4}{*}{\rotatebox[origin=c]{90}{Incidence}}} 
& $0.75$ & -360.92 & -363.55 & -361.80 & -187.87 & -188.29 & -246.81 & \bf -618.48  \\ 
& $1.00$ & -2023.22 & -2179.46 & -2100.24 & -1235.39 & -1287.21 & -1559.19 & \bf -3352.64  \\ 
& $1.25$ & -1564.24 & -4249.68 & -2919.86 & -2715.05 & -3811.97 & -3133.51 & \bf -21499.37  \\ 
& $1.50$ & 4631.92 & -2430.08 & 1071.72 & -1535.43 & -4697.56 & -1508.73 & \bf -40664.22  \\ 
\bottomrule
\end{tabular}
\end{adjustwidth}
\end{table}





\section*{Discussion and conclusions}

In this work we have constructed an epidemic modelling framework which allows to evaluate various adaptive strategies for allocating vaccines based on static demographic data and dynamic evolution of the epidemic situation across different geographical regions. We investigated various heuristic strategies for allocating more  vaccines to regions with higher incidence and hospital load, together with optimized strategies which may flexibly allocate vaccines to different age groups and regions in parallel.
Our numerical results, conducted for scenarios adjusted to the recent COVID-19 epidemic situation in Finland, show that optimized vaccination strategies can reduce the death toll and significantly mitigate the disease burden of the epidemic.
The relative advantage of different adaptive strategies over the static baseline is influenced by the overall epidemic situation. Also, whatever strategy is chosen, a trade-off between different regions is inevitable due to limited supply of vaccine doses and daily vaccination capacity. Nevertheless, the results provide valuable insights for designing efficient vaccination strategies: In general, using hospital loads as basis in allocating vaccine doses tends to lead to worse performance compared to the static baseline. The optimized strategy appears to achieve a good balance between short-term benefits of adaptive strategies and the long-term robustness gained by the uniform vaccine allocation. Further, even though we optimize mortality, there is a delicate balance between favoring individuals with higher direct risk of death as opposite to individuals at risk of getting infected and causing large outbreaks.

As with all modelling, there are several factors and phenomena that are not included, and the results can change if these factors turn out to be important. Typically this would imply that the actual numbers in a modelling study might be subject to change, but the overall phenomena that are observed here are relatively robust. Such numbers would be the exact number of infected, hospitalized, and deceased individuals, and the phenomena the relative order of the different strategies. The only real way of knowing which factors are important is to include them in a model, but in practice the choice of relevant factors is informed by the reliability of the model. This is why we have chosen to start with a model benchmarked in another study related to Sweden \cite{Sjodin_Rocklov_Britton_2021-04}, and modify it by making it more accurate by including geographical information.

There are several factors which we believe that are missing in our model and are important for both the accuracy of the results and important to consider when optimizing vaccination strategies. First is the need for more than a single vaccine dose needed by many of the currently used vaccines, which is not modelled here. Including this in the model would allow one to optimize the vaccination strategy further by finding an optimal strategy for giving the second dose with relation to vaccinating different age groups and geographical locations. 
This could have an impact on the benefits of regional targeting strategies, because the regional differences might even out during the time it takes to build immunity with multiple vaccines.
Second, one should allow the infectious contact rates to change across geographical regions and time. As the public is informed of the current pandemic situation their behavior, and therefore the transmission rate, is bound to change. This induces a feedback loop which makes a large difference especially for long-term predictions, but also makes modeling more difficult as one needs to model the public response to various pandemic situations \cite{gozzi2020towards, gozzi2020self}. In addition, the governments will take actions given that the situation is sufficiently critical \cite{perra2021non}, and these decisions might depend on several hard-to-model factors related to politics.

Studying the effects of cross-region mobility were not at the main focus of this study, but the sensitivity analysis that we performed for the overall mobility factor has interesting implications. It turned out that cross-region mobility can be an important factor even in this relatively advanced state of the epidemic where all regions have some incidence, but there is still a geographical imbalance in the relative incidences. These results are especially striking considering that the mobility factor $\tau$ only controls for cross-region mobility but not the overall contact rates of the  individuals. That is, decreasing $\tau$ decreases the cross-region contacts but increases the inside-region contacts, and the total rate of contacts in the country remains the same but the large-scale geographical mixing patterns changed. This is in contrast to conventional models which assume full mixing across the country. Further, these findings could have implications on interventions that limit long-range mobility. Further research in this direction would be needed for concluding about these type of interventions.

Our analysis reveals that designing efficient vaccination strategies at a level of a country is highly nontrivial. As seen from our results in \figref{Fig_3}, the details of optimized strategies can be complicated and their faithful implementation difficult, and could lead in a slower overall vaccine delivery. However, it should be possible to simplify the strategies and try to follow the main principles of parallel vaccination and geographic distribution of vaccines with as much detail as practically possible.
It is important to note that carefully analyzed and executed strategies can potentially save lives even if the strategy is changed after most of the risk groups are already vaccinated. Much larger effects could potentially be obtained if the planning were done before vaccinations started, but in this case the problem is that the various parameters related to vaccination efficiency might not be known. In any case, the relative performance of different strategies can depend on the effective reproduction number, which means that the vaccination strategy should be chosen in conjunction with non-pharmaceutical intervention strategies of the country.

\section*{Acknowledgments}
This work has been supported in part by the project 105572 NordicMathCovid as part of the Nordic Programme on Health and Welfare funded by NordForsk, and by the Academy of Finland through its PolyDyna grant no. 307806 (T.A-N.). We thank
Kari Auranen
at the Finnish Institute of Health and Welfare and Tom Britton at Stockholm University for helpful discussions.

\section*{Author Contributions}
\paragraph{\bf Conceptualization:} All authors.

\paragraph{\bf Data Curation:} Alejandro Ponce de Le\'on Ch\'avez 

\paragraph{\bf Formal Analysis:} Jeta Molla,  Lasse Leskel\"a

\paragraph{\bf Methodology:} All authors. 

\paragraph{\bf Supervision:} Tapio Ala-Nissila, Lasse Leskel\"a, Mikko Kivel\"a 

\paragraph{\bf Software:} Jeta Molla, Alejandro Ponce de Le\'on Ch\'avez 

\paragraph{\bf Writing – Original Draft Preparation:} All authors. 

\paragraph{\bf Writing – Review \& Editing:}  All authors.

\bibliography{biblio}
\bibliographystyle{plos2015}

\newpage

\begin{center}
{\huge Supplementary  material}
\end{center}

\section*{}

\section{Heuristic vaccination strategies}
\label{sec:sup-heuristic}
We first analyse the different heuristic vaccination strategies to asses their impact on the development of the epidemic. Specifically, we construct different scenarios that determine the number of vaccines $v_k(t)$ that each region $k$ will receive on day $t$ depending on the number of infections and/or hospitalizations. Then, the vaccines $v_k(t)$ are distributed within the region in an age-prioritized strategy from old to younger age groups. We can obtain $v_k(t)$ in the following way
\begin{align*}
 v_k(t)
 = v(t) \Big( w_1 \frac{N_k}{\sum_k N_k} + w_2 \frac{I^D_k(t)}{\sum_k I^D_k(t)} + w_3 \frac{H^D_k(t)}{\sum_k H^D_k(t)} \Big),
\end{align*}
where $v(t)$ is the overall national number of available vaccine doses on day $t$, $w_1$, $w_2$, and $w_3$ are tunable weight parameters of the strategy ($\sum_i w_i = 1$), $I^D_k(t)$ is the number of new infections, and $H^D_k(t)$ is the total hospital occupation in region $k$ over the last $D$ days. In our work we set $D=14$ to capture the changes over two weeks starting from the initial date. The total number of new infections in region $k$ in the last $D$ days is computed by
\begin{align*}
 I^D_k(t) = \int_{t-D}^t \sum_{g=1}^G \frac{1}{T_E} E_{kg}(t) dt,
\end{align*}
and similarly the number of hospitalized individuals is computed by
\begin{align*}
 H^D_k(t)
 = \int_{t-D}^t \sum_{g=1}^G
 \Big( H^w_{kg}(t) + H^c_{kg}(t) + H^r_{kg}(t) \Big) dt.
\end{align*}
Note that, since we want $H^D_k(t)$ to reflect the total hospital occupation at time $t$ for region $k$, an individual may be counted more than once. One at day $t$, another at $t-1$ and so on. As long as the individuals remain in the hospital they will be counted.

Different vaccination strategies can be obtained by changing the weights $w_i$, e.g.\ setting $w_1 = 1$ and $w_2=w_3=0$ corresponds to the baseline strategy \texttt{Pop} where vaccines are equally distributed according to the population density.
%

\section{Numerical discretization of  optimal control problems}
\label{sec:optim}

In this section we describe the numerical approach to solving a general optimal control problem  via Pontryagin’s Maximum Principle \cite{kirk2004optimal, macki2012}.

Let ${\mathbf {x} (t)=\left[x_{1}(t),\, x_{2}(t),\, \ldots , \, x_{n}(t)\right]^{\mathsf {T}}} \in \mathbb{R}^n$ be a state vector and ${\mathbf {u} (t)=\left[u_{1}(t),\, u_{2}(t),\, \ldots , \, u_{r}(t)\right]^{\mathsf {T}}} \in \mathbb{R}^r$ be a control vector. Consider the following optimal control problem: find  ${\bf u}(t)$ to minimize 
\begin{equation}
\label{eq:obfun}
J({\bf u}) = \int_{0}^{t_f} g({\bf x}(t),{\bf u}(t),t)dt 
\end{equation}
subject to the  state equations
\begin{equation}
\dot{\bf x}(t) = f({\bf x}(t), {\bf u}(t),t), \quad {\bf x}(0) = { \bf x}_0 
\end{equation}
and  the  constraint
\[ 
{\bf a} \leq {\bf u} (t) \leq {\bf b}.
\]

In order to state the Maximum Principle  we define a
 Hamiltonian as

\begin{equation}
\label{eq:hamil}
    {\cal H} ({\bf x}(t),{\bf u}(t),t) = g({\bf x}(t),{\bf u}(t),t) + {\bf q}^{\mathsf {T}}(t) f({\bf x}(t), {\bf u}(t),t),
\end{equation}
where ${\bf q}(t)$ are the time-dependent Lagrange multipliers \cite{kirk2004optimal}. The goal now is to find an  optimal trajectory ${\bf x}(t)$, an optimal control ${\bf u}(t)$   and an optimal set of Lagrange multipliers ${\bf q}(t)$ so that to minimize the objective function in \eqref{eq:obfun}. 

From the Hamiltonian \eqref{eq:hamil} and Pontryagin’s Maximum Principle, we obtain the following theorem \cite{kirk2004optimal}. 

\begin{theorem}
If ${\bf x}^*(t), {\bf u}^*(t), t\in[0,t_f]$  is an optimal state-control trajectory  starting at ${\bf x} (0)$, then
there exist Lagrange multipliers ${\bf q}^*(t)$ such that  
\begin{equation}
\label{eq:opsys}
    \begin{split}
        &\dot{\bf x}(t) = \frac{\partial {\cal H}}{\partial {\bf q}} = f({\bf x}(t), {\bf u}(t),t), \quad {\bf x}(0) = {\bf x}_0,\\
        & -\dot{\bf q}(t) = \frac{\partial {\cal H}}{\partial {\bf x}} =   g_x({\bf q}(t), {\bf u}(t),t) + {\bf q}^{\mathsf {T}}(t) f_x({\bf q}(t), {\bf u}(t),t), \quad {\bf q}(t_f) = 0,\\
        & {\cal H}(x^*(t),u^*(t),  q^*(t),t) = \arg \min\limits_{u(t)} {\cal H}(x^*(t),u(t),  q^*(t),t).
    \end{split}
\end{equation}
\end{theorem}
For a given initial value $x_0\in\mathbb{R}^n$, 
 the numerical approach now consists of finding
functions ${\bf x}:[0,t_f]\mapsto \mathbb{R}^n$, ${\bf u}:[0,t_f]\mapsto \mathbb{R}^n$ and ${\bf q}:[0,t_f]\mapsto \mathbb{R}^n$ satisfying the optimality system \eqref{eq:opsys}. The numerical algorithm consists of the following steps: 
\begin{itemize}
    \item[] {\bf Step 1:}  Subdivide the interval $[0, t_f ]$ into $N$ equal sub-intervals and assume
a piecewise-constant control ${\bf u}^{(0)}(t) = {\bf u}^{(0)}(t_k)$, $t\in[t_{k},t_{k+1}]$, $k = 0, 1, \dots, N-1$
    \item[] {\bf Step 2:} Integrate the state equations forward in time for the assumed control ${\bf u}^{(i)}$ and store the trajectory ${\bf x}^{(i)}$
    \item[] {\bf Step 3:} Compute  ${\bf q}^{(i)}$ by solving the second equation in \eqref{eq:opsys} backwards in time 
    \item[] {\bf Step 4:} Compute a new control ${\bf u}^{i+1}$ by solving a  finite-dimensional nonlinear optimization problem  using a sequential least squares programming algorithm
    \item[] {\bf Step 5:} Compute ${\bf x}^{(i+1)}$ and ${\bf q}^{(i+1)}$ for the new control variable as in Steps 2 and 3
    \item[] {\bf Step 6:} Compute the values $J^{(i)}({\bf u}^{(i)}, {\bf x}^{(i)})$ and $J^{(i+1)}({\bf u}^{(i+1)}, {\bf x}^{(i+1)})$  
    \item[] {\bf Step 7:} If 
    \begin{equation} 
    \label{eq:tol}
     | J^{(i+1)} - J^{(i)}| \leq \epsilon
    \end{equation}
    stop  the iterative procedure. Here $\epsilon$  is a small positive constant used as a tolerance. 
    \\
    If \eqref{eq:tol} is not satisfied, replace  ${\bf u}^{(i)}$ with ${\bf u}^{(i+1)}$, ${\bf x}^{(i)}$ with ${\bf x}^{(i+1)}$, ${\bf q}^{(i)}$ with ${\bf q}^{(i+1)}$ and return to Step 4.
\end{itemize}

\section{Different type of vaccines} 
In this section we investigate the sensitivity of the optimization algorithm to different levels of the vaccine efficacy $e$, reduction in susceptibility $\omega$ and protection against developing severe illness $\pi$. We set $e= 0.5, 0.7, 0.9$, $\omega = 0,0.2, 0.6$ and $\pi =0, 0.2, 0.6$ and test the robustness of the optimized strategies with respect to different combinations of these parameters. Let 
\begin{equation}
\begin{split}
P =  \{ 0.5, 0.7,0.9\}\times \{ 0.0, 0.2. 0.6\}\times\{ 0.0, 0.2. 0.6\} 
\end{split}
\end{equation}
be a Cartesian product representing the set of all combinations for the different values of the parameters $e,\omega,\pi$. Let $S_k$ be the optimized strategy obtained for  $P_k \in P$, $k = 1, 2, \dots, 27$. We  investigate the difference between the optimized vaccination strategies, i.e., 
\[ 
\sigma_{kl} =D(P_k,S_k) - D(P_k,S_l), \quad k,l = 1, 2, \dots, 27,
\]
where $D$ is the total number of deaths. Further, we choose the value of the mobility parameter $\tau = 0.5$.Taking the $\max$ norm of the ${\bf \Sigma} = (\sigma_{kl})\in \mathbb{R}^{27\times 27}$  matrix, we get 
\[
\| \Sigma\|_{\max} = \max_{kl}(|\sigma_{kl}|) = 0.82.
\]
Hence the difference in the total number of deaths is less than 1 individual and the performance of the optimized strategies is the same for combinations for the different values of the parameters. 
Further, Figures 17 and 18 in  Section 5 of the Supplementary material verify that the optimization algorithm is robust to different values of the parameters $e,\omega, \pi$ since the optimized strategies are similar. The algorithm is expected to be robust   against changes in the model parameters concerning the efficacy of the vaccine or protection of the vaccine against severe illness since the gradient of the optimization algorithm does not  explicitly depend on these parameters.

\section{Data and parameters}
\label{sec:sup-data}

\subsection{Demographic data for Finland}


\begin{table}[h]
\centering
\tiny
\begin{tabular}{lrrrrrrrrrrrr}
\toprule
 & 0--9 & 10--19 & 20--29 & 30--39 & 40--49 & 50--59 & 60--69 & 70--79 & 80+ & Total\\
\midrule
HYKS & 221613 & 238313 & 272674 & 316173 & 285988 & 289128 & 256006 & 212089 & 106198 & 2198182 \\
TYKS & 82812 & 93001 & 103572 & 106093 & 101979 & 111874 & 113383 & 99917 & 56373 & 869004 \\
TAYS & 88071 & 100864 & 105275 & 112809 & 106951 & 115157 & 117896 & 100045 & 55613 & 902681 \\
KYS & 71910 & 84213 & 92466 & 91390 & 85302 & 103387 & 119723 & 95591 & 53252 & 797234 \\
OYS & 80308 & 91471 & 84511 & 88448 & 82348 & 91225 & 100322 & 75669 & 42261 & 736563 \\
\midrule
Total & 544714 & 607862 & 658498 & 714913 & 662568 & 710771 & 707330 & 583311 & 313697 & 5503664\\
\bottomrule
\end{tabular}
\caption{\label{tab:FinPopAgeErva} Population size by region and age in mainland Finland on 31 Dec 2020. Obtained from \cite{stats_finland_api}.}
\end{table}


\begin{table}[h]
\centering
\scriptsize
\begin{tabular}{c|ccccccccc}
\toprule
 & 0--9 & 10--19 & 20--29 & 30--39 & 40--49 & 50--59 & 60--69 & 70--79 & 80+\\
\midrule
0--9 & 4.61 & 1.24 & 0.81 & 1.71 & 1.08 & 0.63 & 0.58 & 0.15 & 0.08 \\
10--19 & 1.10 & 7.83 & 0.97 & 1.02 & 1.83 & 0.71 & 0.35 & 0.13 & 0.07 \\
20--29 & 0.71 & 0.95 & 3.87 & 1.84 & 1.51 & 1.41 & 0.67 & 0.19 & 0.10 \\
30--39 & 1.51 & 1.01 & 1.86 & 3.25 & 2.24 & 1.97 & 1.18 & 0.21 & 0.12 \\
40--49 & 0.82 & 1.57 & 1.33 & 1.94 & 3.18 & 2.24 & 1.02 & 0.29 & 0.16 \\
50--59 & 0.45 & 0.57 & 1.16 & 1.60 & 2.09 & 2.91 & 1.71 & 0.26 & 0.14 \\
60--69 & 0.62 & 0.42 & 0.83 & 1.43 & 1.42 & 2.56 & 2.19 & 0.72 & 0.40 \\
70--79 & 0.22 & 0.21 & 0.32 & 0.36 & 0.58 & 0.55 & 1.01 & 1.09 & 0.60 \\
80+ & 0.22 & 0.21 & 0.32 & 0.36 & 0.58 & 0.55 & 1.01 & 1.09 & 0.60 \\
\bottomrule
\end{tabular}
\caption{\label{tab:polymod} Finnish age contact matrix with 9 age groups and 10y age resolution. The entry on row $g$ and column $h$ indicates the estimated daily number of contacts made by a typical individual in age group $g$ to individuals in age group $h$ \cite{Mossong_etal_2008}.
}
\end{table}


\begin{table}[h]
\centering
\scriptsize
\begin{tabular}{lrrrrrr}
\toprule
 & HYKS & TYKS & TAYS & KYS & OYS \\ 
\midrule
HYKS & 1 389 016 & 7 688	& 16 710	& 7 789	& 1 774	\\ 
TYKS & 11 316    & 518 173 &	14 139	& 562	& 2 870	\\ 
TAYS & 22 928	& 12 404  & 511 506	& 4 360	& 1 675	\\ 
KYS &  8 990	& 365	  & 4 557	& 459 867 &  3 286 \\ 
OYS &  1 798	& 2 417	& 1 592	& 3 360	& 407 636 \\ 
\bottomrule
\end{tabular}
\caption{\label{tab:MobilityMatrixFIN2019} Finnish regional morning (between 6:00--11:59) mobility, averaged over March--May 2019. Rows represent origins and columns represent destinations. (Source: Telia Crowd Insights) 
}
\end{table}

\subsection{Initial conditions}

\begin{table}[h]
\caption{\label{tab:Compartments} { \bf Epidemiological compartments.} There are $KG$ copies of each compartment, denoted $S^u_{kg}, S^v_{kg}, \dots, V_{kg}$ for regions $k=1,\dots,K$ and age groups $g=1,\dots,G$.}
\centering
\small
\begin{tabular}{ll}
\toprule
\bf Symbol & \bf Description \\
\midrule
$S^u$ & Susceptible, unvaccinated \\
$S^v$ & Susceptible, invited for vaccination \\
$S^x$ & Susceptible, vaccinated with no immunity or declined vaccination \\
$E$ & Infected but not yet infectious \\
$I$ & Infected and infectious \\
$Q^0$ & Quarantined at home, mild disease \\
$Q^1$ & Quarantined at home, severe disease \\
$H^w$ & hospitalized, in general ward \\
$H^c$ & hospitalized, in critical care \\
$H^r$ & hospitalized, in recovery ward \\
$D$   & Deceased \\
$R$   & Recovered with full immunity \\
$V$   & Vaccinated with full immunity \\
\bottomrule
\end{tabular}
\end{table}

We obtain the initial conditions  from data trying to mimic the pandemic situation in Finland as of 18 April 2021. 
More specifically, we calculate the initial conditions  for  the compartments in \tableref{Compartments} as follows 
\begin{align*}
    S^u_{kg} &= N_{kg} - S^v_{kg} - S^x_{kg} - E_{kg}  - I_{kg}  - Q_{kg}^0 - Q_{kg}^1 - H_{kg}^w - H_{kg}^c - H_{kg}^r - R_{kg} - D_{kg} - V_{kg} \\
    S^v_{kg} &= 0 \\
    S^x_{kg} &= (1-e) v_{kg} \\
    E_{kg} &= \frac{T_E}{T_I+T_E} i^r_{kg} \\
    I_{kg} &= \frac{T_I}{T_I+T_E} i^r_{kg} \\
    Q_{kg}^0 &= 0 \\
    Q_{kg}^1 &= 0 \\
    H_{kg}^w &= h_k \mathcal{H}_g \\
    H_{kg}^c &= c_k \mathcal{C}_g \\
    H_{kg}^r &= 0 \\
    R_{kg} &= r^r_{kg} \\
    D_{kg} &= 0 \\
    V_{kg} &= e\ v_{kg},
\end{align*}
where $v_{kg}$ is the cumulative number of people who have received the first dose of any vaccine until 18 April 2021,  $i^r_{kg}$ stands for the estimated number of real infectious, $r^r_{kg}$ represents the number of  real recovered people as of 18 April,  $h_k$  is the reported individuals in hospital ward, $c_k$ is the reported individuals in critical care units, and $\mathcal{H}_g$ and $\mathcal{C}_g$ are the proportions of people at ward and critical care, respectively. 
The estimation of real  infectious individuals at any day $t$ is derived directly from data as follows:
\begin{align*}
    i^r_{kg}(t) &= i^d_{kg}(t) + i^u_{kg}(t),
\end{align*}
where the number of undetected infectious people $i^u_{kg}(t)$ come from upscaling the number of  detected individuals $i^d_{kg}(t)$ by a factor that depends on the index of age group $g$, i.e.,
\begin{align*}
    i^u_{kg}(t) &= (1 + 9g^{-2.46}) i^d_{kg}(t).
\end{align*}
The number of detected infectious people is calculated by summing the reported cases over the last $T_I + T_E$ days,
\begin{align*}
    i^d_{kg}(t) &= \sum_{\omega = \omega_0}^t i^{d}_k(\omega) \mathcal{I}^w_g(\omega),
\end{align*}
where $\omega_0 = t - T_I - T_E$, $i^d_k(t)$ is the number of cases in region $k$ reported by THL (Finnish Institute of Health and Welfare) at day $t$, and $\mathcal{I}^w_g(t)$ is the proportion of infected people in age group $g$.
We do not have daily counts as THL does not provide these on infected people per age group. We have chosen 18 April 2021 as the start day since it is a Sunday, and the weekly proportion  $\mathcal{I}^w_g(t)$ is the same for all the sums ($T_I + T_E = 7$ days, 1 week), which gives
\begin{align*}
    i^d_{kg}(t) &= \mathcal{I}^w_g \sum_{\omega = \omega_0}^t i^d_k(\omega).
\end{align*}
The numerical value for $\mathcal{I}^w_g$ can be found in Table \ref{tab:ParametersAge} and for the result of the summation $\sum_{\omega}i^d_k(\omega)$ see Table \ref{tab:InitialCondition}. The estimation of real recovered people at day $t$ is similar,
\begin{align*}
    r^r_{kg}(t) &= r^d_{kg}(t) + r^u_{kg}(t) \\
    r^u_{kg}(t) &= (1 + 9g^{-2.46}) r^d_{kg}(t) \\
    r^d_{kg}(t) &= \sum_{\omega = 0}^{t-(T_I+T_E)} i^d_k(\omega) \mathcal{I}^w_g(\omega)
\end{align*}
in which $\omega = 0$ marks the beginning of the coronavirus epidemic in Finland. The estimated values at 18 April 2021 of the real infected people $i^r_{kg}$ and real recovered people  $r^r_{kg}$ can be found in Tables \ref{tab:InitialInfected} and \ref{tab:InitialRecovered}, respectively.

\begin{table}[h]
\centering
\tiny
\begin{tabular}{clrrrrrrrrr}
\toprule
\bf Parameter & \bf Description & 0--9 & 10--19 & 20--29 & 30--39 & 40--49 & 50--59 & 60--69 & 70--79 & 80+ \\
\midrule
$\mathcal{H}_g$*     & Proportion in ward & 0.0058 & 0.0107 & 0.0467 & 0.0605 & 0.0911 & 0.1450 & 0.1547 & 0.2008 & 0.2847 \\
$\mathcal{C}_g$*     & Proportion in critical care & 0.0038 & 0.0069 & 0.0301 & 0.0390 & 0.0978 & 0.2231 & 0.2891 & 0.2448 & 0.0655 \\
$\mathcal{I}^w$**     & Infections & 240 & 310 & 354 & 355 & 294 & 200 & 101 & 44 & 31 \\
$\mathcal{I}^w_g$   & Normalized $\mathcal{I}^w$ & 0.1244 & 0.1607 & 0.1835 & 0.1840 & 0.1524 & 0.1037 & 0.0524 & 0.0228 & 0.0161 \\
\bottomrule
\end{tabular}
\caption{\label{tab:ParametersAge} Parameters for age compartments. \\
* From \cite{Sjodin_Rocklov_Britton_2021-04}. \\
** Reported number of infected people in Finland by age group during 12--18 April 2021 \cite{thl_api}.}.
\end{table}

\begin{table}[h]
\centering
\small
\begin{tabular}{clrrrrrr}
\toprule
\bf Parameter & \bf Description & HYKS & TYKS & TAYS & KYS & OYS & Total\\
\midrule
$h_k$*     & Ward & 88 & 11 & 17 & 5 & 11 & 132\\
$c_k$*     & Critical care & 21 & 6 & 2 & 5 & 0 & 34\\
$\sum_{\omega}i^d_k(\omega)$** & Infectious & 1179 & 347 & 225 & 80 & 76  & 1907\\
\bottomrule
\end{tabular}
\caption{\label{tab:InitialCondition} Parameters estimated from data \\
* Numbers reported by \cite{helsingi_sanomat_api} on 19 April 2021.\\
** Sum of reported number of infected people by region from 12--18 April 2021 \cite{thl_api}.
}
\end{table}

\begin{table}[h]
\centering
\scriptsize
\begin{tabular}{c|rrrrrrr}
\toprule
 $v_{kg}$ & HYKS & TYKS & TAYS & KYS & OYS & Total  & Total/$N_g\ (\%)$ ) \\
\midrule
0--9    & 0 & 0 & 0 & 0 & 0 & 0 & 0 \\
10--19  & 1802 & 895 & 647 & 467 & 397 & 4208 & 0.69 \\
20--29  & 14326 & 6391 & 4806 & 4111 & 3570 & 33204 & 5.04 \\
30--39  & 22284 & 8958 & 7639 & 6314 & 5640 & 50835 & 7.11 \\
40--49  & 32713 & 12418 & 11718 & 8261 & 7842 & 72952 & 11.01 \\
50--59  & 53123 & 20671 & 20143 & 15545 & 14676 & 124158 & 17.47 \\
60--69  & 111319 & 46461 & 47329 & 40640 & 33953 & 279702 & 39.54 \\
70--79  & 184419 & 87350 & 85498 & 79872 & 63631 & 500770 & 85.85 \\
80+     & 94809 & 50321 & 49239 & 47561 & 37125 & 279055 & 88.96 \\
\midrule
Total & 514795 & 233465 & 227019 & 202771 & 166834 & 1344884 & 24.44 \\
Total/$N_k\ (\%)$ & 23.42 & 26.87 & 25.15 & 25.43 & 22.65 & 24.44 &   \\
\bottomrule
\end{tabular}
\caption{\label{tab:InitialVaccination} Number of vaccinated people in Finland by region with 9 age groups and 10y age resolution as of 18 April 2021. The entry on row $g$ and column $k$ indicates the number of individuals who have received the first dose in age group $g$ and region $k$. Data from \cite{thl_api}.}
\end{table}

\begin{table}[h]
\centering
\tiny
\begin{tabular}{c|rrrrrrr}
\toprule
 $S^u_{kg}$ & HYKS & TYKS & TAYS & KYS & OYS & Total  & Total/$N_g\ (\%)$  \\
\midrule
0--9    & 171581.02 & 70816.88 & 81903.12 & 67257.71 & 76718.73 & 468277.46 & 85.97 \\
10--19  & 209955.74 & 85708.42 & 96837.14 & 80990.48 & 88887.55 & 562379.33 & 92.52 \\
20--29  & 228534.80 & 89876.46 & 96665.64 & 85189.05 & 78392.50 & 578658.45 & 87.88 \\
30--39  & 270855.73 & 91783.82 & 102310.14 & 82761.97 & 80860.11 & 628571.77 & 87.92 \\
40--49  & 235429.06 & 85427.65 & 93018.80 & 75258.19 & 73005.88 & 562139.57 & 84.84 \\
50--59  & 221934.99 & 88013.76 & 93279.21 & 86411.78 & 75323.60 & 564963.33 & 79.49 \\
60--69  & 137785.90 & 65351.37 & 69714.24 & 78372.62 & 65760.56 & 416984.69 & 58.95 \\
70--79  & 24360.73 & 11823.42 & 14133.96 & 15379.26 & 11743.93 & 77441.30 & 13.28 \\
80+     & 8689.93 & 5505.76 & 6064.27 & 5424.06 & 4885.78 & 30569.80 & 9.75 \\
\midrule
Total & 1509127.92 & 594307.53 & 653926.52 & 577045.10 & 555578.63 & 3889985.71 & 70.68 \\
Total/$N_k\ (\%)$ & 68.65 & 68.39 & 72.44 & 72.38 & 75.43 & 70.68 & \\
\bottomrule
\end{tabular}
\caption{\label{tab:InitialSusceptible} Number of susceptible people in Finland by region with 9 age groups and 10y age resolution as of 18 April 2021. The entry on row $g$ and column $k$ indicates the number of individuals who are susceptible in age group $g$ and region $k$.}
\end{table}

\begin{table}[h]
\centering
\tiny
\begin{tabular}{c|rrrrrrr}
\toprule
 $i^r_{kg}$ & HYKS & TYKS & TAYS & KYS & OYS & Total  & Total/$N_g\ (\%)$  \\
\midrule
0--9    &  1613.56 & 473.53 & 307.93 & 110.86 & 102.64 & 2608.52 & 0.48 \\
10--19  &  688.86 & 202.16 & 131.46 & 47.33 & 43.82 & 1113.63 & 0.18 \\
20--29  &  563.26 & 165.30 & 107.49 & 38.70 & 35.83 & 910.58 & 0.14 \\
30--39  &  498.45 & 146.28 & 95.12 & 34.24 & 31.71 & 805.81 & 0.11 \\
40--49  &  390.24 & 114.52 & 74.47 & 26.81 & 24.82 & 630.87 & 0.10 \\
50--59  &  257.88 & 75.68 & 49.21 & 17.72 & 16.40 & 416.90 & 0.06 \\
60--69  &  128.09 & 37.59 & 24.45 & 8.80 & 8.15 & 207.08 & 0.03 \\
70--79  &  55.24 & 16.21 & 10.54 & 3.80 & 3.51 & 89.30 & 0.02 \\
80+     &  38.66 & 11.35 & 7.38 & 2.66 & 2.46 & 62.50 & 0.02 \\
\midrule
Total  & 4234.25 & 1242.62 & 808.06 & 290.90 & 269.35 & 6845.19 & 0.12 \\
Total/$N_k\ (\%)$ & 0.19 & 0.14 & 0.09 & 0.04 & 0.04 & 0.12 & \\
\bottomrule
\end{tabular}
\caption{\label{tab:InitialInfected} Estimated number of real infectious people in Finland by region with 9 age groups and 10y age resolution as of 18 April 2021. The entry on row $g$ and column $k$ indicates the number of individuals who are infectious in age group $g$ and region $k$.}
\end{table}

\begin{table}[h]
\centering
\tiny
\begin{tabular}{c|rrrrrrr}
\toprule
 $r^r_{kg}$ & HYKS & TYKS & TAYS & KYS & OYS & Total  & Total/$N_g\ (\%)$  \\
\midrule
0--9    &  48404.26 & 11519.76 & 5855.94 & 4539.03 & 3486.14 & 73805.15 & 13.55 \\
10--19  &  25865.12 & 6195.55 & 3247.81 & 2708.32 & 2142.57 & 40159.38 & 6.61 \\
20--29  &  29244.78 & 7138.85 & 3694.80 & 3127.12 & 2512.22 & 45717.78 & 6.94 \\
30--39  &  22527.03 & 5203.81 & 2763.23 & 2279.31 & 1915.45 & 34688.83 & 4.85 \\
40--49  &  17445.50 & 4017.41 & 2137.80 & 1755.16 & 1474.33 & 26830.20 & 4.05 \\
50--59  &  13794.60 & 3110.75 & 1682.52 & 1410.74 & 1207.43 & 21206.04 & 2.98 \\
60--69  &  6751.83 & 1529.35 & 824.89 & 699.27 & 598.56 & 10403.91 & 1.47 \\
70--79  &  3231.19 & 723.73 & 398.56 & 333.74 & 288.36 & 4975.57 & 0.85 \\
80+     &  2633.95 & 531.39 & 297.37 & 262.54 & 244.63 & 3969.89 & 1.27 \\
\midrule
Total & 169898.27 & 39970.61 & 20902.92 & 17115.24 & 13869.71 & 261756.74 & 4.76 \\
Total/$N_k\ (\%)$ & 7.73 & 4.60 & 2.32 & 2.15 & 1.88 & 4.76 & \\
\bottomrule
\end{tabular}
\caption{\label{tab:InitialRecovered} Estimated number of real recovered people in Finland by region with 9 age groups and 10y age resolution as of 18 April 2021. The entry on row $g$ and column $k$ indicates the number of individuals who are recovered in age group $g$ and region $k$.}
\end{table}

\section{Time plots for different values of mobility levels}

\begin{figure}[p]
     \centering
     \begin{subfigure}[b]{0.3\paperheight}
         \centering
         \includegraphics[width=\textwidth]{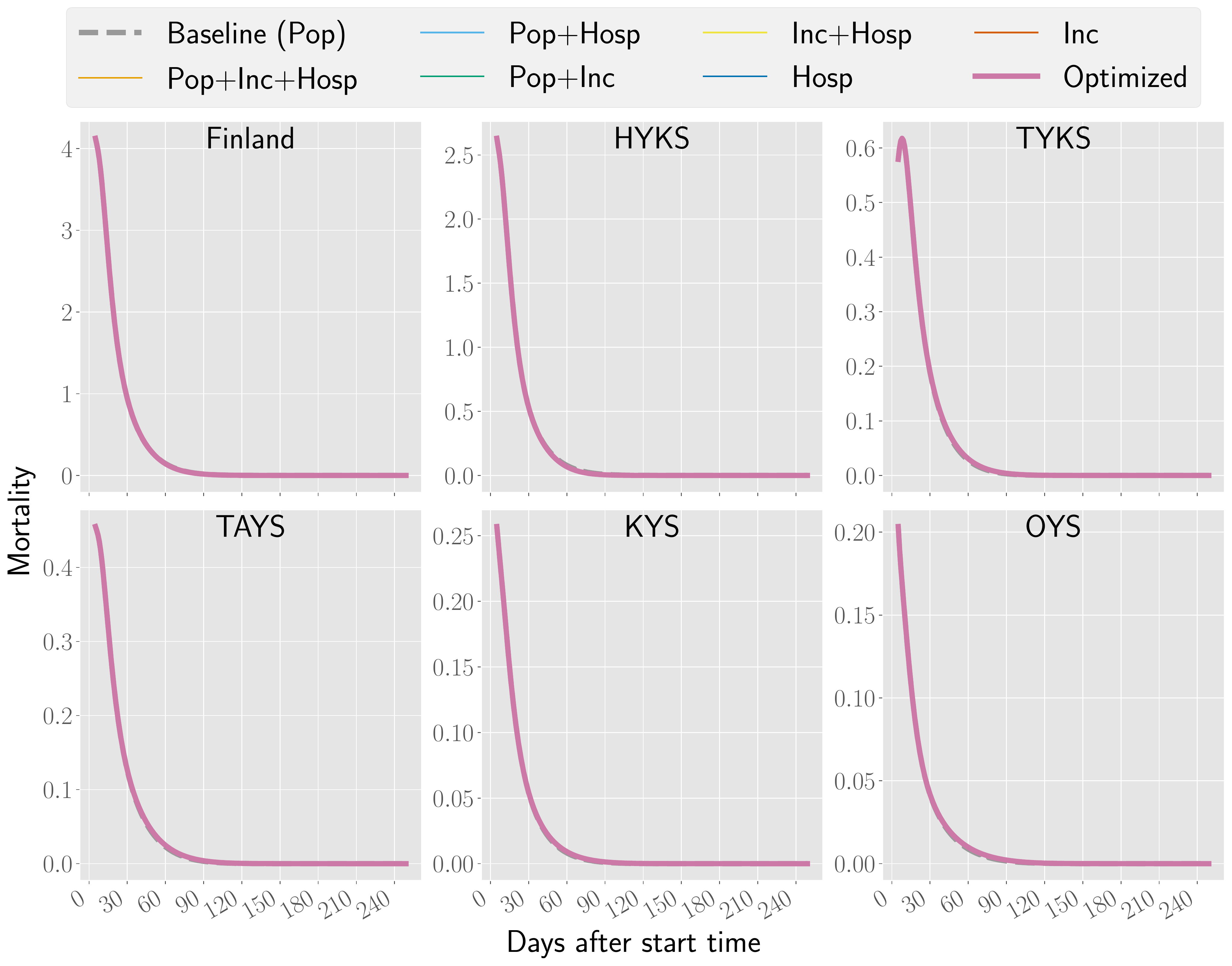}
         \label{fig:r0.75_tau0.0_metric_mortality}
     \end{subfigure}
     
     \begin{subfigure}[b]{0.3\paperheight}
         \centering
         \includegraphics[width=\textwidth]{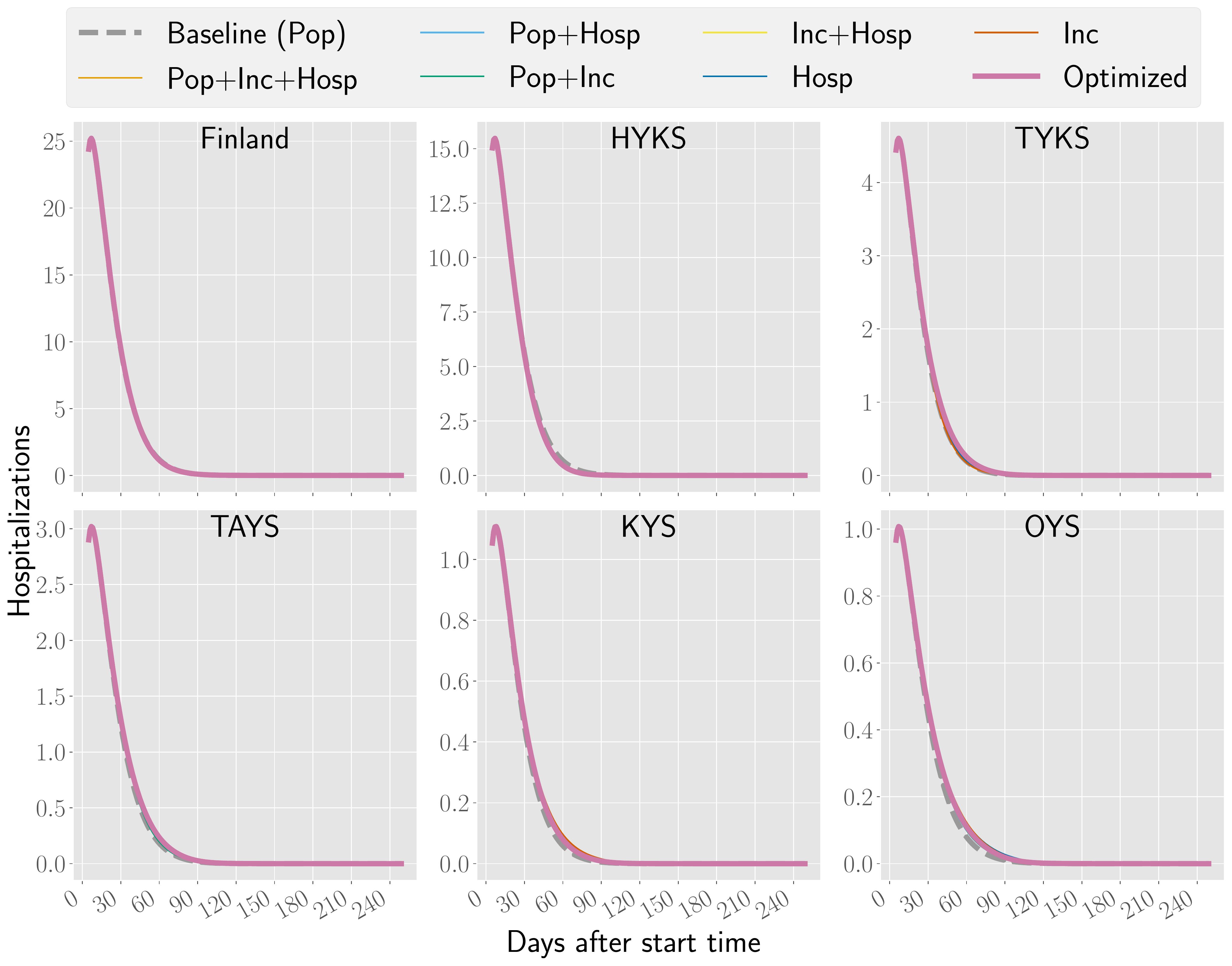}
         \label{fig:r0.75_tau0.0_metric_new hospitalizations}
     \end{subfigure}

     \begin{subfigure}[b]{0.3\paperheight}
         \centering
         \includegraphics[width=\textwidth]{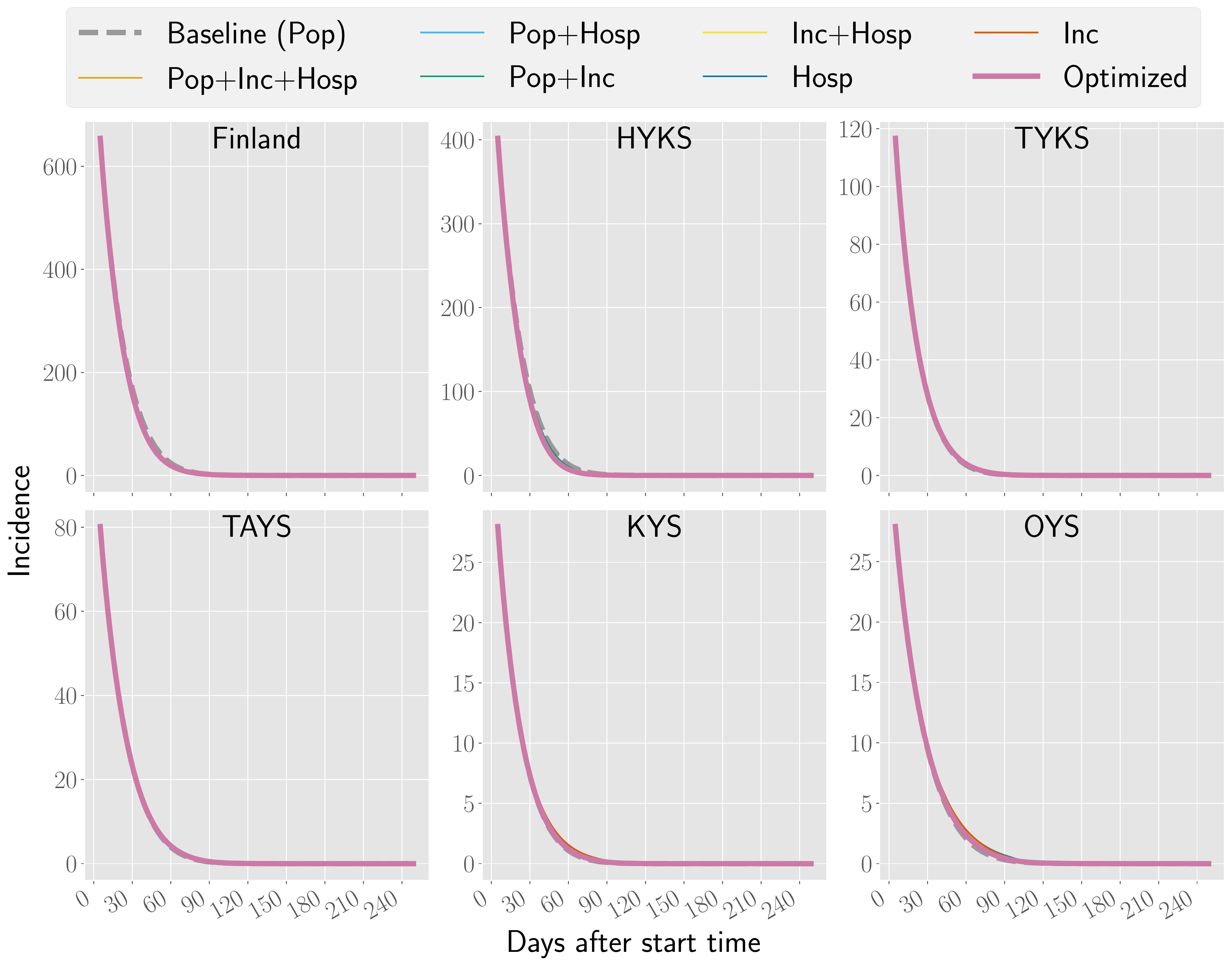}
         \label{fig:r0.75_tau0.0_metric_incidence}
     \end{subfigure}

     \caption{Different metrics in Finland and the five hospital catchment areas included here
      for all the vaccination strategies. For these scenarios, 
      the basic reproduction number $\Reff = 0.75$ and the mobility value $\tau = 0.0$.}
    \end{figure}

    \begin{figure}[p]
     \centering
     \begin{subfigure}[b]{0.3\paperheight}
         \centering
         \includegraphics[width=\textwidth]{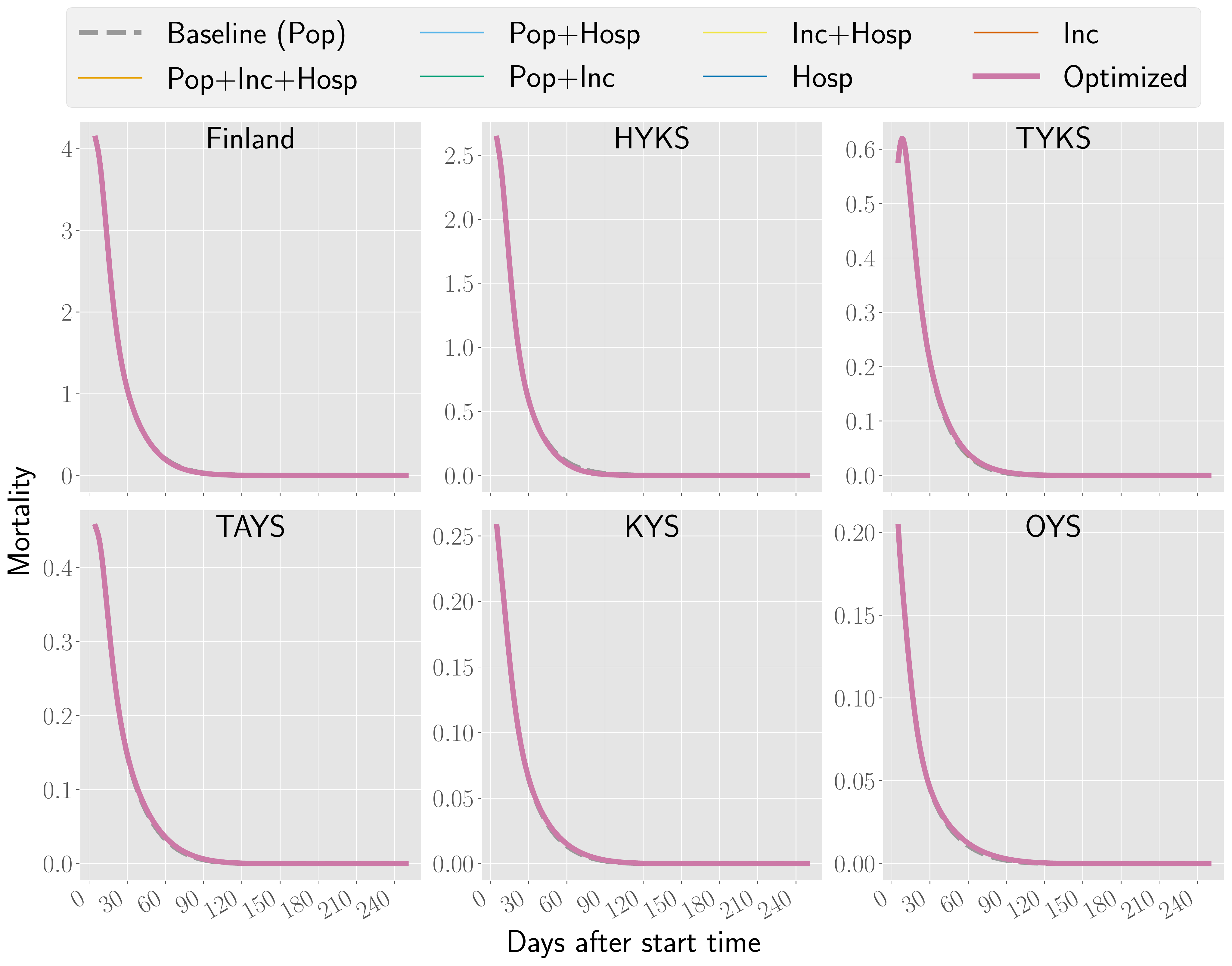}
         \label{fig:r0.75_tau0.5_metric_mortality}
     \end{subfigure}
     
     \begin{subfigure}[b]{0.3\paperheight}
         \centering
         \includegraphics[width=\textwidth]{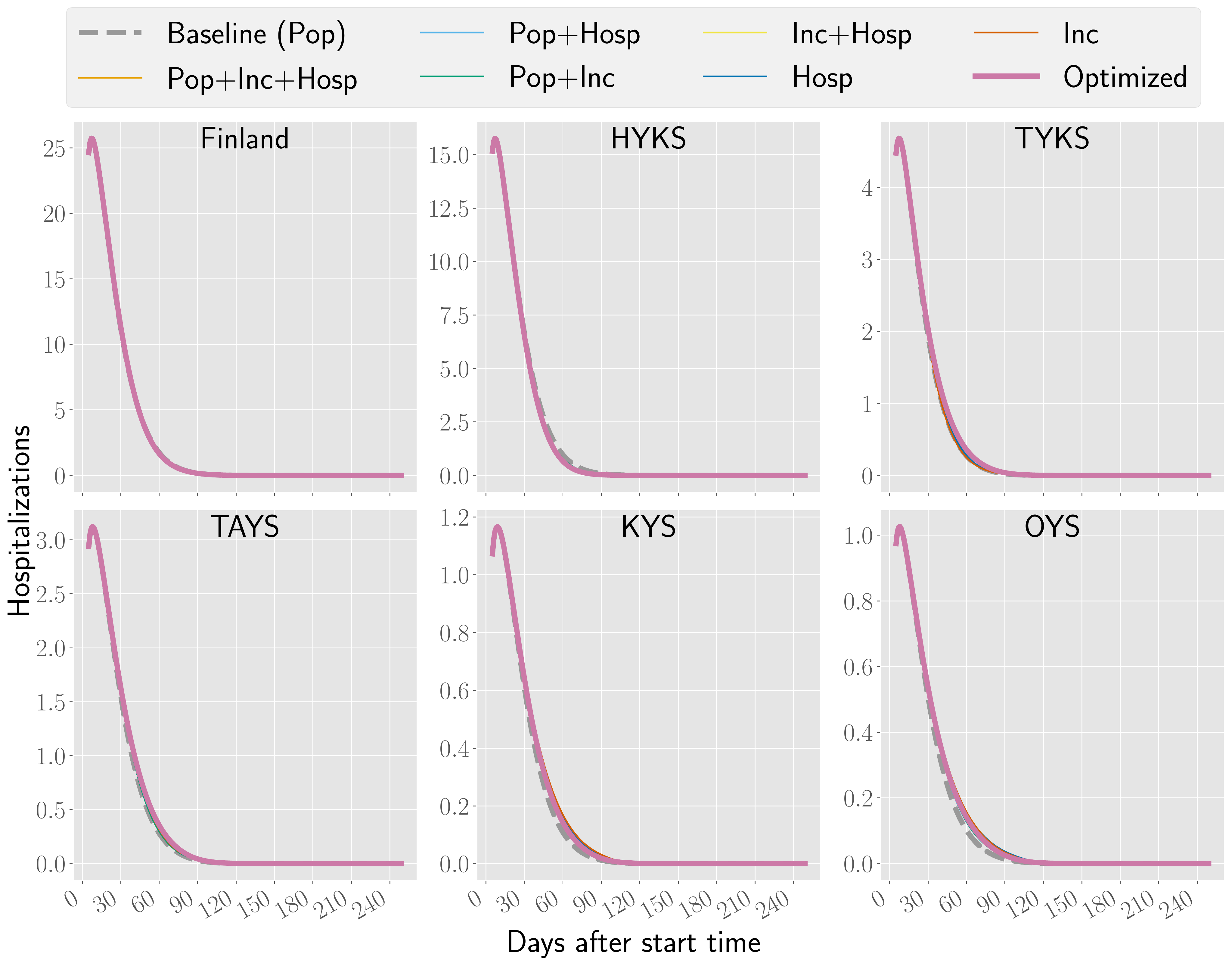}
         \label{fig:r0.75_tau0.5_metric_new hospitalizations}
     \end{subfigure}

     \begin{subfigure}[b]{0.3\paperheight}
         \centering
         \includegraphics[width=\textwidth]{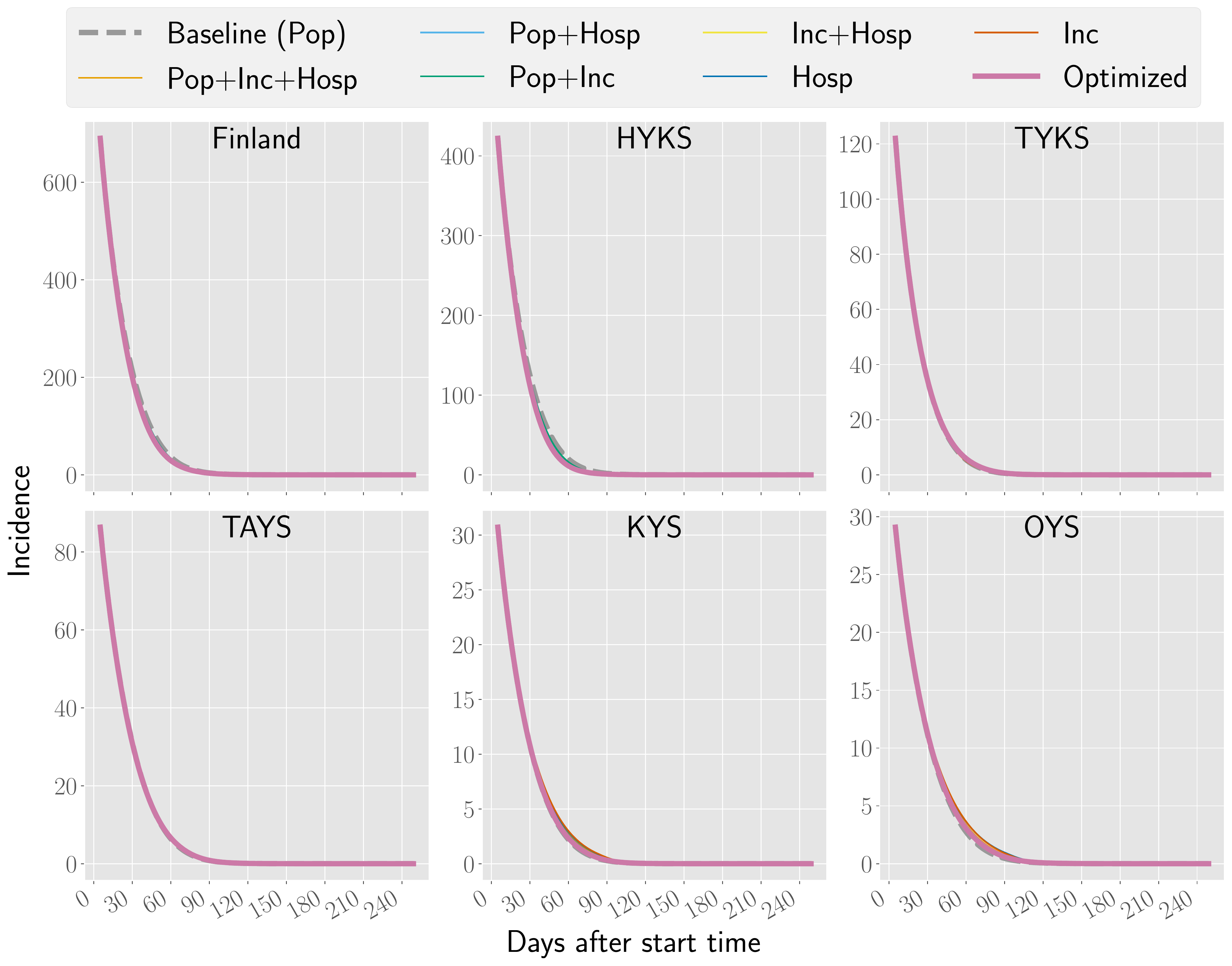}
         \label{fig:r0.75_tau0.5_metric_incidence}
     \end{subfigure}

     \caption{Different metrics in Finland and the five hospital catchment areas included here
      for all the vaccination strategies. For these scenarios, 
      the basic reproduction number $\Reff = 0.75$ and the mobility value $\tau = 0.5$.}
    \end{figure}

    \begin{figure}[p]
     \centering
     \begin{subfigure}[b]{0.3\paperheight}
         \centering
         \includegraphics[width=\textwidth]{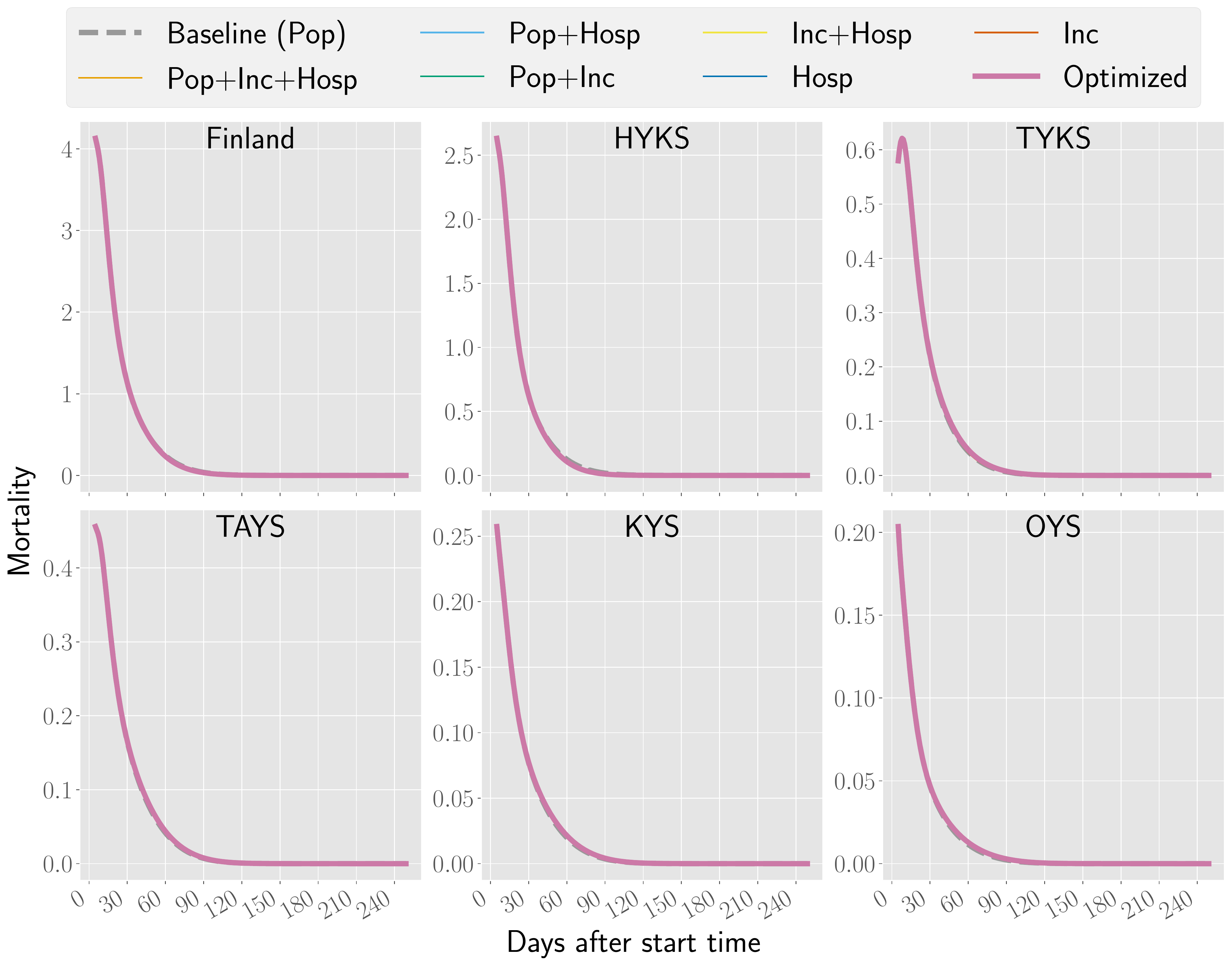}
         \label{fig:r0.75_tau1.0_metric_mortality}
     \end{subfigure}
     
     \begin{subfigure}[b]{0.3\paperheight}
         \centering
         \includegraphics[width=\textwidth]{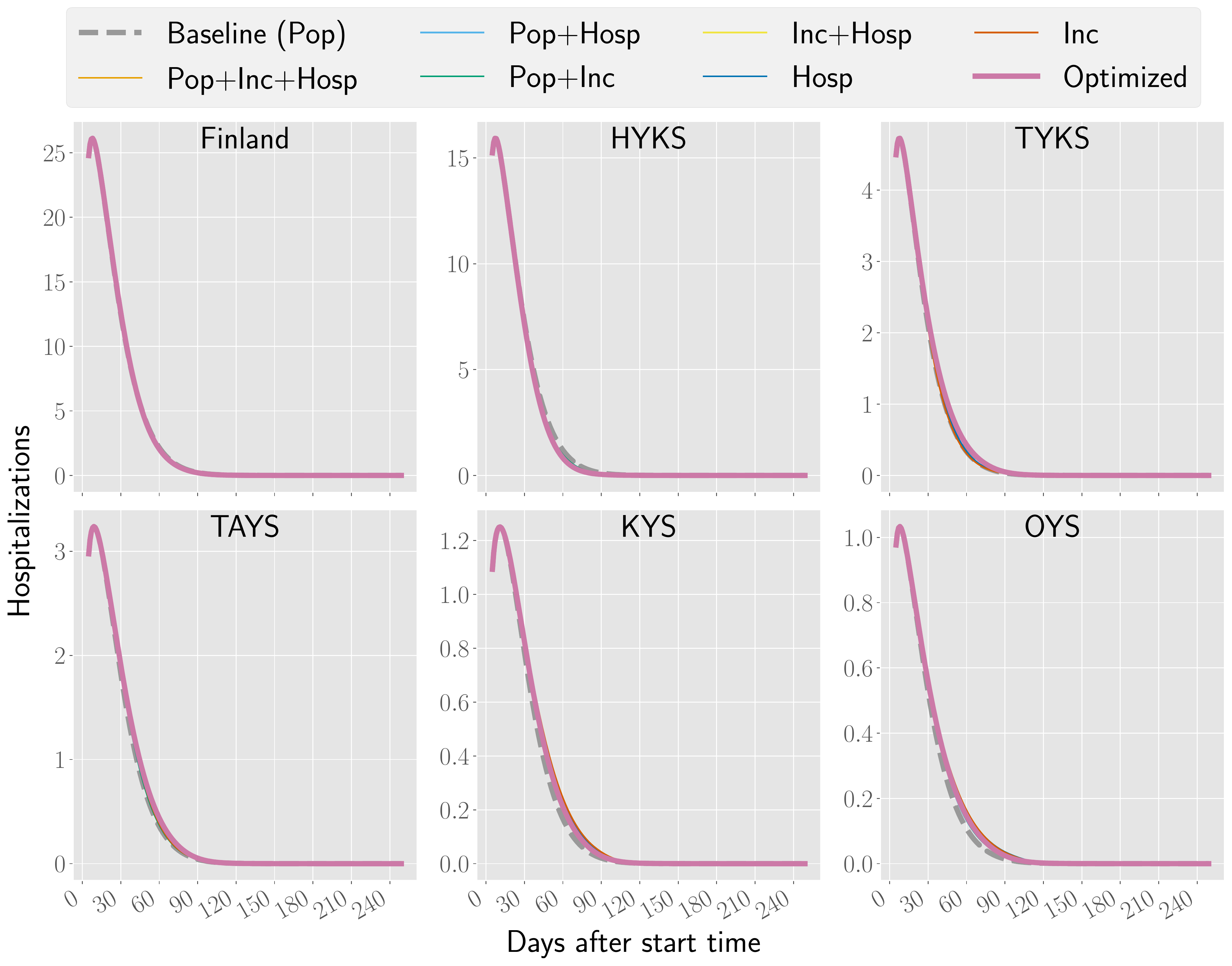}
         \label{fig:r0.75_tau1.0_metric_new hospitalizations}
     \end{subfigure}

     \begin{subfigure}[b]{0.3\paperheight}
         \centering
         \includegraphics[width=\textwidth]{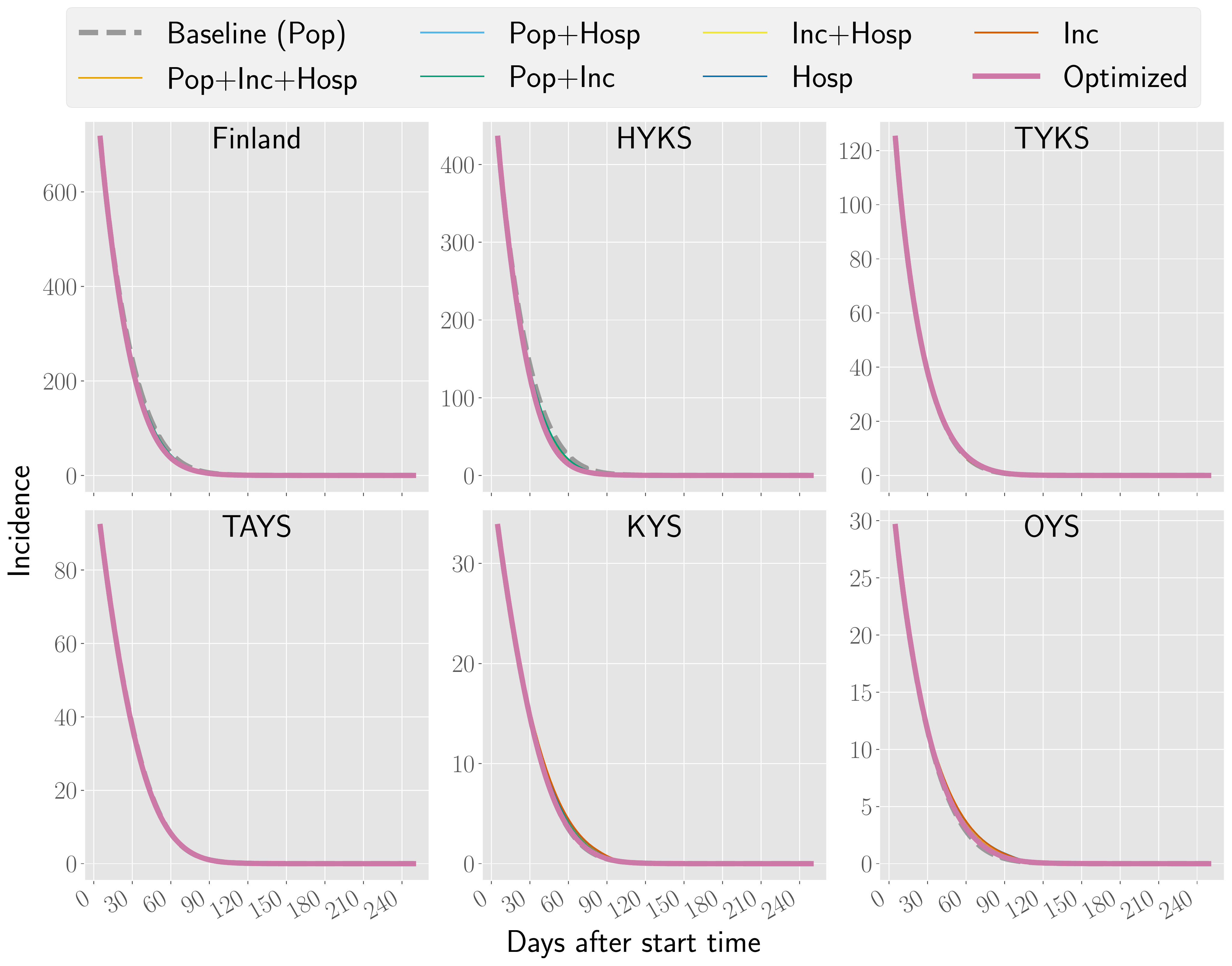}
         \label{fig:r0.75_tau1.0_metric_incidence}
     \end{subfigure}

     \caption{Different metrics in Finland and the five hospital catchment areas included here
      for all the vaccination strategies. For these scenarios, 
      the basic reproduction number $\Reff = 0.75$ and the mobility value $\tau = 1.0$.}
    \end{figure}

    \begin{figure}[p]
     \centering
     \begin{subfigure}[b]{0.3\paperheight}
         \centering
         \includegraphics[width=\textwidth]{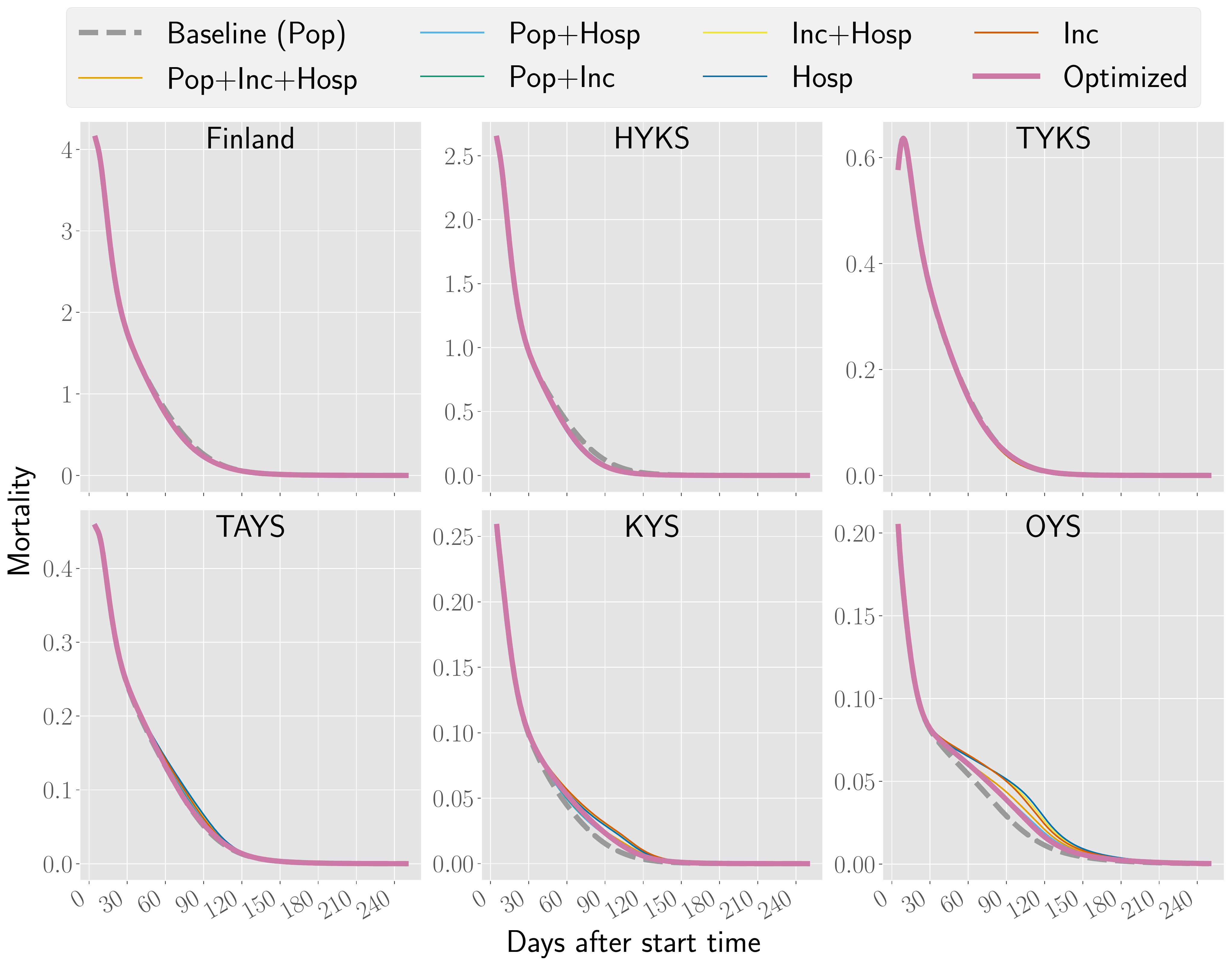}
         \label{fig:r1.0_tau0.0_metric_mortality}
     \end{subfigure}
     
     \begin{subfigure}[b]{0.3\paperheight}
         \centering
         \includegraphics[width=\textwidth]{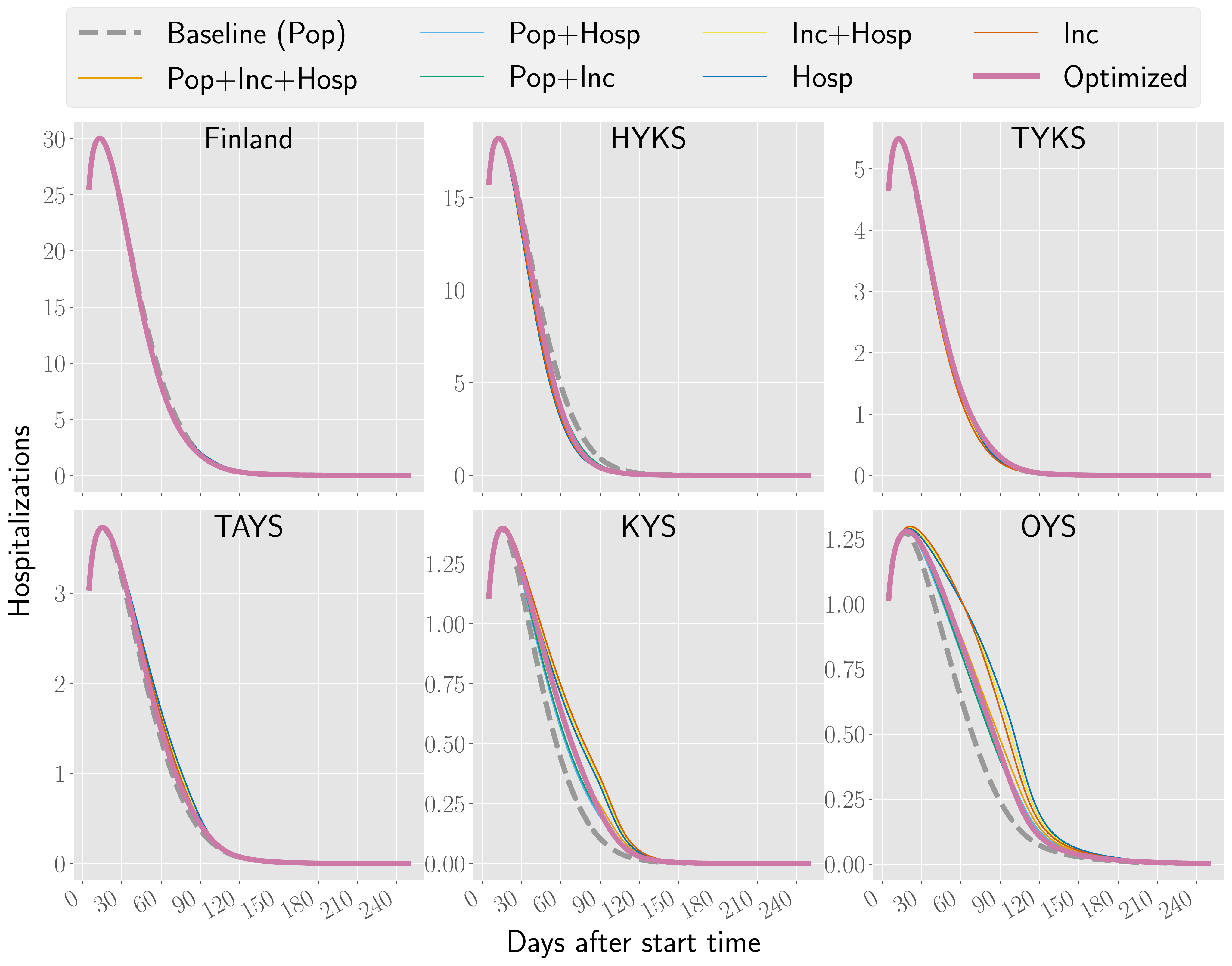}
         \label{fig:r1.0_tau0.0_metric_new hospitalizations}
     \end{subfigure}

     \begin{subfigure}[b]{0.3\paperheight}
         \centering
         \includegraphics[width=\textwidth]{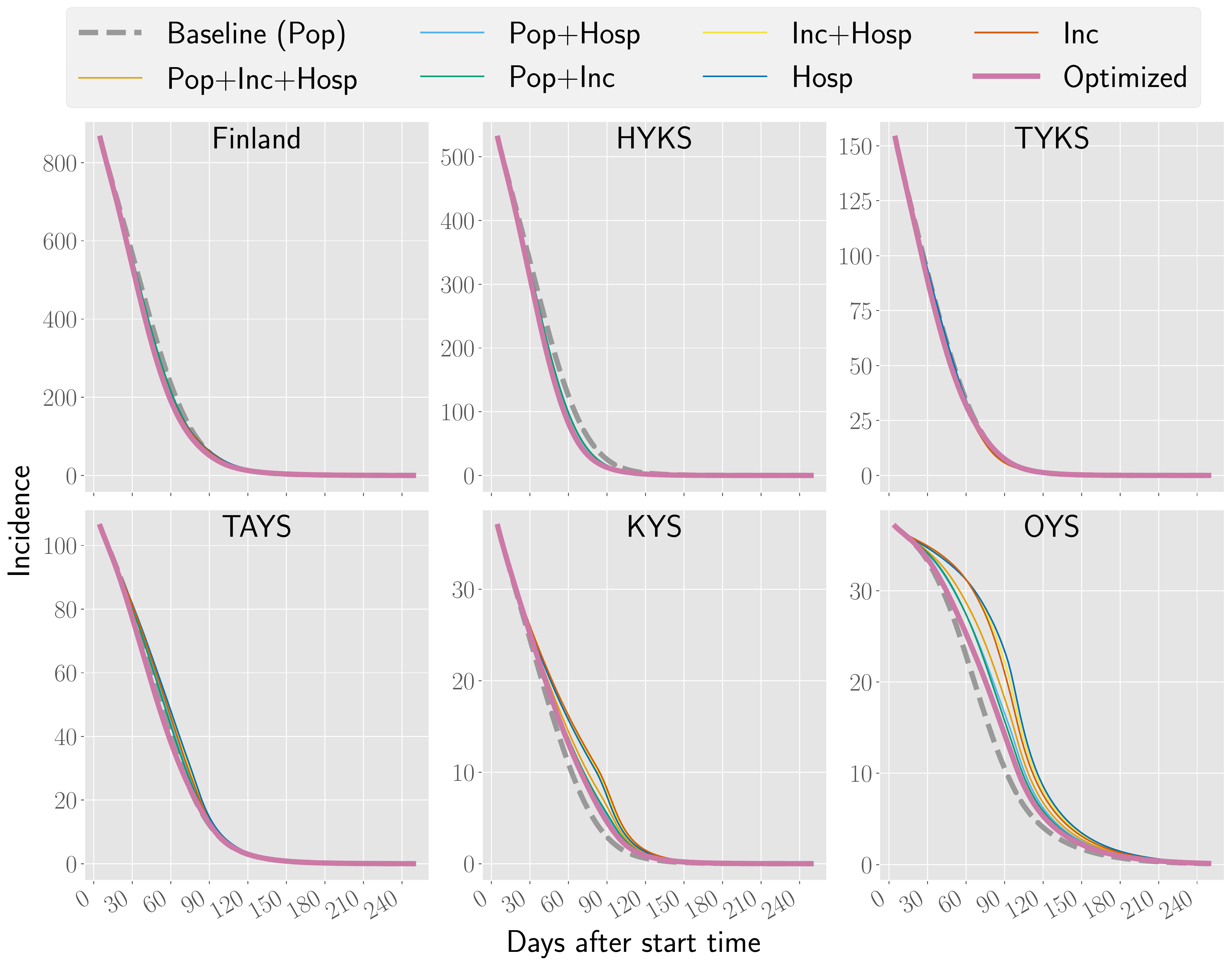}
         \label{fig:r1.0_tau0.0_metric_incidence}
     \end{subfigure}

     \caption{Different metrics in Finland and the five hospital catchment areas included here
      for all the vaccination strategies. For these scenarios, 
      the basic reproduction number $\Reff = 1.0$ and the mobility value $\tau = 0.0$.}
    \end{figure}

    \begin{figure}[p]
     \centering
     \begin{subfigure}[b]{0.3\paperheight}
         \centering
         \includegraphics[width=\textwidth]{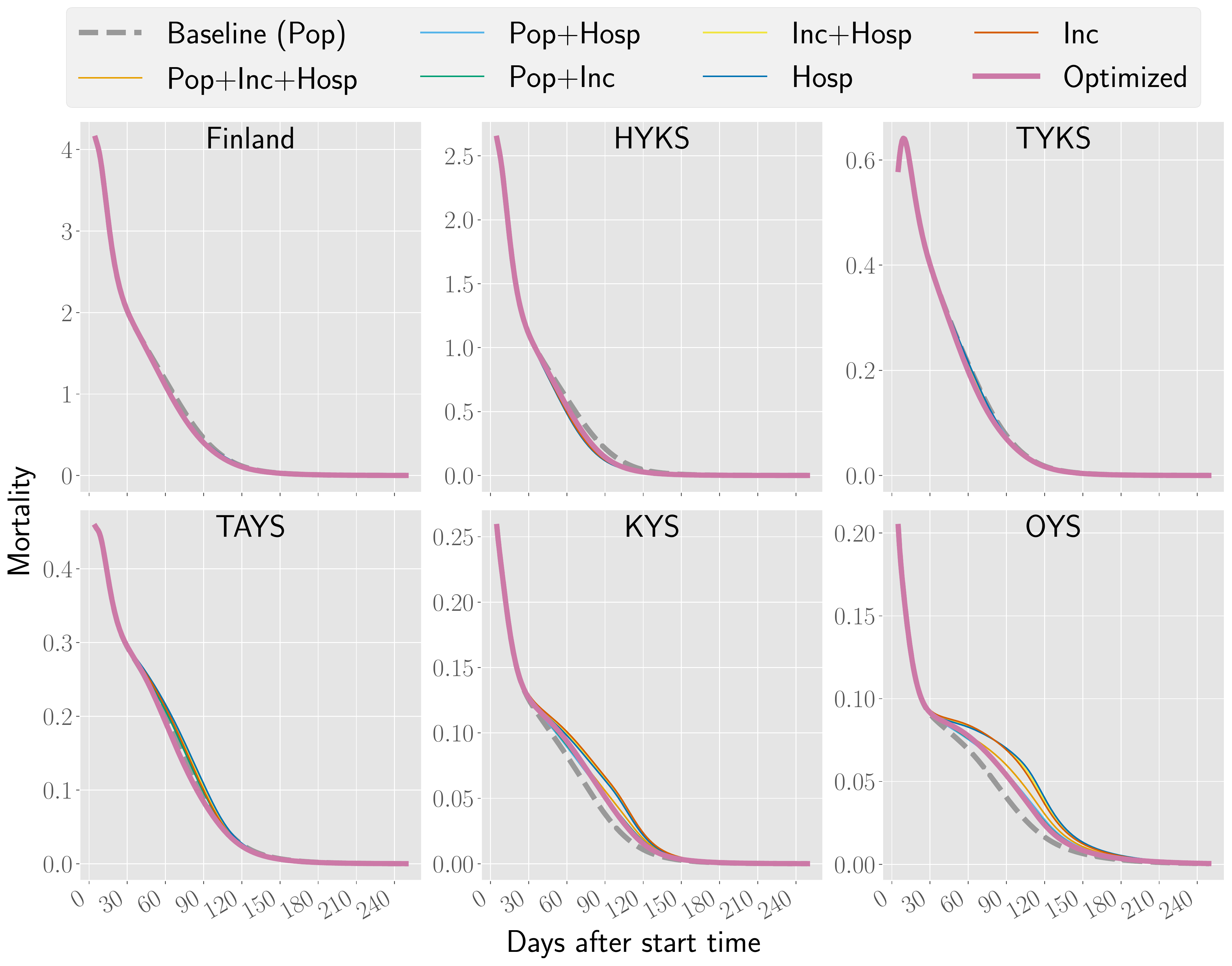}
         \label{fig:r1.0_tau0.5_metric_mortality}
     \end{subfigure}
     
     \begin{subfigure}[b]{0.3\paperheight}
         \centering
         \includegraphics[width=\textwidth]{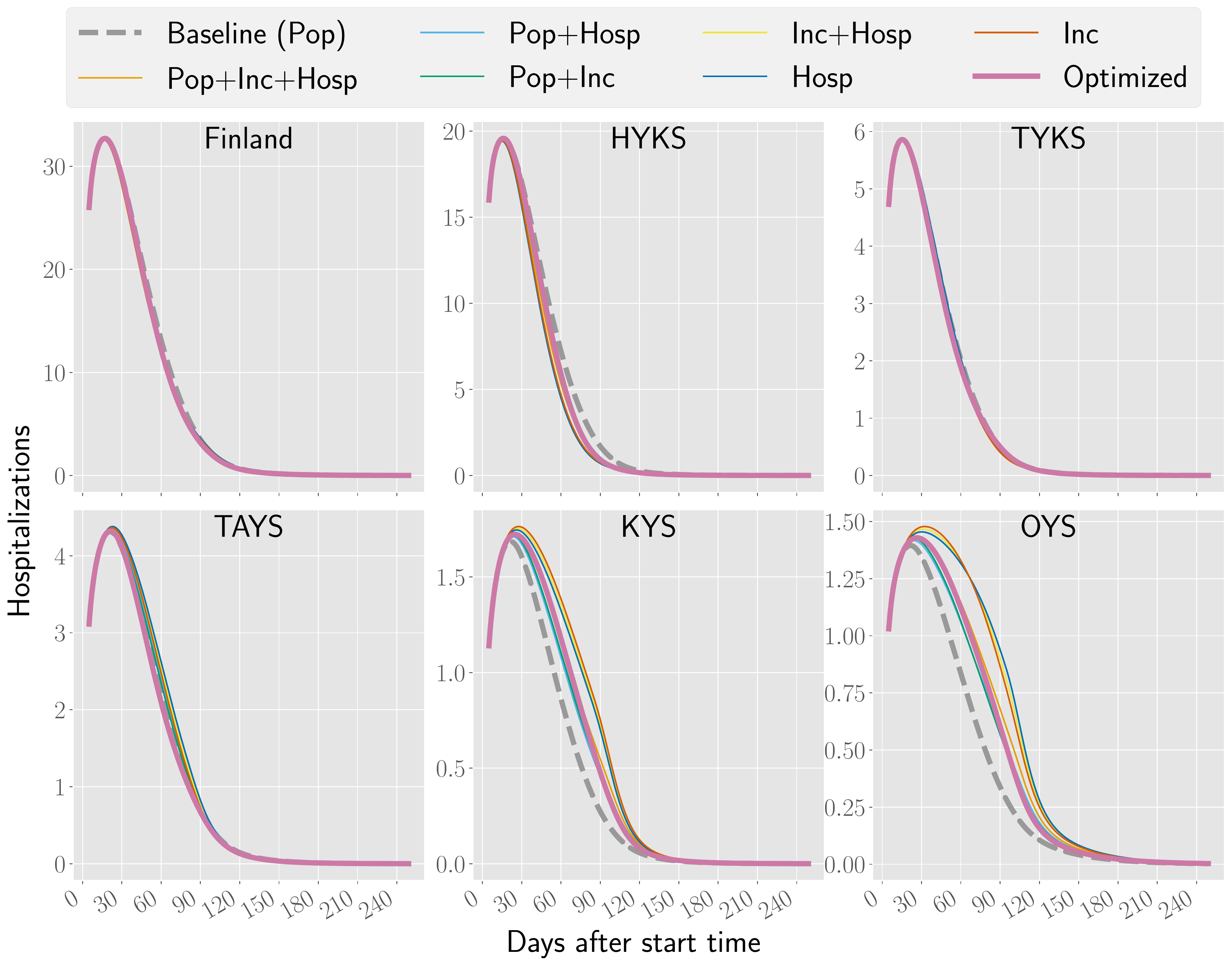}
         \label{fig:r1.0_tau0.5_metric_new hospitalizations}
     \end{subfigure}

     \begin{subfigure}[b]{0.3\paperheight}
         \centering
         \includegraphics[width=\textwidth]{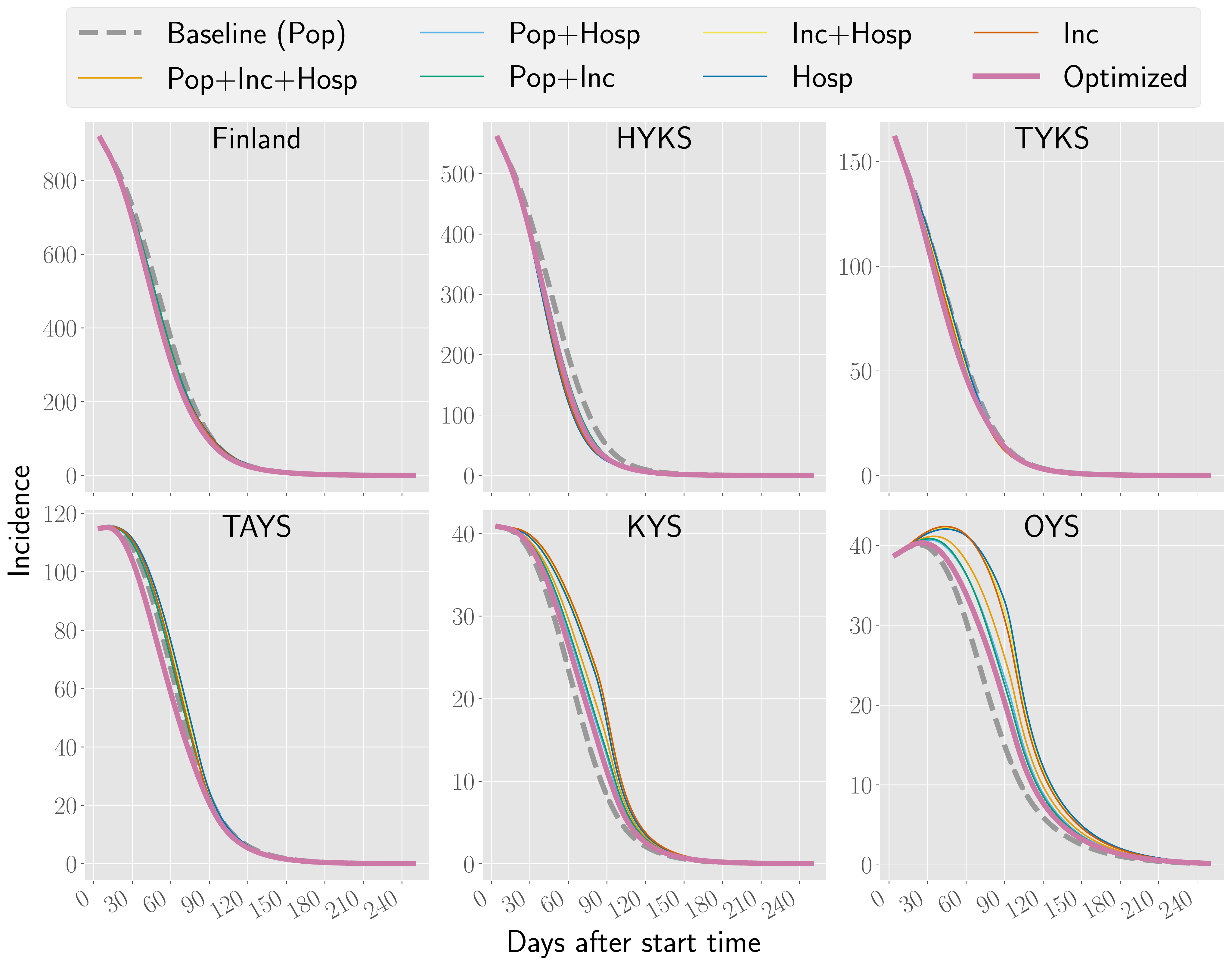}
         \label{fig:r1.0_tau0.5_metric_incidence}
     \end{subfigure}

     \caption{Different metrics in Finland and the five hospital catchment areas included here
      for all the vaccination strategies. For these scenarios, 
      the basic reproduction number $\Reff = 1.0$ and the mobility value $\tau = 0.5$.}
    \end{figure}

    \begin{figure}[p]
     \centering
     \begin{subfigure}[b]{0.3\paperheight}
         \centering
         \includegraphics[width=\textwidth]{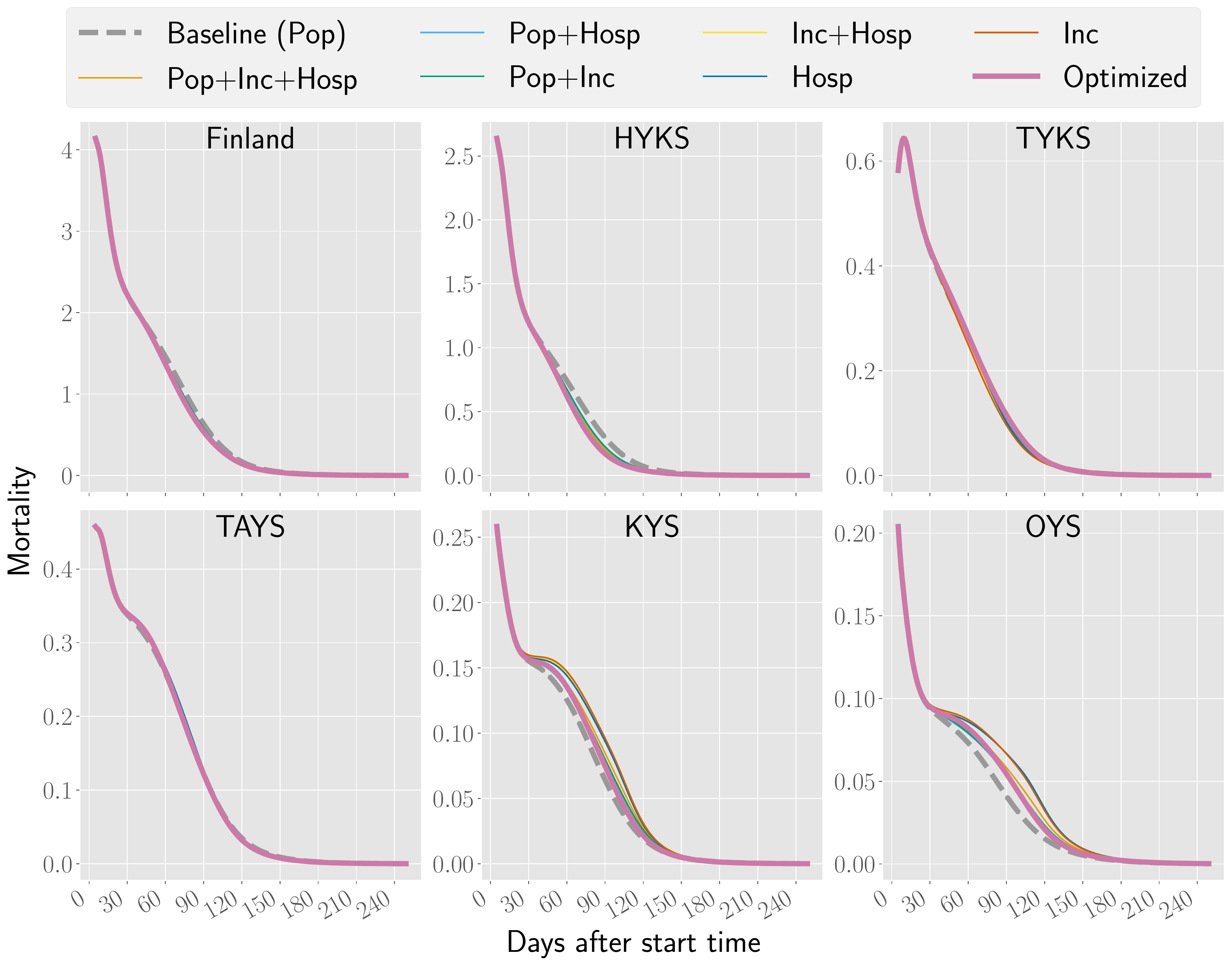}
         \label{fig:r1.0_tau1.0_metric_mortality}
     \end{subfigure}
     
     \begin{subfigure}[b]{0.3\paperheight}
         \centering
         \includegraphics[width=\textwidth]{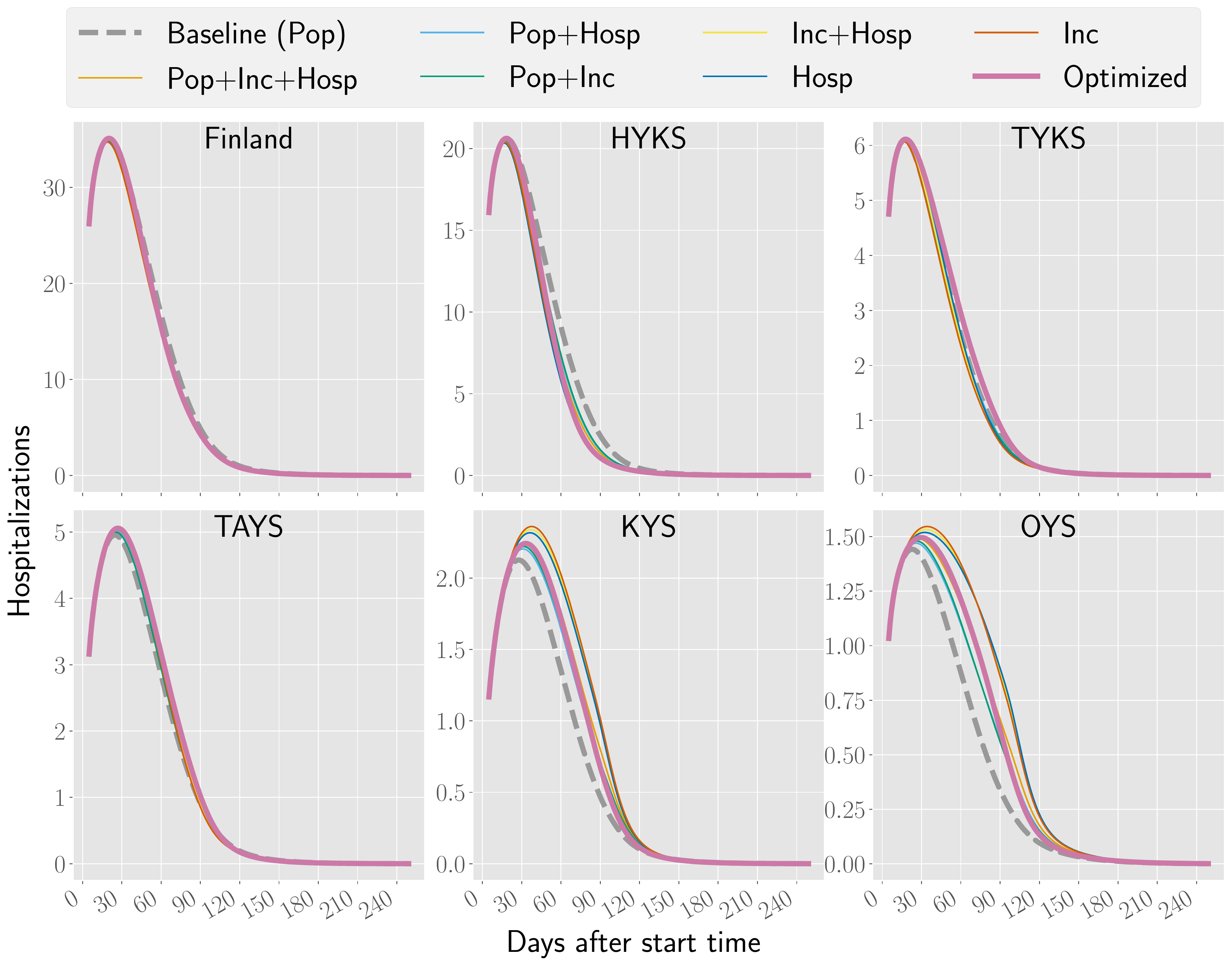}
         \label{fig:r1.0_tau1.0_metric_new hospitalizations}
     \end{subfigure}

     \begin{subfigure}[b]{0.3\paperheight}
         \centering
         \includegraphics[width=\textwidth]{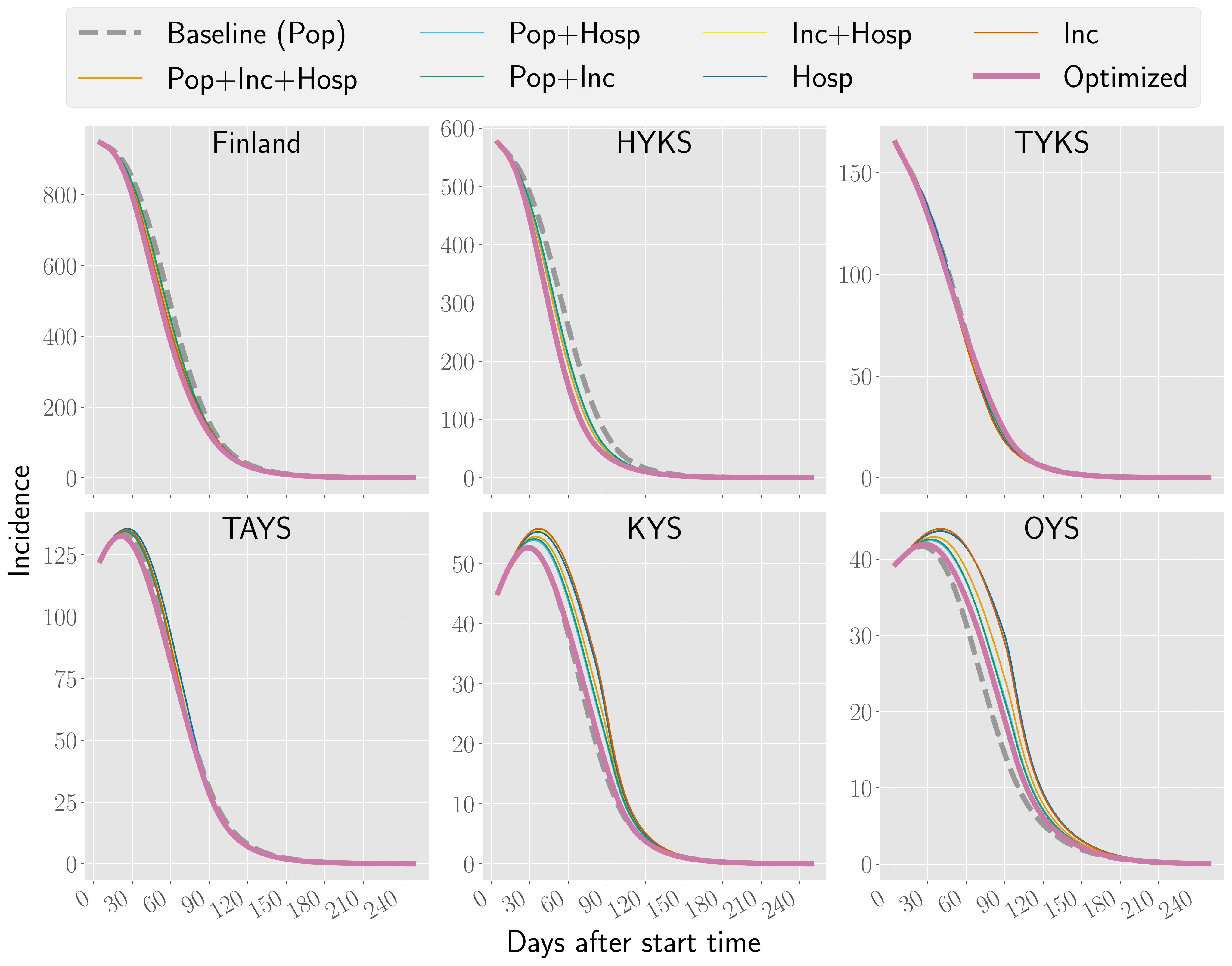}
         \label{fig:r1.0_tau1.0_metric_incidence}
     \end{subfigure}

     \caption{Different metrics in Finland and the five hospital catchment areas included here
      for all the vaccination strategies. For these scenarios, 
      the basic reproduction number $\Reff = 1.0$ and the mobility value $\tau = 1.0$.}
    \end{figure}

    \begin{figure}[p]
     \centering
     \begin{subfigure}[b]{0.3\paperheight}
         \centering
         \includegraphics[width=\textwidth]{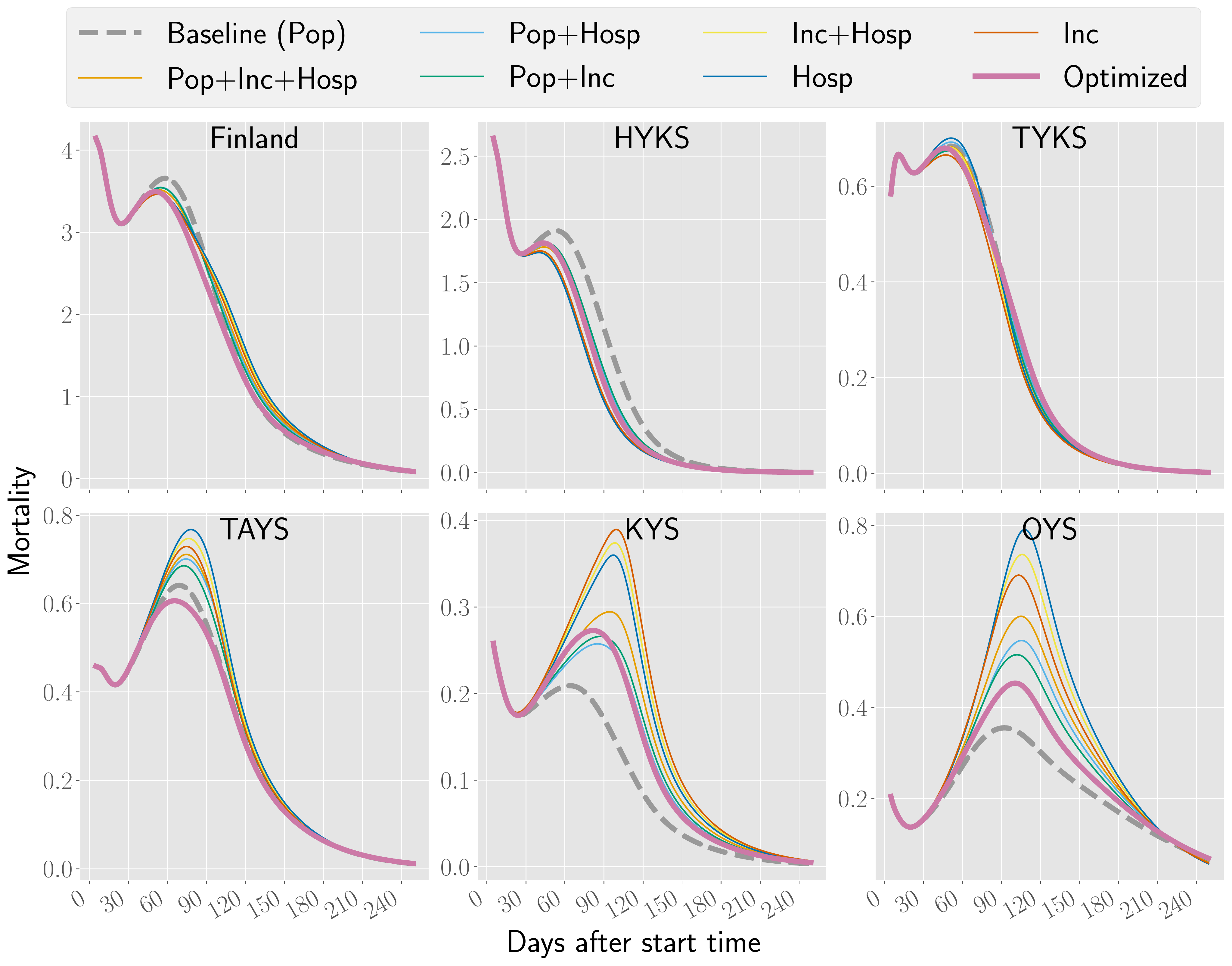}
         \label{fig:r1.25_tau0.0_metric_mortality}
     \end{subfigure}
     
     \begin{subfigure}[b]{0.3\paperheight}
         \centering
         \includegraphics[width=\textwidth]{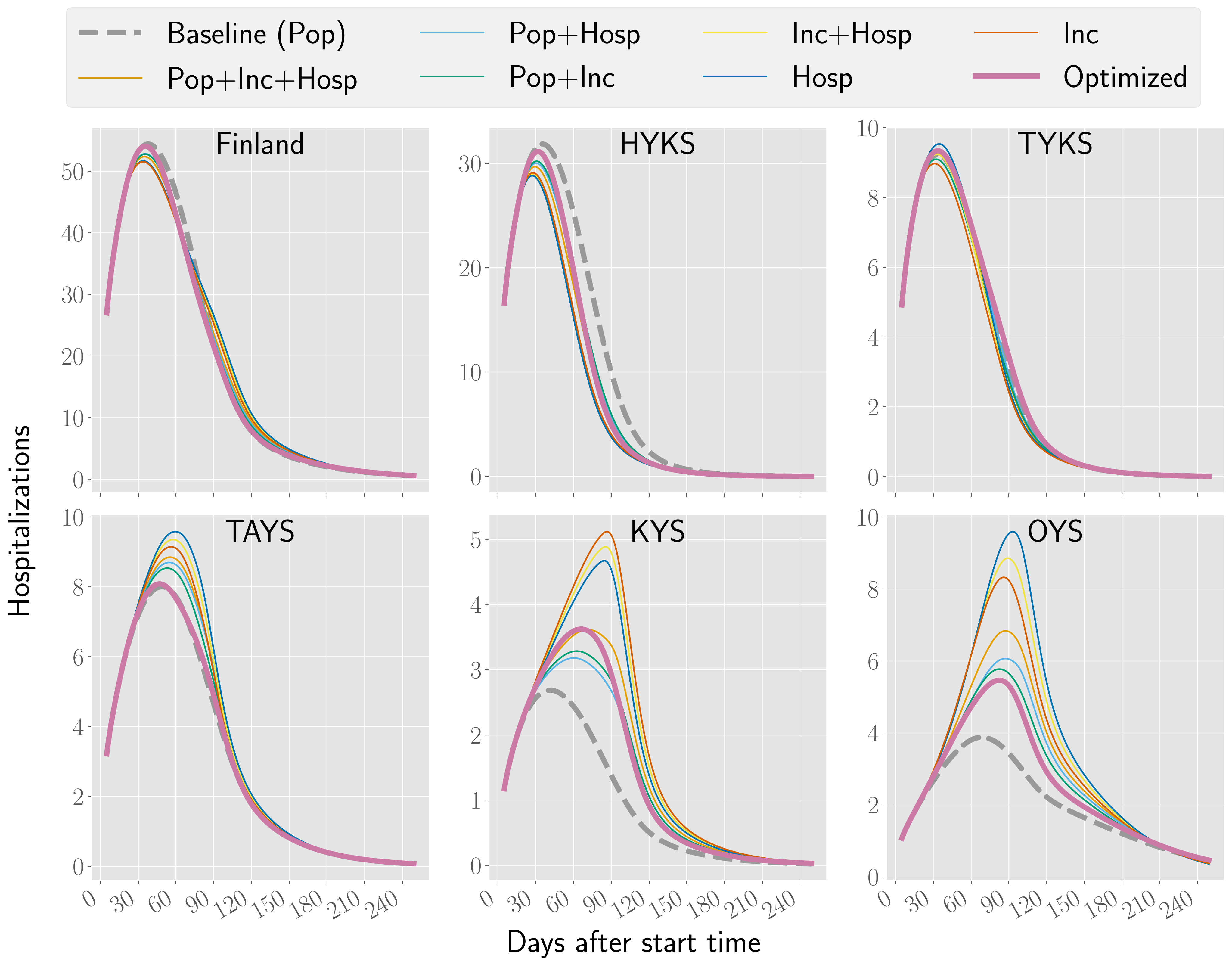}
         \label{fig:r1.25_tau0.0_metric_new hospitalizations}
     \end{subfigure}

     \begin{subfigure}[b]{0.3\paperheight}
         \centering
         \includegraphics[width=\textwidth]{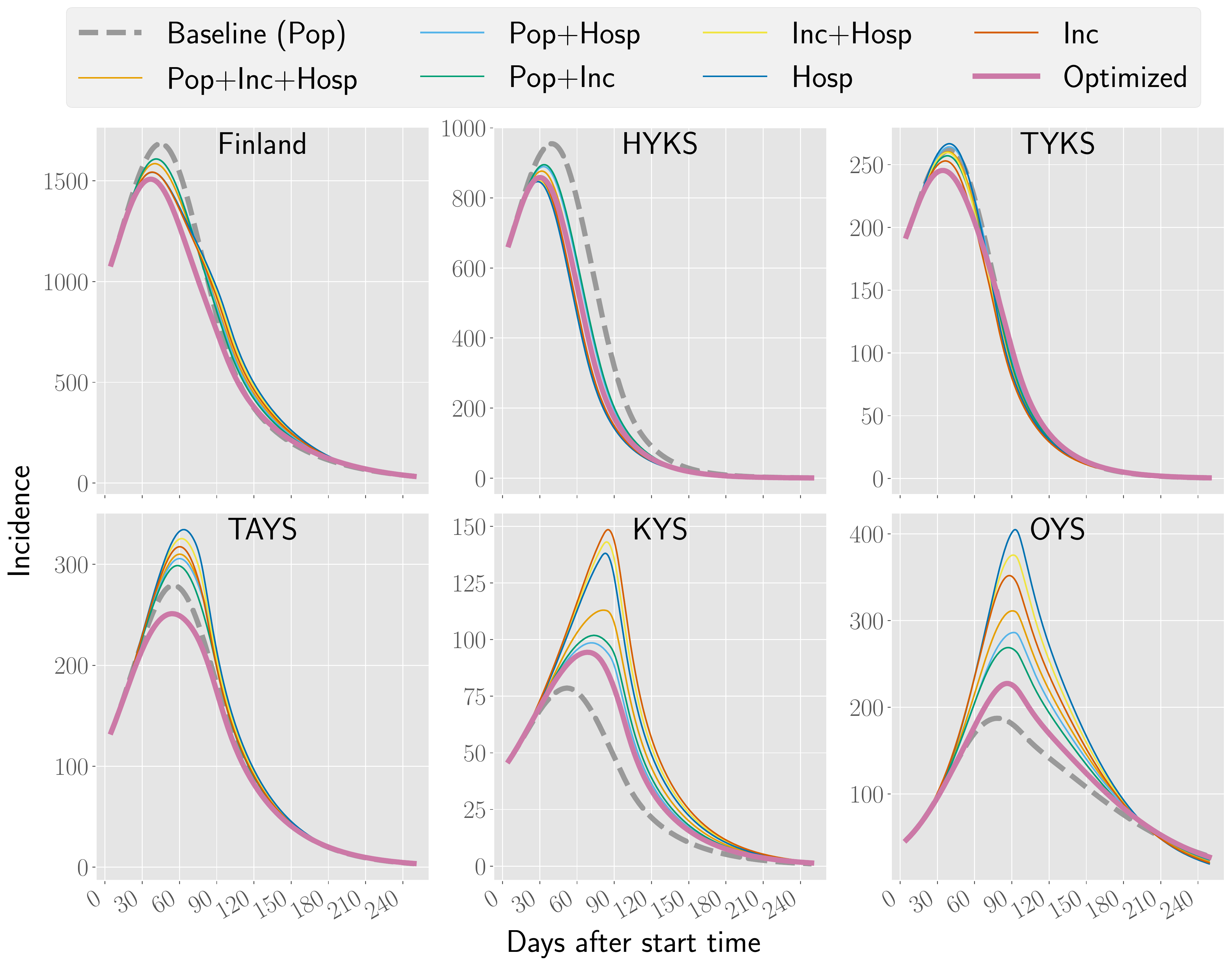}
         \label{fig:r1.25_tau0.0_metric_incidence}
     \end{subfigure}

     \caption{Different metrics in Finland and the five hospital catchment areas included here
      for all the vaccination strategies. For these scenarios, 
      the basic reproduction number $\Reff = 1.25$ and the mobility value $\tau = 0.0$.}
    \end{figure}

    \begin{figure}[p]
     \centering
     \begin{subfigure}[b]{0.3\paperheight}
         \centering
         \includegraphics[width=\textwidth]{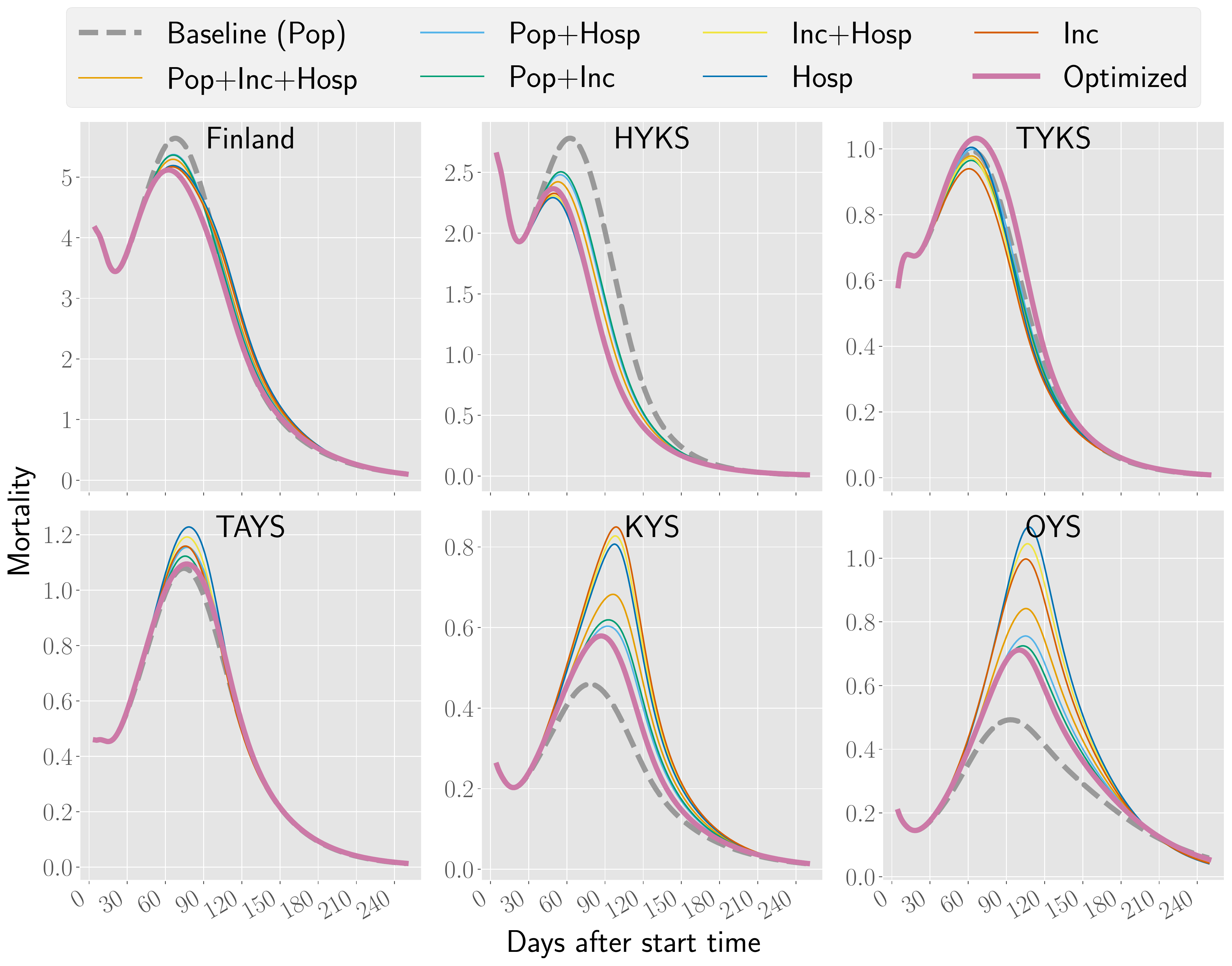}
         \label{fig:r1.25_tau0.5_metric_mortality}
     \end{subfigure}
     
     \begin{subfigure}[b]{0.3\paperheight}
         \centering
         \includegraphics[width=\textwidth]{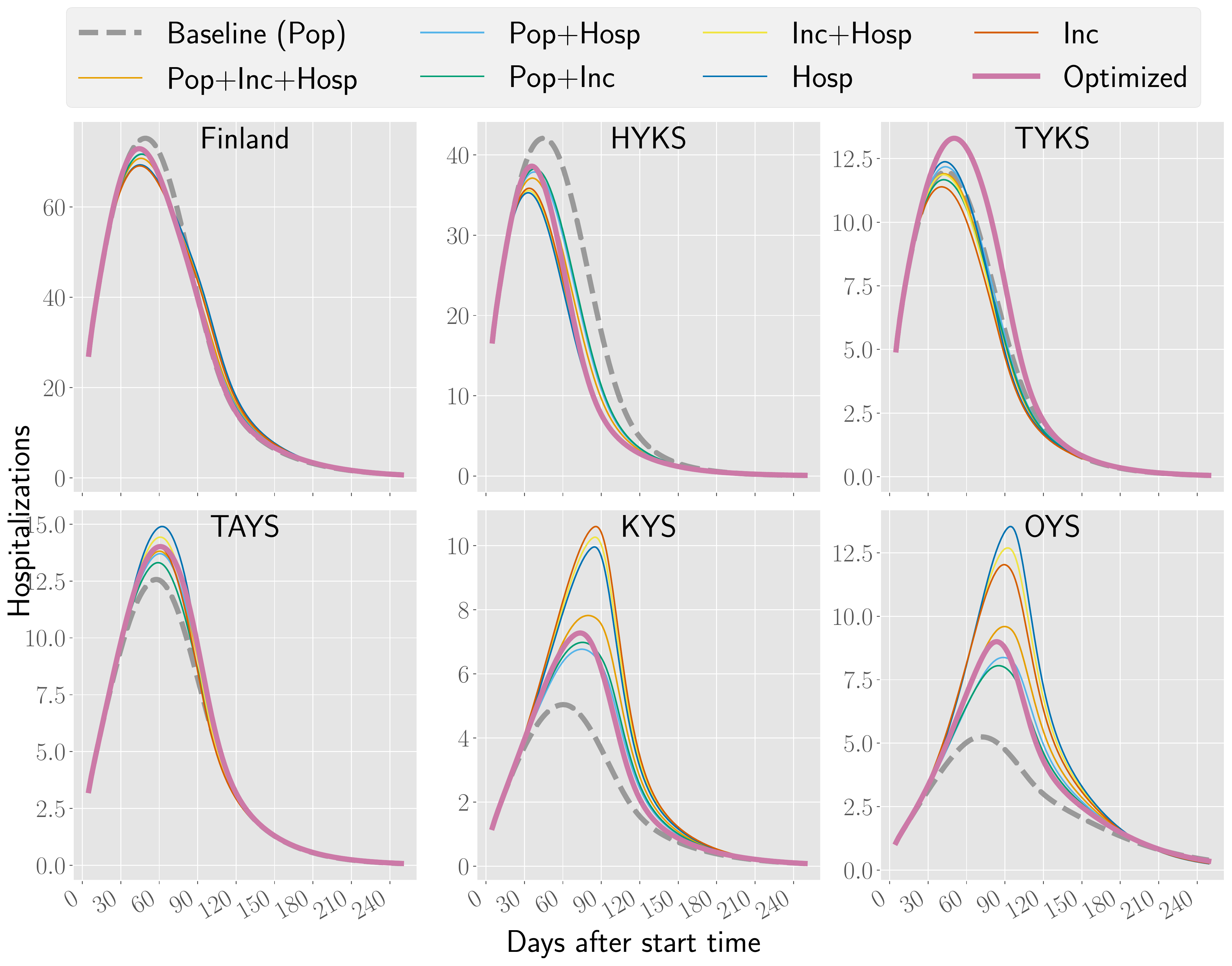}
         \label{fig:r1.25_tau0.5_metric_new hospitalizations}
     \end{subfigure}

     \begin{subfigure}[b]{0.3\paperheight}
         \centering
         \includegraphics[width=\textwidth]{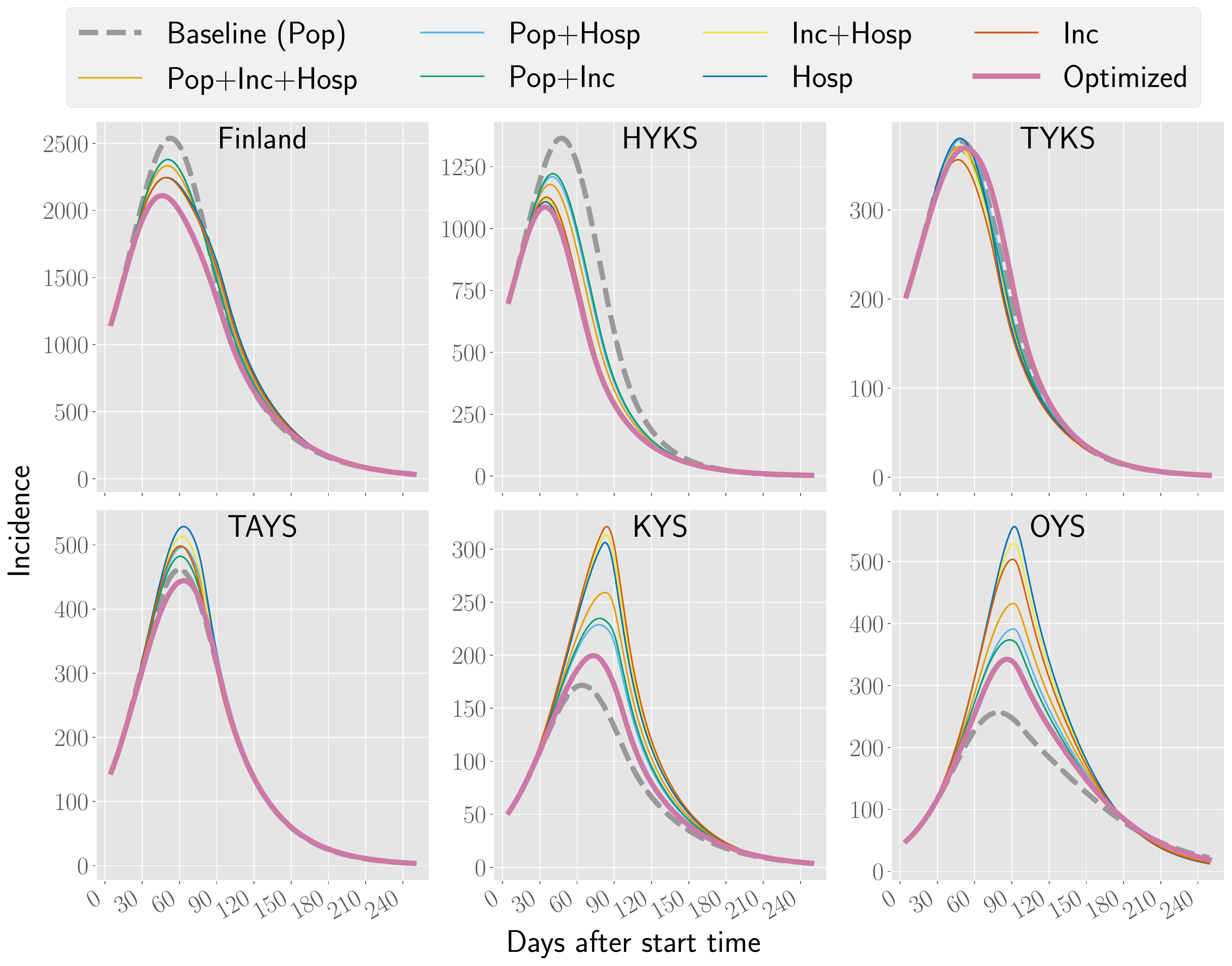}
         \label{fig:r1.25_tau0.5_metric_incidence}
     \end{subfigure}

     \caption{Different metrics in Finland and the five hospital catchment areas included here
      for all the vaccination strategies. For these scenarios, 
      the basic reproduction number $\Reff = 1.25$ and the mobility value $\tau = 0.5$.}
    \end{figure}

    \begin{figure}[p]
     \centering
     \begin{subfigure}[b]{0.3\paperheight}
         \centering
         \includegraphics[width=\textwidth]{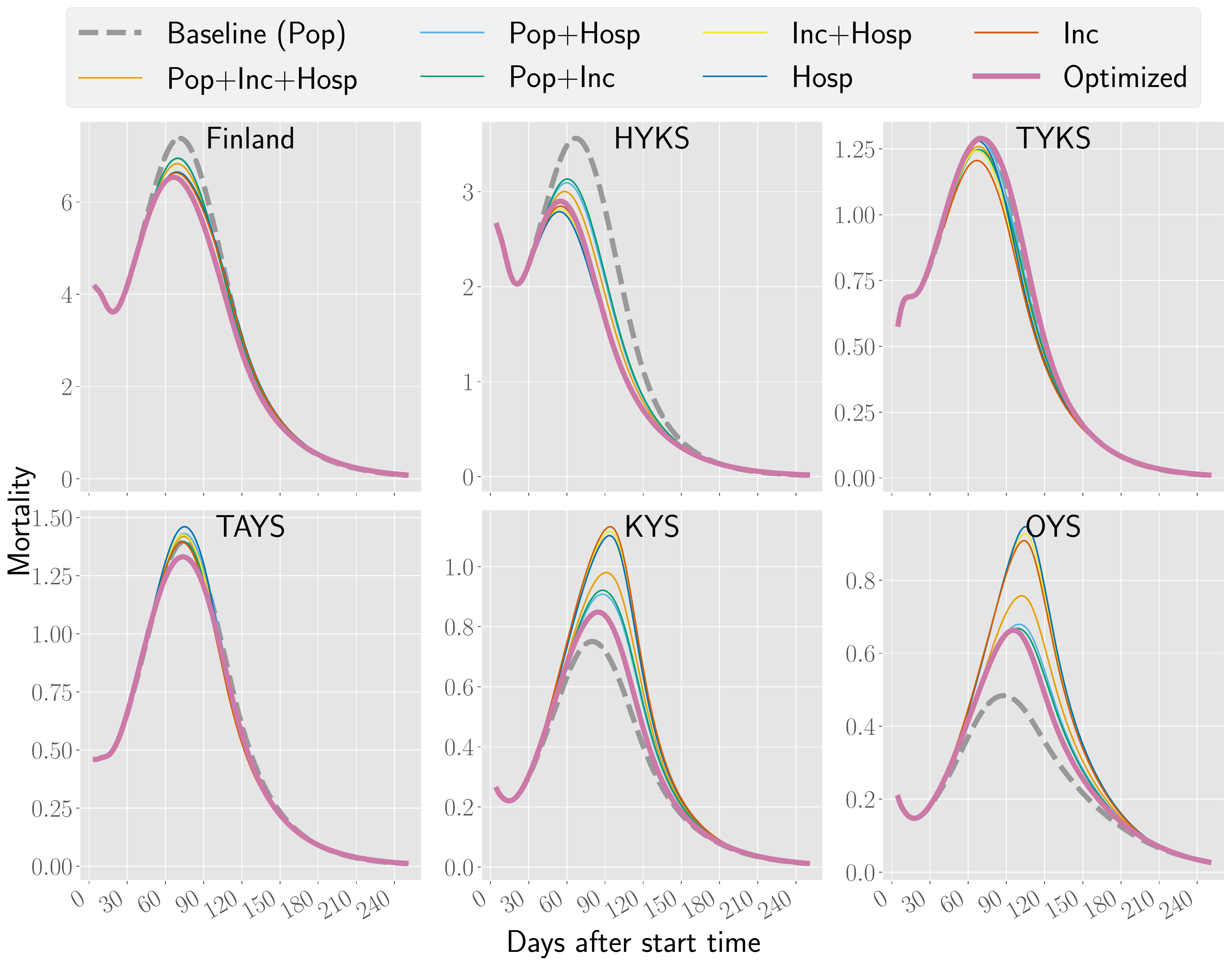}
         \label{fig:r1.25_tau1.0_metric_mortality}
     \end{subfigure}
     
     \begin{subfigure}[b]{0.3\paperheight}
         \centering
         \includegraphics[width=\textwidth]{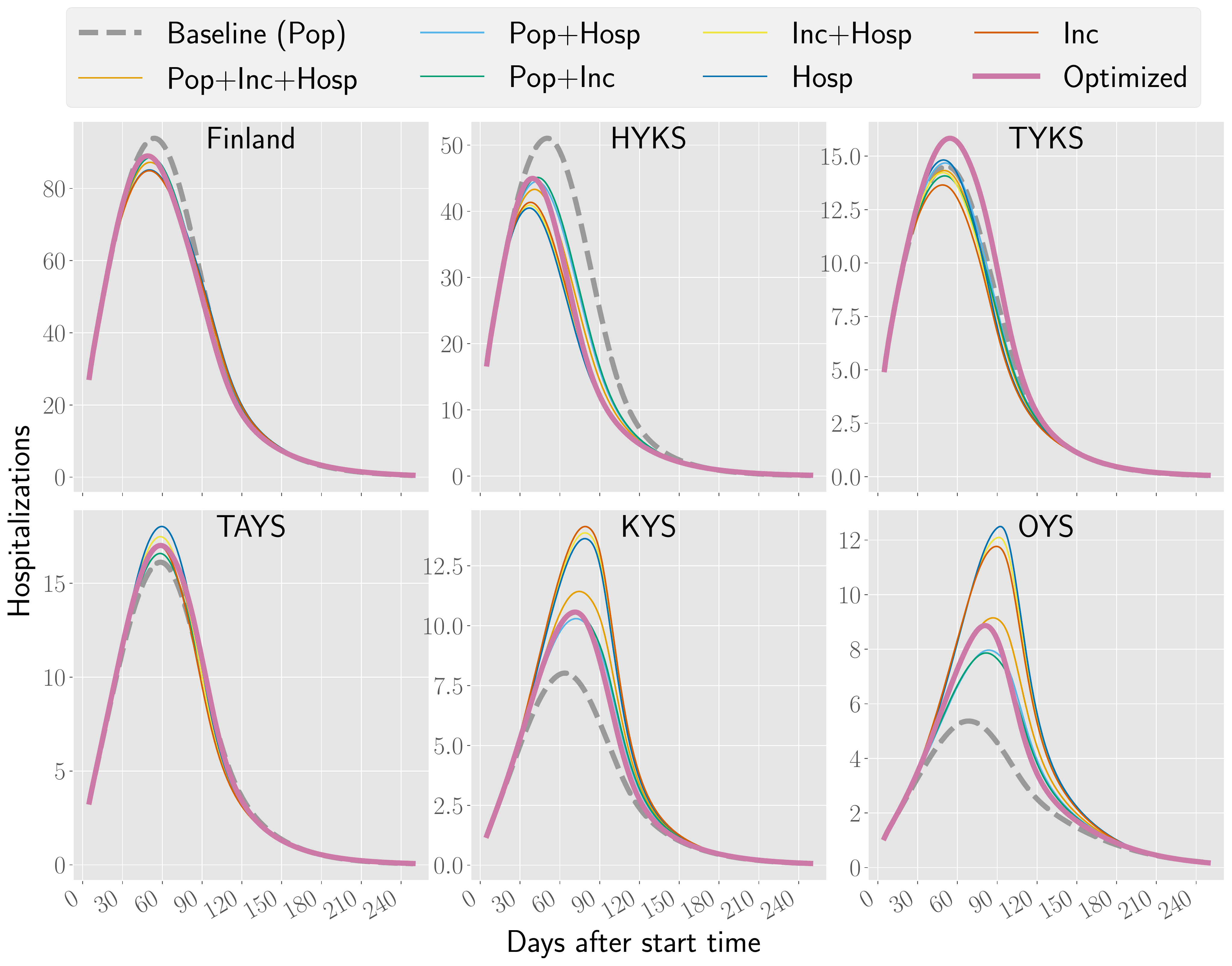}
         \label{fig:r1.25_tau1.0_metric_new hospitalizations}
     \end{subfigure}

     \begin{subfigure}[b]{0.3\paperheight}
         \centering
         \includegraphics[width=\textwidth]{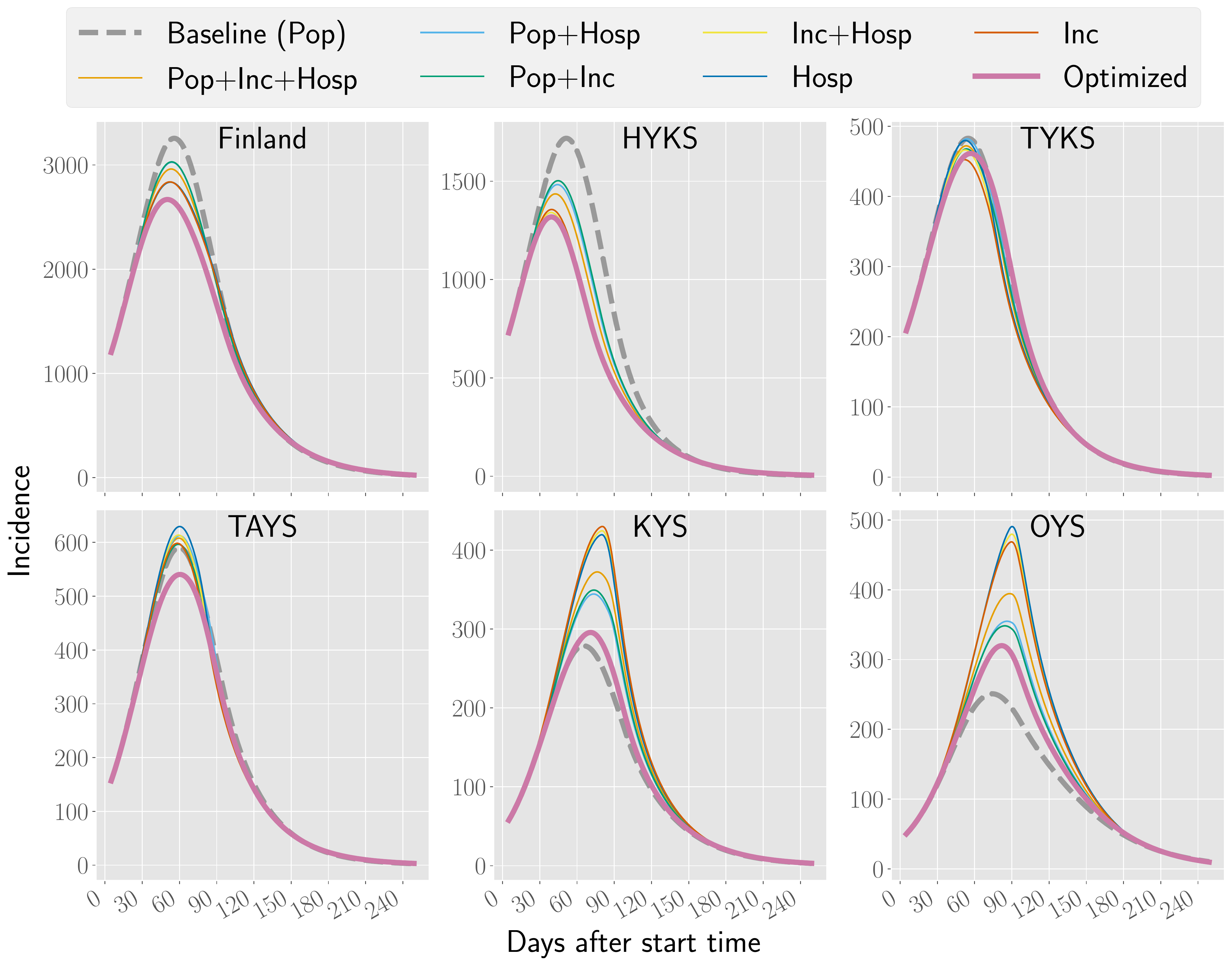}
         \label{fig:r1.25_tau1.0_metric_incidence}
     \end{subfigure}

     \caption{Different metrics in Finland and the five hospital catchment areas included here
      for all the vaccination strategies. For these scenarios, 
      the basic reproduction number $\Reff = 1.25$ and the mobility value $\tau = 1.0$.}
    \end{figure}

    \begin{figure}[p]
     \centering
     \begin{subfigure}[b]{0.3\paperheight}
         \centering
         \includegraphics[width=\textwidth]{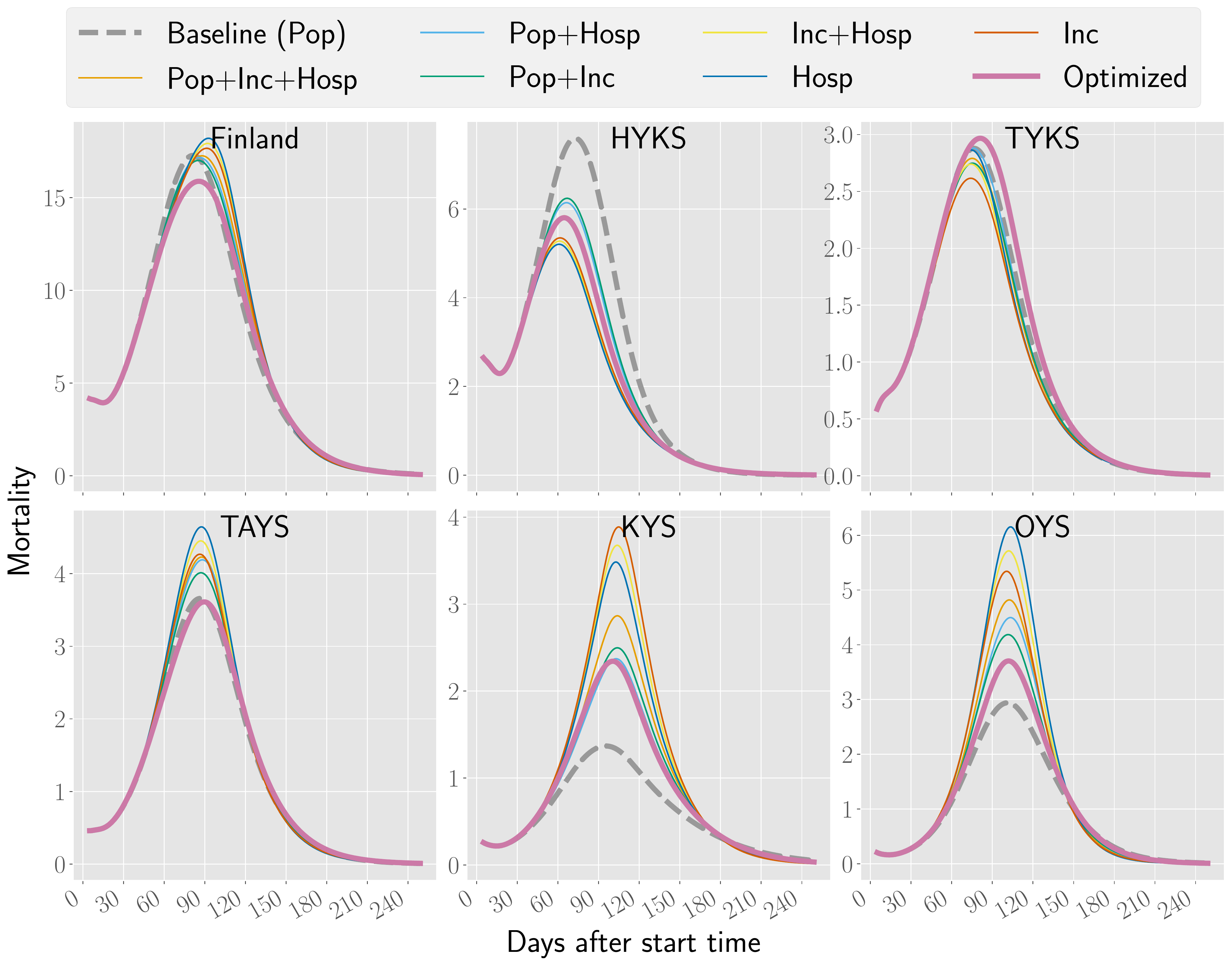}
         \label{fig:r1.5_tau0.0_metric_mortality}
     \end{subfigure}
     
     \begin{subfigure}[b]{0.3\paperheight}
         \centering
         \includegraphics[width=\textwidth]{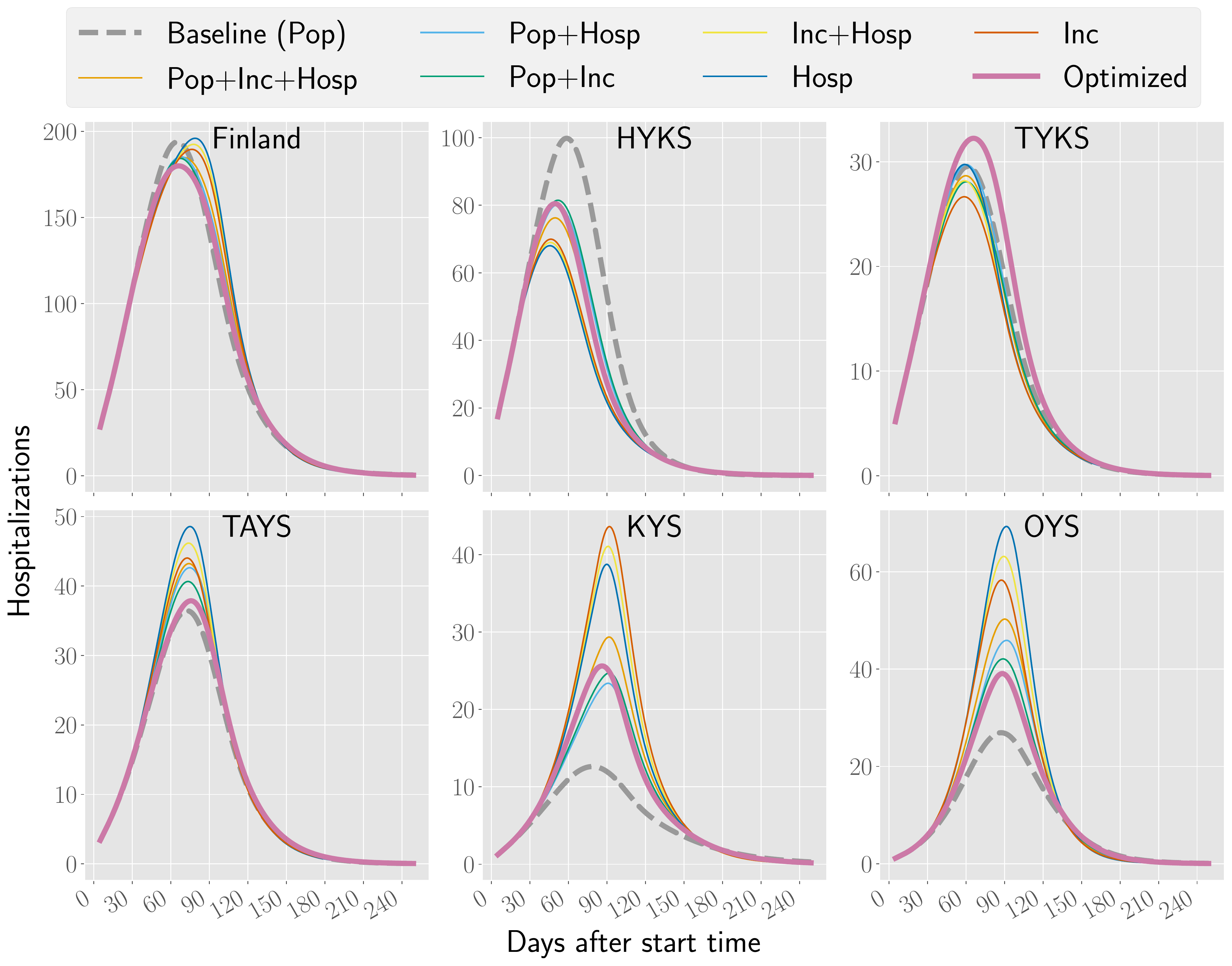}
         \label{fig:r1.5_tau0.0_metric_new hospitalizations}
     \end{subfigure}

     \begin{subfigure}[b]{0.3\paperheight}
         \centering
         \includegraphics[width=\textwidth]{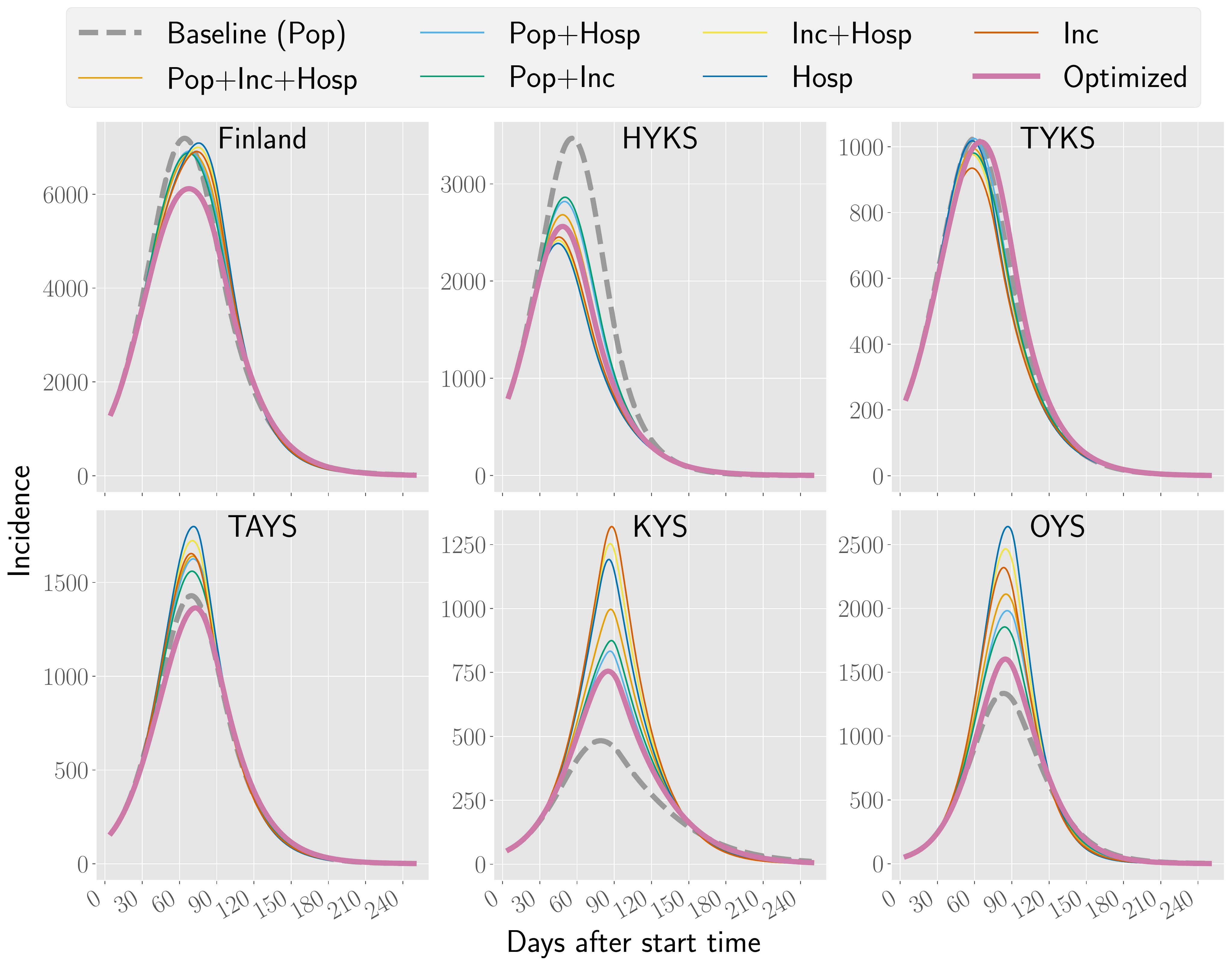}
         \label{fig:r1.5_tau0.0_metric_incidence}
     \end{subfigure}

     \caption{Different metrics in Finland and the five hospital catchment areas included here
      for all the vaccination strategies. For these scenarios, 
      the basic reproduction number $\Reff = 1.5$ and the mobility value $\tau = 0.0$.}
    \end{figure}

    \begin{figure}[p]
     \centering
     \begin{subfigure}[b]{0.3\paperheight}
         \centering
         \includegraphics[width=\textwidth]{img/r1.5_tau0.5_metric_mortality.pdf}
         \label{fig:r1.5_tau0.5_metric_mortality}
     \end{subfigure}
     
     \begin{subfigure}[b]{0.3\paperheight}
         \centering
         \includegraphics[width=\textwidth]{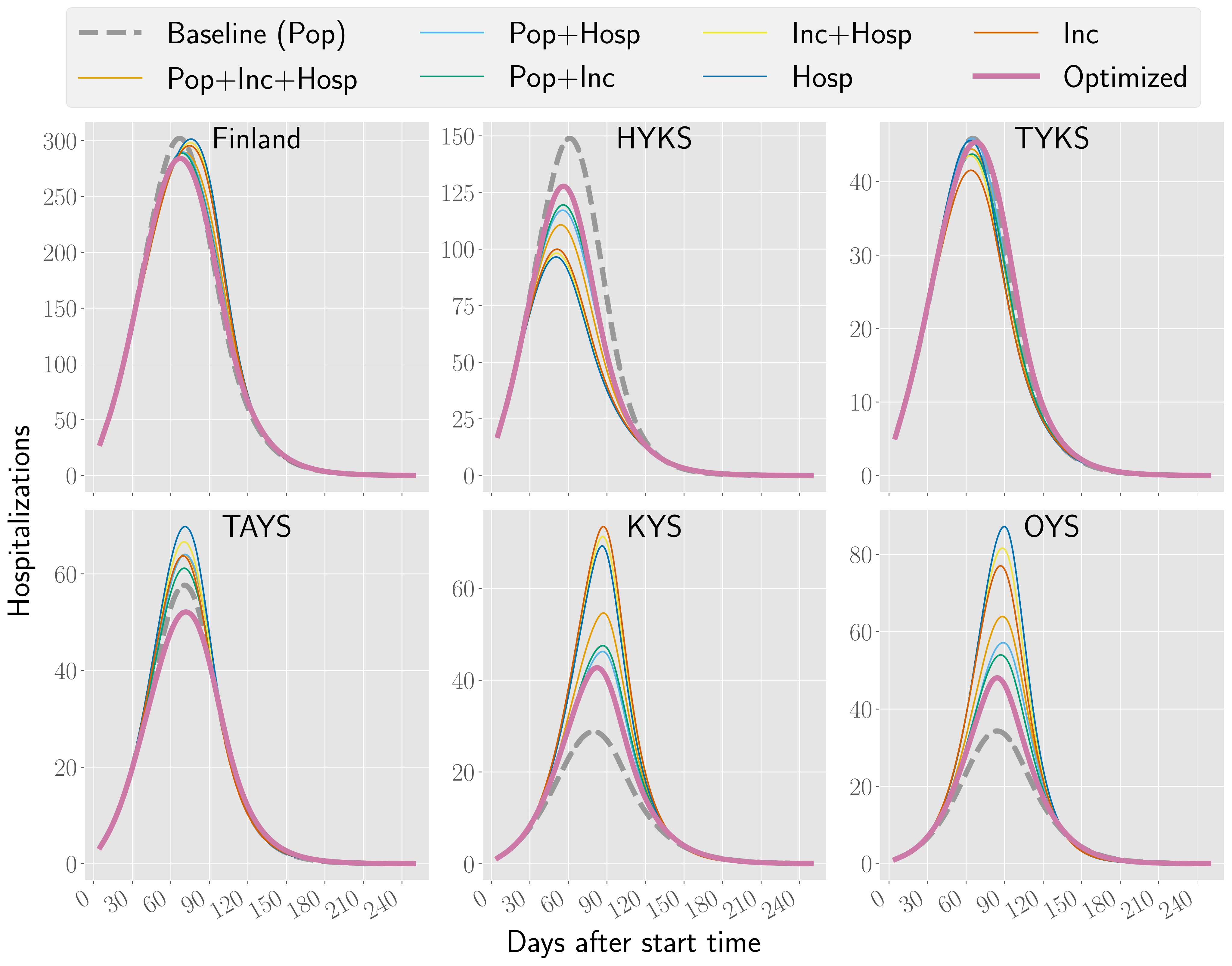}
         \label{fig:r1.5_tau0.5_metric_new hospitalizations}
     \end{subfigure}

     \begin{subfigure}[b]{0.3\paperheight}
         \centering
         \includegraphics[width=\textwidth]{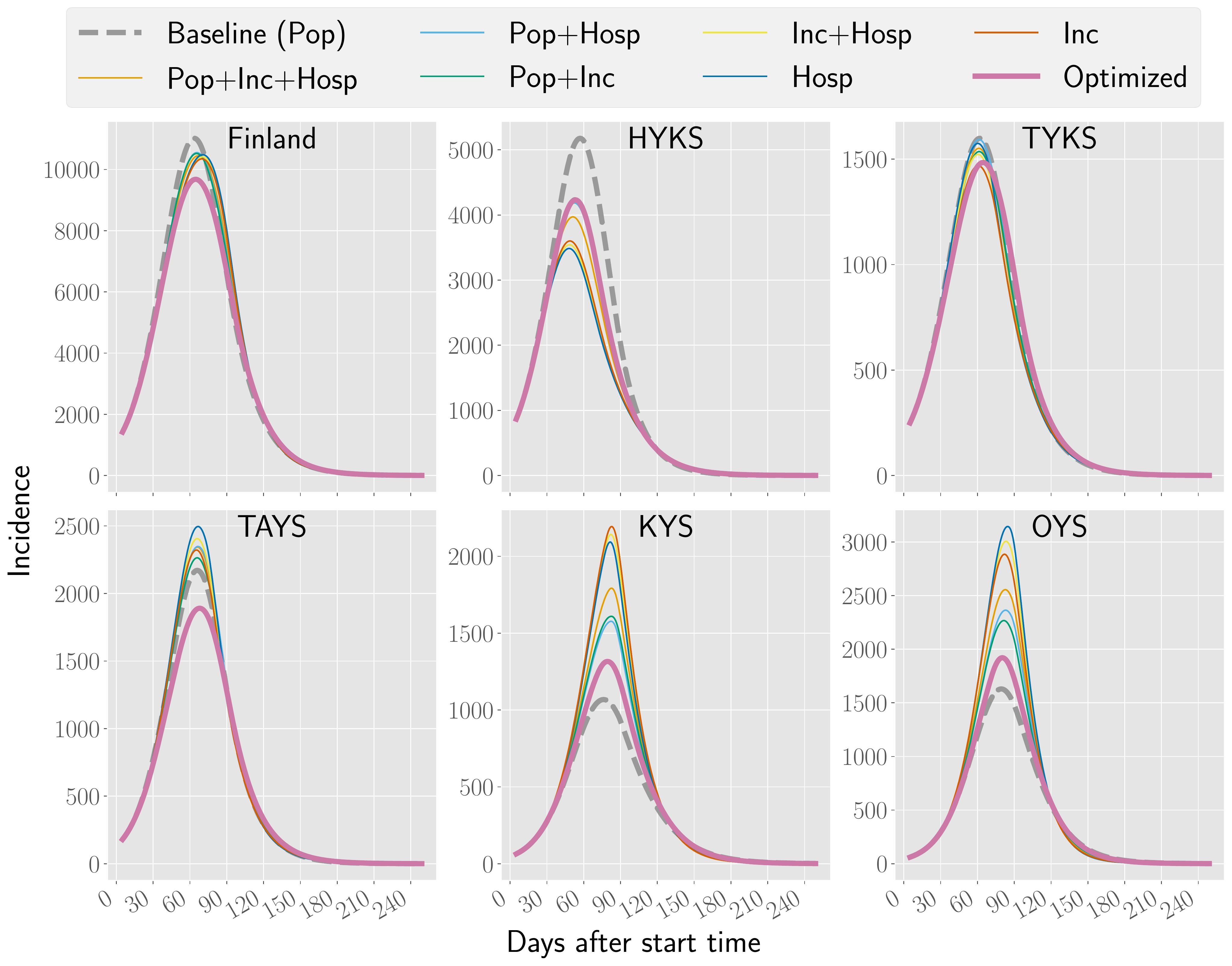}
         \label{fig:r1.5_tau0.5_metric_incidence}
     \end{subfigure}

     \caption{Different metrics in Finland and the five hospital catchment areas included here
      for all the vaccination strategies. For these scenarios, 
      the basic reproduction number $\Reff = 1.5$ and the mobility value $\tau = 0.5$.}
    \end{figure}

    \begin{figure}[p]
     \centering
     \begin{subfigure}[b]{0.3\paperheight}
         \centering
         \includegraphics[width=\textwidth]{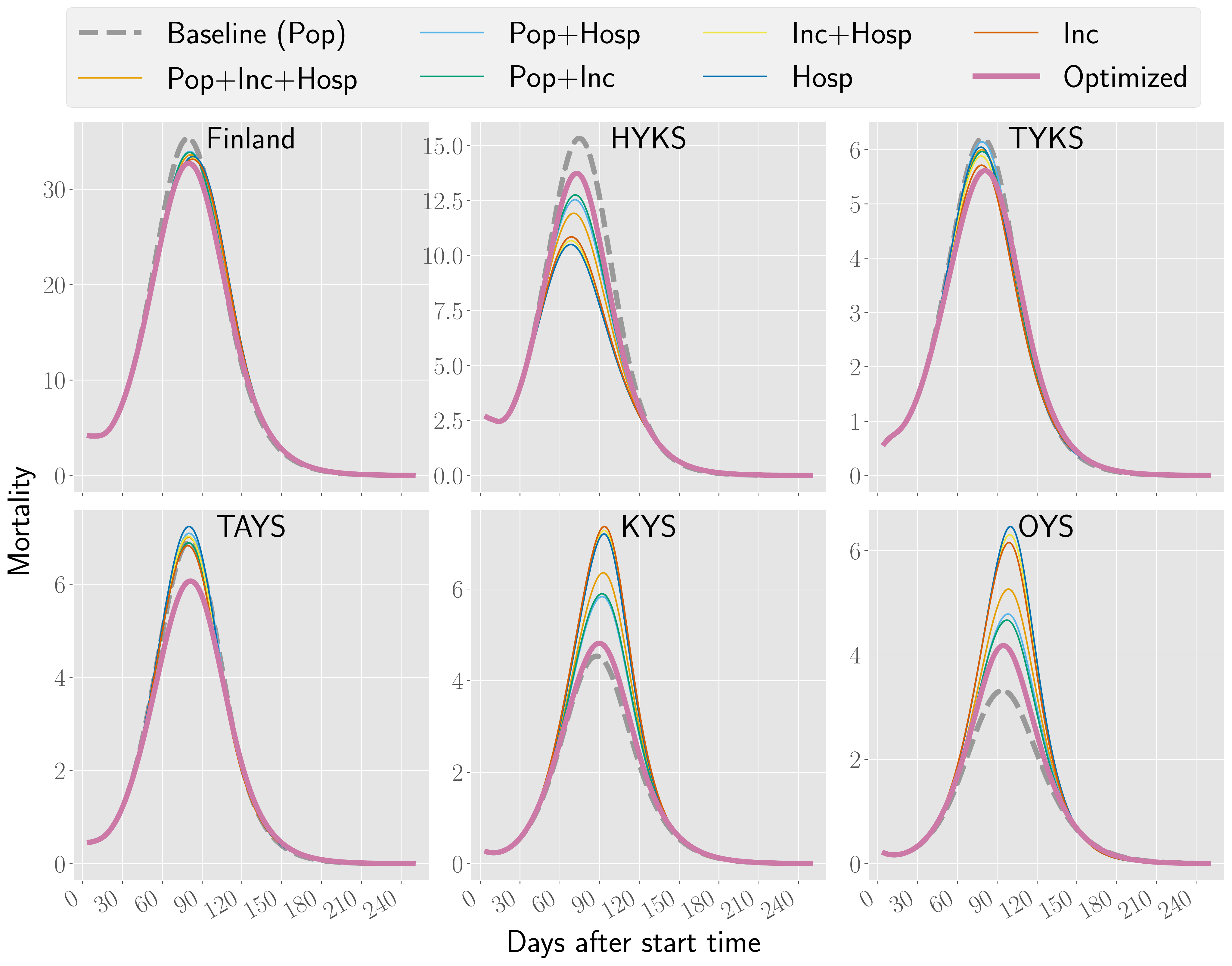}
         \label{fig:r1.5_tau1.0_metric_mortality}
     \end{subfigure}
     
     \begin{subfigure}[b]{0.3\paperheight}
         \centering
         \includegraphics[width=\textwidth]{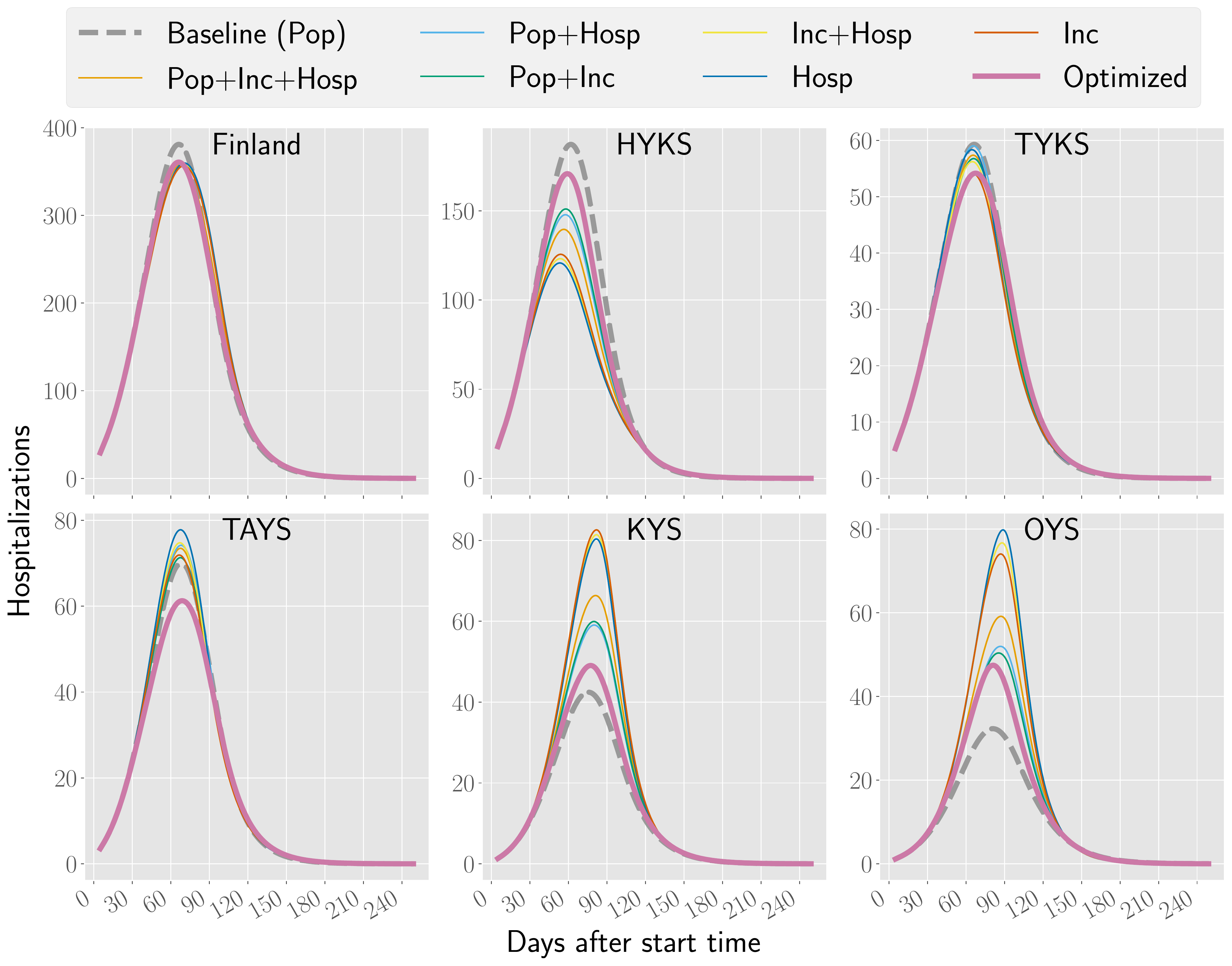}
         \label{fig:r1.5_tau1.0_metric_new hospitalizations}
     \end{subfigure}

     \begin{subfigure}[b]{0.3\paperheight}
         \centering
         \includegraphics[width=\textwidth]{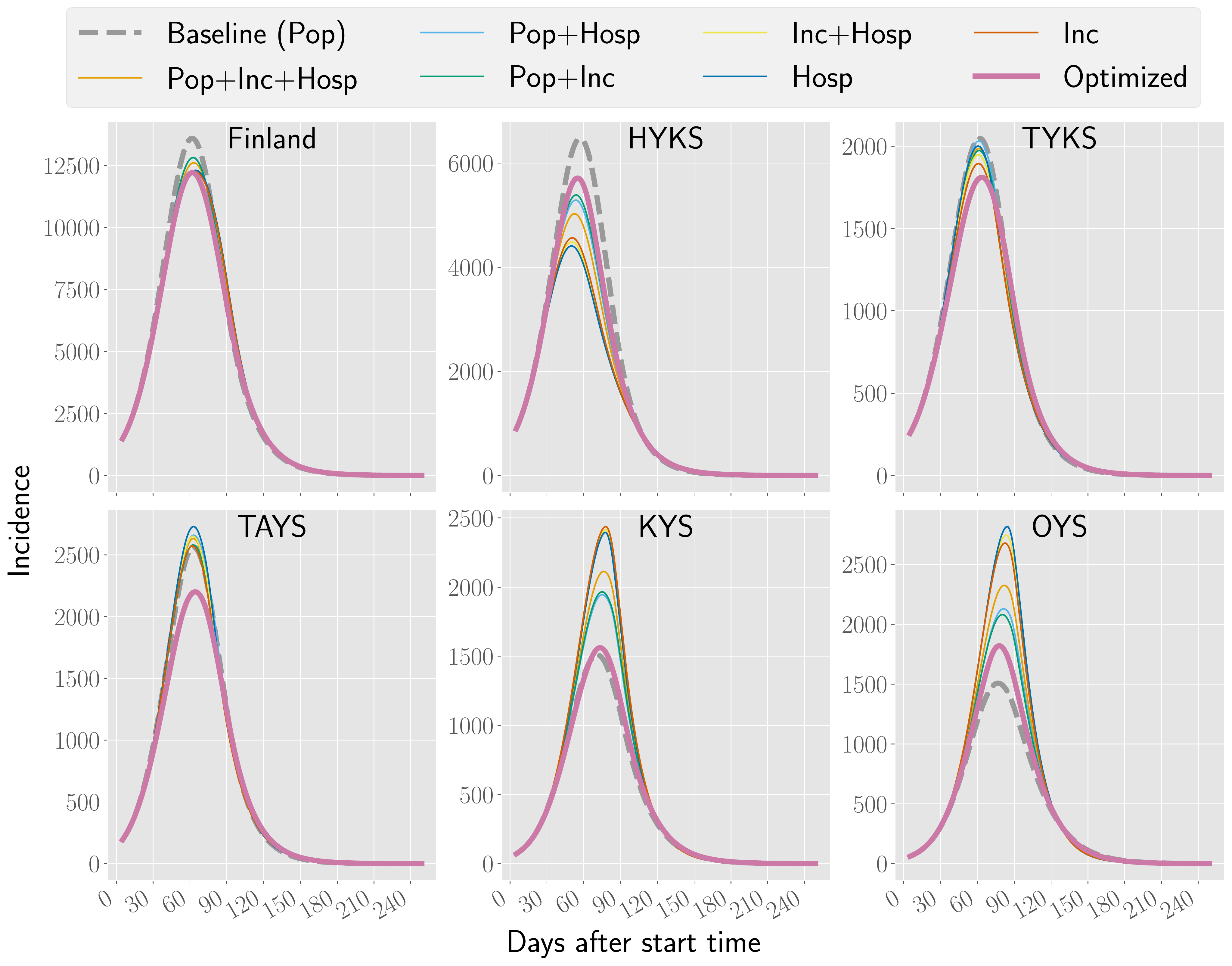}
         \label{fig:r1.5_tau1.0_metric_incidence}
     \end{subfigure}

     \caption{Different metrics in Finland and the five hospital catchment areas included here
      for all the vaccination strategies. For these scenarios, 
      the basic reproduction number $\Reff = 1.5$ and the mobility value $\tau = 1.0$.}
    \end{figure}

\begin{figure}[h]
    \centering
    \includegraphics[width=\textwidth]{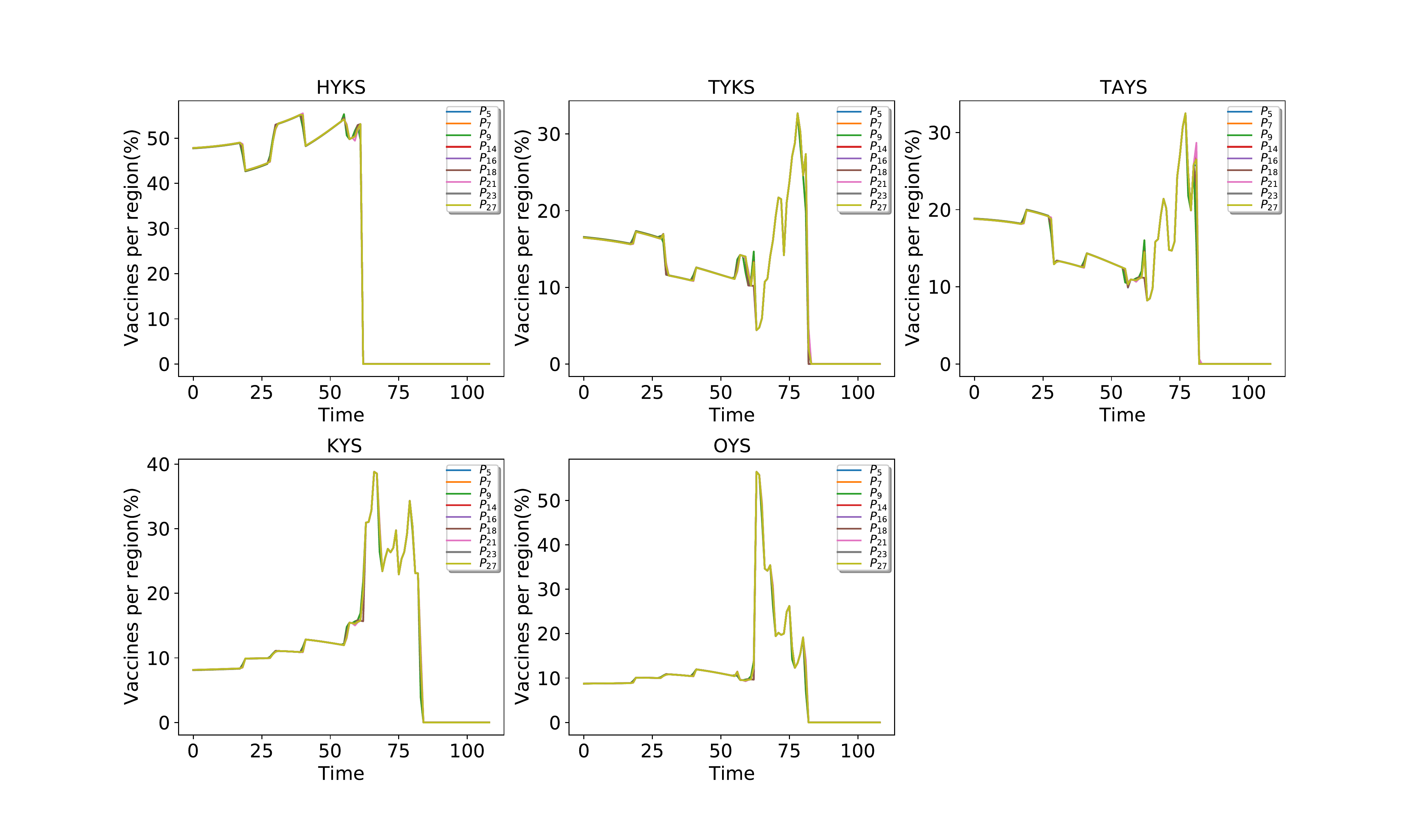}
    \caption{Percentage of vaccine doses allocated by the optimized strategy to
regions for different values of $e, \omega$ and $\pi$.}
    \label{fig:vac_erva}
\end{figure}

\begin{figure}[h]
    \centering
    \includegraphics[width=\textwidth]{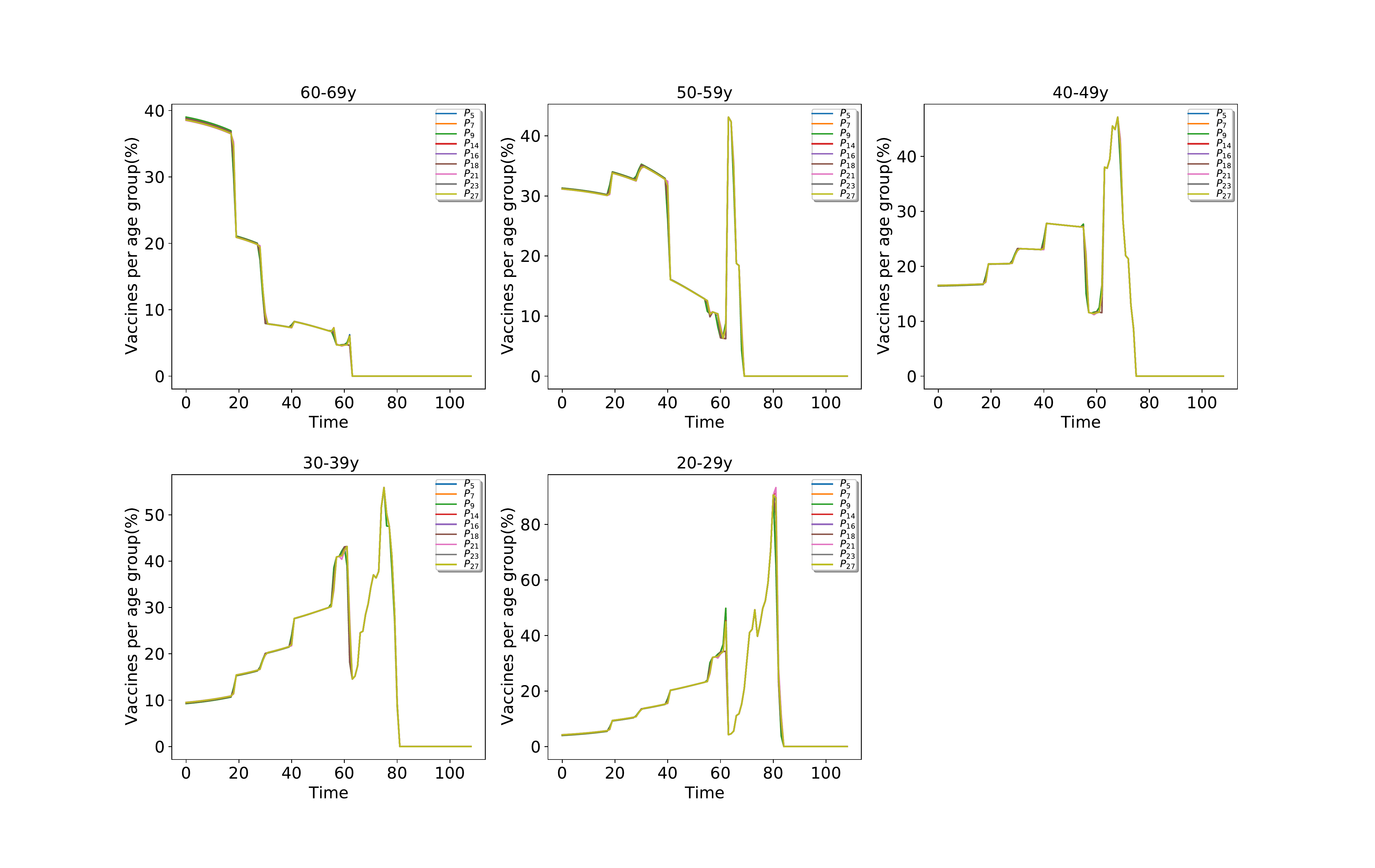}
    \caption{Percentage of vaccine doses allocated by the optimized strategy to
age groups for different values of $e, \omega$ and $\pi$.  Each $P_i, i\in\{5, 7, 9, 14, 16, 18, 21, 23, 27 \}$ represents a different combination of $e, \omega$ and $\pi$ taken from (3.1) in supplementary material.}
    \label{fig:vac_age}
\end{figure}

\begin{figure}[h]
    \centering
    \includegraphics[width=0.7\textwidth]{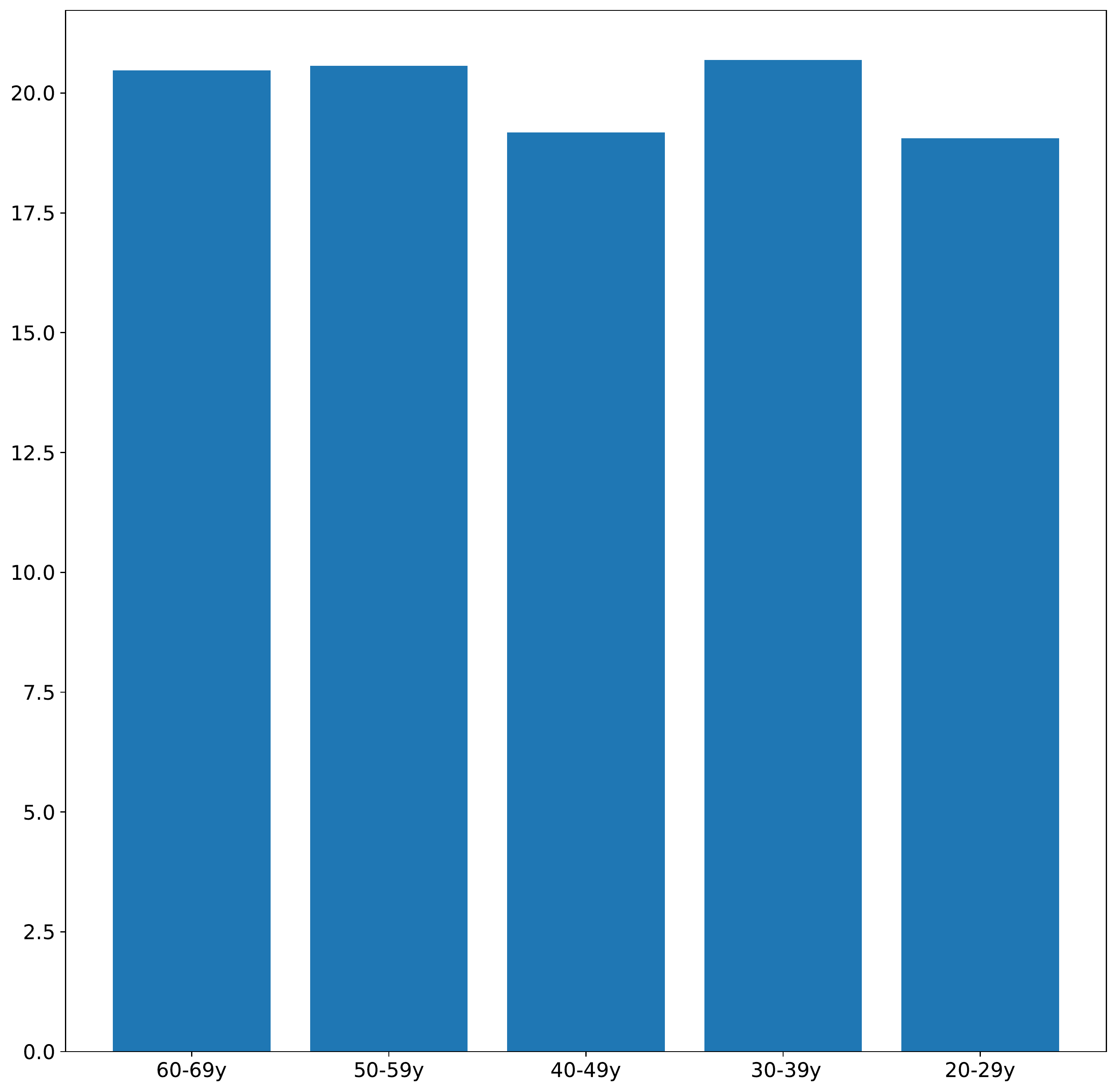}
    \caption{Percentage of vaccines which each age group would receive proportionally to the age group size.  Each $P_i, i\in\{5, 7, 9, 14, 16, 18, 21, 23, 27 \}$ represents a different combination of $e, \omega$ and $\pi$ taken from (3.1) in supplementary material.}
    \label{fig:vac_age}
\end{figure}




\end{document}